%% file: 0_Main.tex
\documentclass{article}

\usepackage[nonatbib,final]{neurips_2024}

\usepackage[utf8]{inputenc} % allow utf-8 input
\usepackage[T1]{fontenc}    % use 8-bit T1 fonts
\usepackage{hyperref}       % hyperlinks
\usepackage{url}            % simple URL typesetting
\usepackage{booktabs}       % professional-quality tables
\usepackage{amsfonts}       % blackboard math symbols
\usepackage{amsmath}
\usepackage{amssymb}
\usepackage{nicefrac}       % compact symbols for 1/2, etc.
\usepackage{microtype}      % microtypography
\usepackage{xcolor}         % colors
\usepackage{graphicx}  % 加载 graphicx 包
\usepackage{array}   % 表格格式支持
\usepackage{tabularx} % 自动调整列宽
\usepackage{multirow} % 合并行
\usepackage{xcolor}   % 颜色支持
\usepackage{subcaption}    % 用于子图环境
\usepackage{calc}          % 用于计算尺寸（可选）
\usepackage{tabularx}
\usepackage{makecell}

\title{AgentSociety: Large-Scale Simulation of\\LLM-Driven Generative Agents Advances\\Understanding of Human Behaviors and Society}

\author{%
 Jinghua Piao$^{1\dagger}$\\
  \And
  Yuwei Yan$^{1\dagger}$\\
  \And
  Jun Zhang$^{1\dagger}$\\
  \And
  Nian Li$^{1}$\\
  \And
  Junbo Yan$^{1}$\\
  \And
  Xiaochong Lan$^{1}$\\
  \And
  Zhihong Lu$^{1}$\\
  \And
  Zhiheng Zheng$^{1}$\\
 \And
  Jing Yi Wang$^{1}$\\
  \And
  Di Zhou$^{2}$\\
  \And
  Chen Gao$^{3}$\\
  \And
  Fengli Xu$^{1}$\\
  \And
  Fang Zhang$^{4*}$\\
  \And
  Ke Rong$^{2*}$\\
  \And
  Jun Su$^{4*}$\\
  \And
  Yong Li$^{1*}$\\
}

\begin{document}

\maketitle

\begin{abstract}
Understanding human behavior and society is a central focus in social sciences, with the rise of generative social science marking a significant paradigmatic shift. By leveraging bottom-up simulations, it replaces costly and logistically challenging traditional experiments with scalable, replicable, and systematic computational approaches for studying complex social dynamics. Recent advances in large language models (LLMs) have further transformed this research paradigm, enabling the creation of human-like generative social agents and realistic simulacra of society. In this paper, we propose AgentSociety, a large-scale social simulator that integrates LLM-driven agents, a realistic societal environment, and a powerful large-scale simulation engine. Based on the proposed simulator, we generate social lives for over 10k agents, simulating their 5 million interactions both among agents and between agents and their environment. Furthermore, we explore the potential of AgentSociety as a testbed for computational social experiments, focusing on five key social issues: polarization, the spread of inflammatory messages, the effects of universal basic income policies, the impact of external shocks such as hurricanes, and urban sustainability. These five issues serve as valuable cases for assessing AgentSociety's support for typical research methods -- such as surveys, interviews, and interventions -- as well as for investigating the patterns, causes, and underlying mechanisms of social issues. The alignment between AgentSociety's outcomes and real-world experimental results not only demonstrates its ability to capture human behaviors and their underlying mechanisms, but also underscores its potential as an important platform for social scientists and policymakers. 

\end{abstract}
\footnote{\\
$^1$ Department of Electronic Engineering, Beijing National Research Center for Information Science and Technology (BNRist), Tsinghua University\\
$^2$ Institute of Economics, School of Social Sciences, Tsinghua University\\
$^3$ BNRist, Tsinghua University\\
$^4$ School of Public Policy and Management, Tsinghua University\\
$^\dagger$ These authors contributed equally to this work. \\
$^*$ Corresponding authors. E-mail: liyong07@tsinghua.edu.cn.
}
\clearpage

\input{1_Introduction}

\input{1_Overview}
\input{3_Agents}

\input{4_Environment}
\input{5_Archtecture}

\input{6_Experiments}

\input{2_Related}
\input{7_Conclusion}

\bibliographystyle{plain}
\bibliography{sn-bibliography}% common bib file

\end{document}

%% file: 1_Introduction.tex
\newcommand{\rvs}[1]{{\textcolor{black}{#1}}}
\section{Introduction}

% 如何理解世界 understand，explain， predict, generate
% \textit{``What I cannot create, I do not understand''} -- Richard Feynman

Over the past decades, researchers across various fields -- spanning social science, physical science, and computational science -- have made significant efforts to understand the functioning and development of society along two dimensions: explanation and prediction~\cite{hofman2021integrating,lazer2009computational,lazer2020computational}. Explanation seeks to identify the causal mechanisms and underlying factors that drive observed social patterns, aiming to offer a deeper understanding of why certain outcomes occur in society~\cite{hofman2021integrating,hedstrom2010causal}. \rvs{Gaining such an understanding often requires conducting social experiments,  which can be costly to implement and pose substantial practical and ethical challenges.} On the other hand, prediction focuses on using data to forecast future events or emergent behaviors, often without delving into the causal factors, but instead concentrating on the accuracy of anticipating future trends~\cite{breiman2001statistical,hofman2021integrating}. The framework that combines explanation and prediction methods, forms the foundation of \textit{computational social science}~\cite{hofman2021integrating}. However, as the physicist Richard Feynman famously stated, \textit{``What I cannot create, I do not understand''}, suggesting that true understanding goes beyond merely observing, explaining, or predicting human behavior~\cite{epstein2012generative,epstein1999agent}; instead, it requires the ability to generate the systems we study~\cite{epstein2012generative,epstein1999agent}. In this vein, a new paradigm of ``generative social science'' is rapidly emerging, which emphasizes the bottom-up simulation of social systems to gain in-depth insights into their underlying mechanisms and predict future outcomes~\cite{epstein2012generative,epstein1999agent}. A well-established method in this paradigm is agent-based modeling, which aims to model complex social dynamics by simulating the actions and interactions of agents~\cite{epstein1999agent,macal2005tutorial,wilensky2015introduction}. This method, \rvs{compensating for the limitations of social experiments}, is widely applied in studies across social science~\cite{gilbert2000build,berry2002adaptive,gao2023s}, political science~\cite{de2014agent,laver2011party,li2024econagent}, economics~\cite{arthur2006out,feng2012linking}, and other interdisciplinary fields~\cite{an2021challenges,gao2024large,piao2023human,piao2025emergence}, advancing their understanding of human behaviors and society through various simulations. However, the broader impact of these studies has been hindered by the same long-standing issue: to what extent can these simulations authentically replicate the complexities of real human society?

Indeed, the authenticity of these simulations depends on to what extent the most basic components -- i.e., the agents -- behave like humans. However, most existing agents, driven by rules~\cite{epstein1996growing}, equations~\cite{helbing1995social}, or even machine learning models~\cite{zheng2022ai}, are limited in their ability to generate human-like behaviors. For example, when simulating opinion dynamics, people's opinions are often represented as scalars or vectors, and their interactions as equations~\cite{baumann2021emergence}. While this modeling approach offers valuable insights, it is still far from reality, as people typically communicate using natural language, rather than numeric values. Fortunately, recent advances in large language models (LLMs) have shown promise in creating human-like agents~\cite{gao2024large,wang2024survey,xi2023rise}. Numerous studies have pointed out that after being empowered by LLMs, these agents have generated human-like ``minds''~\cite{strachan2024testing,li2023camel,kosinski2024evaluating}. They not only possess basic cognition abilities, such as learning~\cite{xi2023rise,wang2024survey}, reasoning~\cite{wei2022chain}, and decision-making~\cite{li2024econagent,gao2024large}, but also demonstrate the capability to understand and predict the thoughts and intentions of others~\cite{kosinski2024evaluating,strachan2024testing}. Furthermore, beyond exploring these agents' minds, some researchers have investigated their potential to mimic human behaviors~\cite{li2024econagent,piao2025emergence,shao2024beyond,yan2024opencity,feng2024agentmove,horton2023large,gao2023s,park2023generative}. Their investigations have revealed that, through elaborate designs incorporating domain knowledge, these LLM-driven agents can generate social behaviors, such as mobility~\cite{shao2024beyond,yan2024opencity,feng2024agentmove}, employment~\cite{li2024econagent,horton2023large}, consumption~\cite{li2024econagent,horton2023large}, and social interactions~\cite{gao2023s,park2023generative}. While great efforts have been made to examine specific facets of these agents, simulating a comprehensive social being remains largely underexplored.

% more is different, 仅仅依靠单个智能体还是不行，需要多个智能体做social simulation。在他们的交互中evolve 出新的东西
As the famous sociologist George Herbert Mead stated, ``The self is something which has a development; it is not initially there, at birth, but arises in the process of social experience and activity.''~\cite{mead1934mind} Therefore, the mere incorporation of minds and behaviors into these generative agents is insufficient to create a social being; instead, social experience and activity, emerging from interactions with other agents and the environment, are crucial. Several recent studies have provided substantial empirical evidence supporting this point. Some have discovered that the collaboration of multiple agents can generate believable social organizing behaviors~\cite{park2023generative} and solve complex tasks~\cite{li2023camel,cheng2024sociodojo}. Moreover, as the number of agents further scales up, large-scale interactions among them can lead to the emergence of social norms and collectives~\cite{lai2024evolving,piao2025emergence}. Meanwhile, the environment not only serves as the ground for interactions among agents, but also provides critical feedback that guides their behaviors~\cite{al2024project,li2024econagent,wang2023voyager,zhu2023ghost}. For example, the widely-adopted gaming environment ``Minecraft'' provides feedback, such as crafting materials, tools, or resources, enabling agents to adapt their behaviors, solve tasks, and gain civilizational progression~\cite{wang2023voyager,zhu2023ghost,al2024project}. Overall, as highlighted by these studies, a scalable framework supporting large-scale interactions and a realistic environment is foundational to simulating a comprehensive social being and society. However, the current investigation of both remains limited.

To address the above gaps, we propose \textit{AgentSociety}, a large-scale social generative simulator that incorporates \textit{LLM-driven social generative agents}, \textit{a realistic societal environment}, and \textit{large-scale interactions} both among agents and between agents and the environment. Specifically, following social theories from a broad range of fields, including psychology~\cite{maslow1943theory,ajzen1991theory}, economics~\cite{christiano2005nominal} and behavioral sciencel~\cite{zipf1946p}, we first design a framework for LLM-driven social agents. These agents are endowed with human-like ``minds'', which include emotions, needs, motivations, and cognition of the external world. Their behaviors such as mobility, employment, consumption, and social interactions are dynamically driven by these internal mental states. Beyond individual agents, we construct a realistic societal environment that seamlessly integrates urban, social, and economic spaces, providing a rich foundation for agent interactions and self-evolution. At its core, society emerges from the bottom-up interactions among individuals, where agent-level interactions collectively give rise to complex social structures and phenomena. Recognizing that social systems exhibit emergent behaviors shaped by scale, we develop a large-scale social simulation engine equipped with distributed computing and an MQTT-powered high-performance messaging system. This enables simulations with up to 10k agents, each engaging in an average of 500 interactions per day, capturing the intricate dynamics of large-scale social systems. Based on the proposed large-scale social simulator, we successfully reproduce behaviors, outcomes, and patterns observed in five real-world social experiments, including polarization, inflammatory message spread, the effects of universal basic income policies, the impact of external shocks like hurricanes, and urban sustainability. These experiments not only cover social research methods, such as surveys, interviews, and interventions, but also demonstrate the simulator's ability to replicate social dynamics, unlocking new possibilities for social scientists and policymakers. Overall, AgentSociety marks a paradigm shift in AI for social science, enabling large-scale, high-fidelity simulations that overcome traditional experimental limitations in costs, scalability, and feasibility. By leveraging LLM-driven social generative agents, it facilitates deeper analysis, prediction, and intervention in complex social systems, laying the foundation for computational social science 2.0.

% These experiments not only encompass fundamental social research methods, such as surveys, interviews, and interventions, but also showcase the simulator’s ability to replicate social dynamics, opening up new possibilities for social scientists and policymakers.

% This not only demonstrates the capability of the proposed simulator as an authentic replica of social beings and society, but also underscores its potential applications for social scientists and policymakers.
% 做了四个层次的贡献，明天讨论
% 【社会人智能体】 llm驱动的类人模拟 心智->多种社会行为
% 【真实的城市环境】提供真实的城市环境 
% 【大规模交互】 大规模多智能体之间的交互，more is different，涌现
% 【范式改变】应用，改变社会学研究/研究、理解社会社会的式，从explanation，prediction，到generation

%% file: 1_Overview.tex
\section{AgentSociety: Design and Overview}

% 复杂系统的思路 -- 需求
Society is a complex system, characterized by large-scale interactions among individuals with diverse social behaviors, whose nonlinear dynamics often give rise to emergent phenomena and unpredictable collective behaviors in a certain environment~\cite{sawyer2005social,ladyman2013complex,epstein1999agent}. For example, in social networks, interactions between individuals can result in the emergence of polarization~\cite{baumann2021emergence}. Moreover, financial market crashes, a classic phenomenon in economic systems, stem from the collective behavior of market participants and the herding tendencies of individuals~\cite{sornette2009stock}. These emergent phenomena, despite originating from individuals' behaviors, cannot be fully explained or predicted solely based on individual components~\cite{sawyer2005social,ladyman2013complex,epstein1999agent}. Therefore, this requires us to adopt a bottom-up perspective~\cite{epstein1999agent,epstein2012generative,epstein1996growing}: we should begin by simulating \textit{an individual social agent}, and then generate an artificial society by incorporating \textit{a realistic environment} and facilitating \textit{large-scale interactions} among agents as well as between agents and their environment.

% 根据要点梳理，每个维度需要具备哪些进阶型的特征
% 得到过去工作的一个整理

\begin{figure}[t]
\centering
\includegraphics[width=\textwidth]{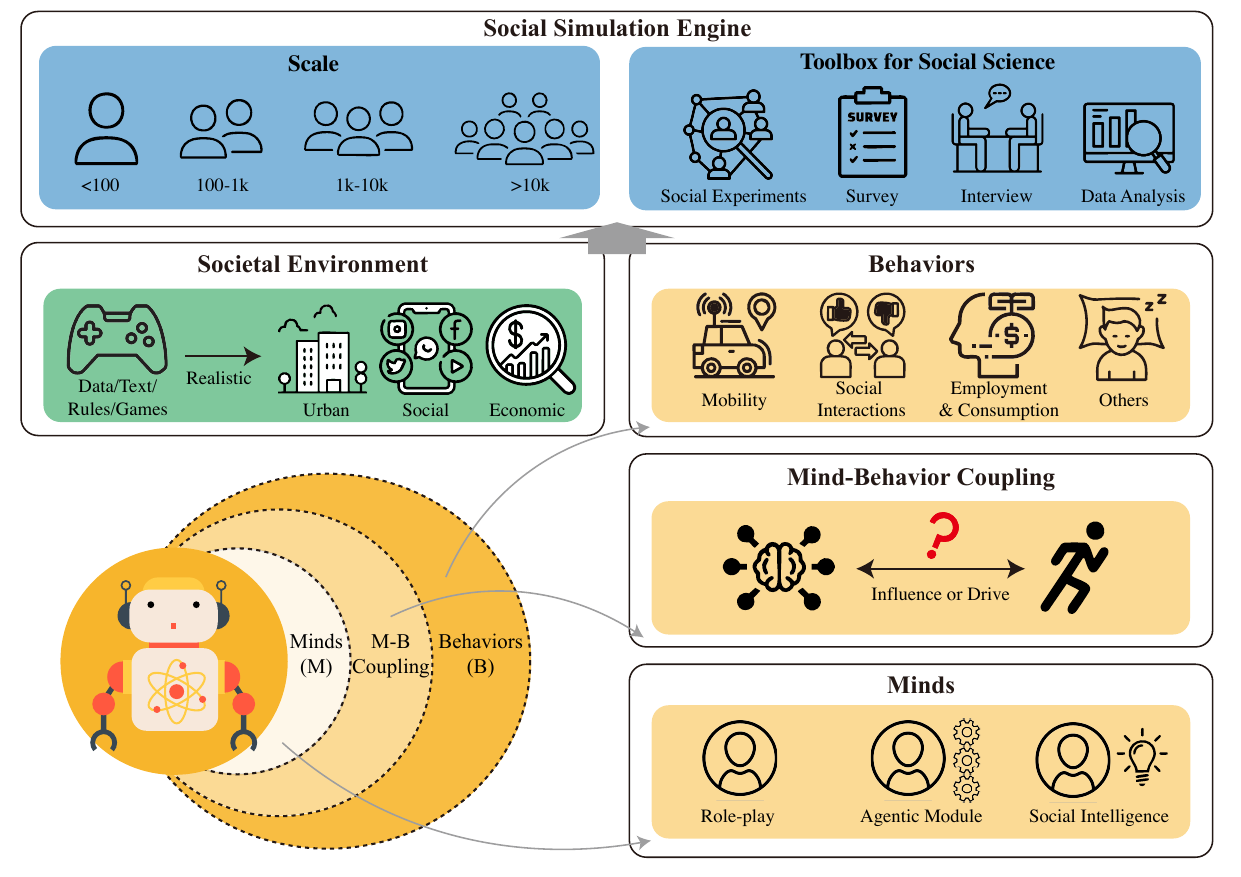}
\caption{Evaluation framework for LLM-driven social simulators.}
\label{fig:evaluation}
\end{figure}

Therefore, we develop an evaluation framework to examine the capabilities of various LLM-driven social simulators along these three key dimensions (Figure~\ref{fig:evaluation}). We first focus on the most basic element of the simulator, i.e., LLM-driven social generative agents. As discussed above, the design of these agents can be divided into three levels: minds, social behaviors, and their coupling methods (M-B coupling). At the mind level, researchers initially input a profile description into LLMs, enabling them to role-play and respond like a real person with a similar profile~\cite{cheng2024sociodojo}. However, such simple role-play cannot guarantee the quality of behavior generation. Consequently, an increasing number of studies, inspired by the pioneering work of Park et al.~\cite{park2023generative}, incorporate agentic module design such as profile, memory, reflection, and action, into their LLM-driven agents~\cite{gao2023s,li2024econagent}. In this way, these agents can exhibit more human-like behaviors and generate responses that are coherent, context-aware, and aligned with their designated profiles. Recently, some researchers have realized that agents designed purely based on the commonsense knowledge of LLMs lack the social intelligence needed to mimic a real social being. To improve this, they have drawn on some theories from psychology to create agents with socially intelligent designs~\cite{wang2024simulating,al2024project}. However, they do not organically integrate theories from multiple social science disciplines, which is central to our design of LLM-driven generative social agents.

At the behavioral level, simulated behaviors can be broadly categorized into two types. The first type includes complex behaviors, which involve multiple intricate steps and cannot be executed solely by the agent itself. These behaviors require interaction with other agents or the environment, such as socializing, engaging in economic activities, or navigating movement. The second type comprises simpler behaviors, such as sleeping, which are relatively straightforward and do not demand external interactions. To systematically evaluate these complex behaviors, such as movement, social interaction, and economic activities, we have developed specific evaluation criteria. For mobility behaviors, we examine whether the simulator simply models the switching of an agent’s position (i.e., relocation)~\cite{wang2024simulating} or incorporates the entire process of mobility trajectory~\cite{shao2024beyond,park2023generative}. For social behaviors, we assess whether the agents merely engage in basic interactions~\cite{pangself} or demonstrate organized social relationships, reflecting more human-like group dynamics~\cite{yang2024oasis,park2023generative,al2024project,mou2024unveiling}. For economic behaviors, we evaluate whether the agents recognize only the concept of ``resources'' (e.g., money in the real world or diamonds in Minecraft)~\cite{cheng2024sociodojo} or perform advanced activities, such as value-based resource exchanges grounded in logical reasoning and strategic decision-making~\cite{al2024project,li2024econagent}. In the case of simpler behaviors, we focus on the level of constraints in the simulated activities. These range from highly restricted tasks, such as choosing a favorite movie~\cite{zhang2024generative}, to more autonomous and creative undertakings, like organizing a party without external prompts~\cite{park2023generative,wang2024simulating}.

After introducing the minds and behaviors of agents, we further focus on understanding how behaviors are generated from their minds, which we refer to as mind-behavior coupling. Some researchers have adopted implicit modeling approaches, relying on the planning, memory, and reasoning capabilities of LLMs to generate plausible behaviors~\cite{gao2023s,li2024econagent,tang2024gensim}. In contrast, others have leveraged established theories, (e.g. Maslow's Hierarchy of Needs~\cite{maslow1943theory} and Theory of Planned Behavior~\cite{ajzen1991theory}) to explicitly model how behaviors are driven by minds~\cite{wang2024simulating}. This explicit modeling aims to create behaviors that are not only plausible but also more closely aligned with human-like patterns~\cite{wang2024simulating}.

As discussed above, a realistic societal environment serves as the foundation for simulating authentic human behaviors and society. Current social simulators employ a range of strategies for environment design, each with its own strengths and limitations. Dataset-based environments~\cite{cheng2024sociodojo,tang2024gensim} rely on pre-existing data but lack the ability to provide dynamic, real-time feedback to agents' behaviors. For example, Sociodojo~\cite{cheng2024sociodojo} For example, Sociodojo~\cite{cheng2024sociodojo} incorporates pre-existing real-world time series data to provide these agents with a sense of the external world. Text-based environments~\cite{pangself,wang2024simulating}, often built based on LLMs, can offer some interactive feedback; however, their realism and objectivity remain questionable, limiting their reliability for simulating complex scenarios. Rule-based virtual environments, like Minecraft, provide richer and more objective feedback, but they still fall short of capturing the intricate complexity of real human social systems~\cite{li2024econagent,al2024project,yang2024oasis}. To advance toward a truly realistic social simulator, it is essential to design an environment that faithfully reflects the multifaceted nature of human society. Such an environment should integrate key dimensions of urban living, economic dynamics, and social relationships, while supporting diverse interactions among agents and providing feedback on their behaviors.

\begin{figure}[t]
\centering
\includegraphics[width=\textwidth]{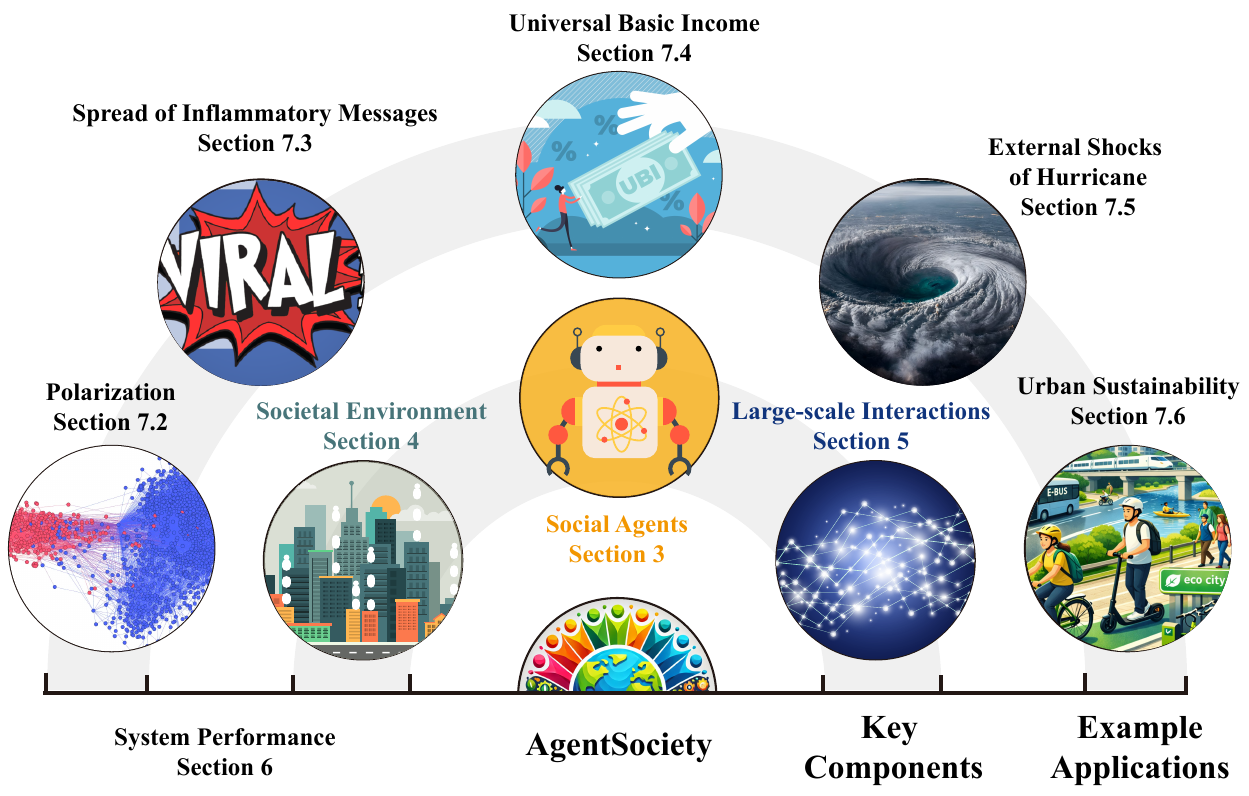}
\caption{Overview of the proposed social simulator AgentSociety. AgentSociety consists of three key components: LLM-driven social generative agents, a realistic societal environment, and a powerful simulation engine that supports large-scale interactions. Based on these components, AgentSociety not only demonstrates superior computational performance but also serves as a valuable testbed for various social experiments.}
\label{fig:overview}
\end{figure}

After evaluating LLM-driven social generative agents and their environments, we further extend our focus to examine the capabilities of the social simulation engine, particularly in terms of its scalability and its potential to support social science research. The scale is a key factor in determining its capacity to support research on complex social systems~\cite{sawyer2005social,ladyman2013complex,epstein1999agent}. We classify the supported scale into four levels: \textless{} 100, 100-1k, 1k-10k, and \textgreater{} 10k agents. Larger scales enable more intricate simulations and provide a platform for studying emergent phenomena in greater detail. Moreover, the engine’s ability to facilitate traditional social science methodologies, such as experiments, surveys, and interviews, is also important. The extent to which the system supports these methods directly influences its applicability across diverse research domains. By accommodating these methodologies, the engine can bridge the gap between simulation-based research and real-world social science, unlocking new opportunities for understanding and addressing societal challenges. Overall, Table~\ref{fig:overview} shows the comparison of different LLM-driven social simulators across the three key dimensions. Existing platforms, although capable of simulating societies and human behaviors to some degree, face substantial limitations in various areas. Since these platforms were not specifically designed for social science research, they lack support for these methods. As a result, this aspect has not been included in the table.

In this paper, we propose AgentSociety, a comprehensive large-scale social simulator designed to integrate LLM-driven social generative agents, a realistic societal environment, and a robust simulation engine. This simulator not only supports large-scale agents and their interactions, but also facilitates advanced social science research. Figure~\ref{fig:overview} provides an overview of AgentSociety and outlines the structure of this paper. AgentSociety consists of three key components: LLM-driven social generative agents, a realistic societal environment, and a powerful simulation engine that supports large-scale interactions. Extensive experiments demonstrate AgentSociety’s superior performance and its potential as a valuable testbed for various social experiments. In particular, we first introduce LLM-driven social generative agents in Section~\ref{sec:social_agents}, which discusses the designs for agents' minds, complex social behaviors, and their coupling in detail. We then demonstrate our real-world societal environment in Section~\ref{sec:environment}, which includes our modeling of urban, social, and economic spaces. Furthermore, we illustrate our large-scale social simulation engine in Section~\ref{sec:engine} and evaluate its performance in Section~\ref{sec:performances}. Finally, we show a typical one day life of our simulated agents in Section~\ref{sec:one_day_life} and launch several social experiments based on our proposed large-scale social simulator in Sections~\ref{sec:polarization} - \ref{sec:hurricane}. These examples, covering polarization (Section~\ref{sec:polarization}), the spread of inflammatory messages (Section~\ref{sec:infl_message}), universal basic income~\ref{sec:ubi}, external shocks of hurricanes (Section~\ref{sec:hurricane}), and urban sustainability(Section~\ref{sec:sust}), demonstrate the validity and authenticity of our proposed simulator.

% 我们具体是怎么做的，如何切分的，文章组织是什么样的？

% 文献对比？？？
\begin{table}[t]
\caption{Comparison of LLM-driven agents and social simulators.}
\hspace*{-1cm}
\resizebox{1.2\linewidth}{!}{%
\begin{tabular}{|p{2cm}|p{0.5cm}|p{0.5cm}|p{0.5cm}|p{0.5cm}|p{0.5cm}|p{0.5cm}|p{0.5cm}|p{0.5cm}|p{0.5cm}|p{0.5cm}|p{0.5cm}|p{0.5cm}|p{0.5cm}|p{1cm}|p{1cm}|}
\hline 
Model & \multicolumn{3}{c|}{Minds} & \multicolumn{2}{c|}{Mobility} & \multicolumn{2}{c|}{Economics} & \multicolumn{2}{c|}{Social} & \multicolumn{2}{c|}{Others} & \multicolumn{2}{c|}{M-B} & Scale &Env. \\ 
\cline{2-15}
 & RP. & AM. & SI. & Relo. & Traj. & Res. & Exc.& Int. & Rel. & Con. & Free & Infl. & Dri. & \# &  \\
 \hline
D2A~\cite{wang2024simulating} & \checkmark & \checkmark & \checkmark & \checkmark &  &  &  &  &  & \checkmark & \checkmark &  &  & \textless{}100 & Text \\
Econagent~\cite{li2024econagent} & \checkmark & \checkmark &  &  &  & \checkmark & \checkmark &  &  &  &  & \checkmark &  & 100-1k & Rules\\
OASIS~\cite{yang2024oasis} & \checkmark & \checkmark &  &  &  &  &  & \checkmark & \checkmark &  &  & \checkmark &  & \textgreater{}10k & Rules\\
GA1000~\cite{park2024generative} & \checkmark & \checkmark & \checkmark &  &  &  &  &  &  &  &  &  &  & 1k-10k &$\times$ \\
MATRIX~\cite{pangself} & \checkmark & \checkmark &  &  &  &  &  & \checkmark &  &  &  &  &  & \textless{}100 & Text\\
Sociodojo~\cite{cheng2024sociodojo} & \checkmark &  &  &  &  & \checkmark &  &  &  &  &  &  &  & \textless{}100 & Data\\
GA~\cite{park2023generative} & \checkmark & \checkmark &  & \checkmark & \checkmark &  &  & \checkmark & \checkmark & \checkmark & \checkmark & \checkmark & \checkmark & \textless{}100 & Rules\\
GenSim~\cite{tang2024gensim}& \checkmark & \checkmark &  &  &  &  &  &  &  & \checkmark &  &  &  & \textgreater{}10k & Data\\
Project Sid~\cite{al2024project} & \checkmark & \checkmark & \checkmark & \checkmark & \checkmark & \checkmark & \checkmark & \checkmark & \checkmark & \checkmark & \checkmark & \checkmark & \checkmark & 1k-10k & Rules \\
AgentScope~\cite{gao2024agentscope} & \checkmark & \checkmark &  &  &  &  &  & \checkmark &  &  &  &  &  & N/A & N/A\\
HiSim~\cite{mou2024unveiling}& \checkmark & \checkmark &  &  &  &  &  & \checkmark & \checkmark &  &  &  &  & 100-1k & Rules\\
S3~\cite{gao2023s} & \checkmark & \checkmark &  &  &  &  &  & \checkmark & \checkmark &  &  &  &  &  1k-10k& Rules\\
Agent4Rec~\cite{zhang2024generative} & \checkmark & \checkmark &  &  &  &  &  &  &  & \checkmark &  &  &  & 1k-10k &Rules\\
RecAgent~\cite{wang2024user} & \checkmark & \checkmark &  &  &  &  &  & \checkmark & \checkmark & \checkmark &  &  &  & 100-1k &Rules \\
Sotopia~\cite{zhou2023sotopia} & \checkmark &  &  &  &  &  &  & \checkmark &  &  &  &  &  & \textless{}100 &$\times$ \\
Casevo~\cite{jiang2024casevo} & \checkmark & \checkmark &  &  &  &  &  & \checkmark & \checkmark &  &  &  &  & 100-1k & Rules\\
 \hline 
 Ours & \checkmark & \checkmark & \checkmark & \checkmark & \checkmark & \checkmark & \checkmark & \checkmark & \checkmark & \checkmark & \checkmark & \checkmark & \checkmark & \textgreater{}10k & Society \\
 \hline
\end{tabular}
}
\end{table}

%% file: 3_Agents.tex
\section{LLM-driven Social Generative Agents}\label{sec:social_agents}

\subsection{Overview}

As discussed above, the rapid development of LLMs allows us to design human-like agents with not only basic psychological states~\cite{abdurahman2024perils,strachan2024testing}, but also complex social behaviors such as mobility~\cite{shao2024beyond,yan2024opencity,feng2024agentmove}, employment~\cite{li2024econagent,horton2023large}, consumption~\cite{li2024econagent,horton2023large}, and social interactions~\cite{gao2023s,park2023generative}. While these efforts in specific areas have shown the human-level intelligence of LLMs, creating LLM-driven social generative agents capable of simulating a comprehensive social being remains difficult. This difficulty primarily lies in two aspects. First, human behaviors are inherently motivated by psychological states~\cite{eysenck2020cognitive,mcleod2007maslow,maslow1943theory,ajzen1991theory}. However, this crucial connection is largely absent in a vanilla LLM or existing agents designed for specific aspects. Second, different types of behaviors are highly interdependent. For example, the decision of when and how people commute to work is shaped by the interplay between their mobility and employment behaviors. Similarly, social interactions among individuals often take place when people go shopping. These examples highlight the crucial interdependence of human behaviors. Despite its significance, current LLMs and agents fail to capture this, limiting their ability to accurately simulate realistic, complex human behaviors. Addressing these two aspects requires deep insights into social science theories of human behavior, as well as advancements in algorithmic design to integrate these insights into LLM-driven social generative agents.

\begin{figure}[t]
\centering
\includegraphics[width=\textwidth]{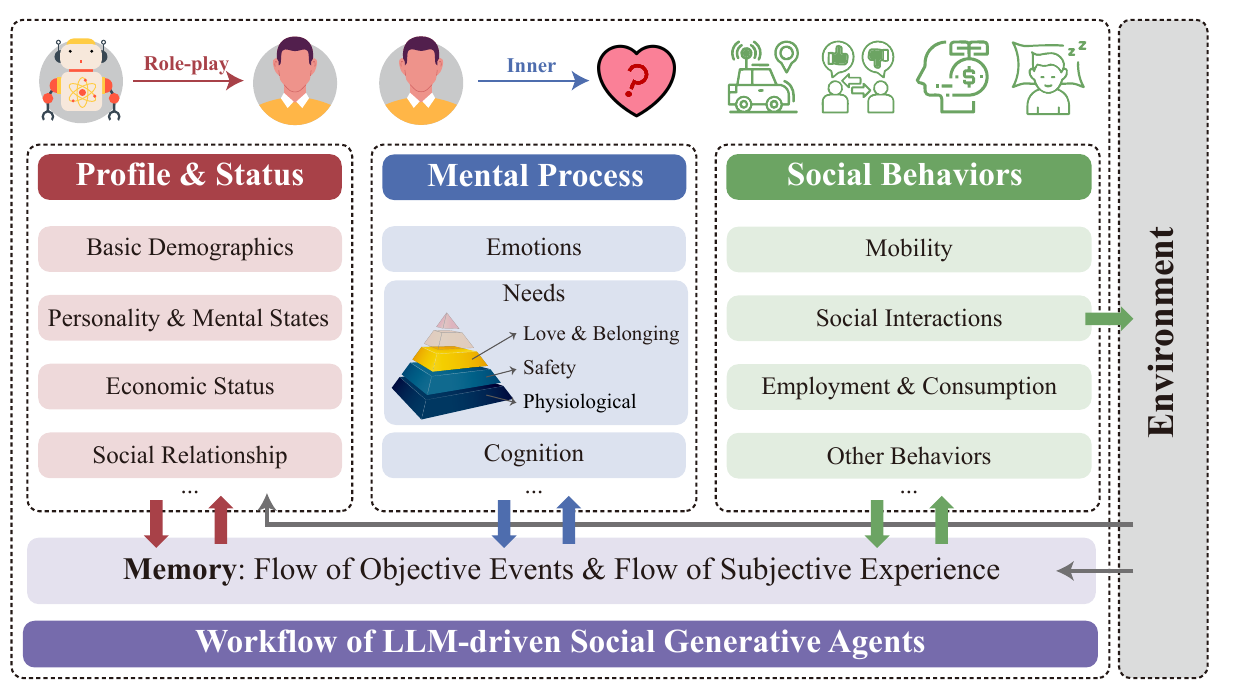}
\caption{Overview of LLM-driven social generative agents.}
\label{fig:overall_agents}
\end{figure}

Therefore, we propose a design for LLM-driven social generative agents, deeply rooted theories from psychology (e.g. Maslow's Hierarchy of Needs~\cite{maslow1943theory} and Theory of Planned Behavior~\cite{ajzen1991theory}), economics (e.g., Dynamic Stochastic General Equilibrium~\cite{christiano2005nominal}), and behavioral science (e.g., Gravity Model~\cite{zipf1946p}). Figure~\ref{fig:overall_agents} provides an overview of the proposed agents, which can be roughly divided into four parts. First, each agent retains their profile, typically regarded as relatively stable (e.g., personality), and status, which is dynamic (e.g., emotion). In particular, the profile includes basic demographics such as name, age, gender, and education, as well as personality. The status comprises three key aspects: the agent’s current mental states, economic status, and social relationships. Mental states reflect the agent’s inner experiences, while economic status and social relationships capture their power and connections in the external world. The integration of the profile and status into these LLM agents enables them to role-play like real people, providing the foundation for simulating complex mental processes and behaviors.

Second, each agent is designed with three levels of mental processes: emotions, needs, and cognition. Emotions reflect the agent’s immediate response to both internal and external stimuli, shaping its behaviors and reactions. Needs serve as the underlying motivational drivers that guide an agent’s actions, ranging from basic survival requirements to higher aspirations such as personal growth and self-fulfillment. Cognition refers to the agent’s understanding of the world, e.g., its attitudes toward climate change and political issues. By incorporating these three levels of mental processes (see the detailed design in Section~\ref{sec:emotion}), agents can autonomously perceive the external environment, ultimately developing their cognition of it. 

Third, social behaviors are the core of LLM-driven social generative agents, which serve as the bridge between their internal mind and external environment. Given the importance and complexity of various human behaviors, we explicitly model three types of social behaviors: mobility, social interactions, as well as employment \& consumption. In Sections~\ref{sec:mobility}-\ref{sec:economy}, we detail the special designs for these three behaviors. Other simple behaviors such as sleeping are directly handled by LLMs. It is worth noting that these behaviors are conditioned by the agents' profile and status, and driven by their mental processes. Finally, we introduce the workflow of the overall LLM-driven social generative agents in Section~\ref{sec:workflow}, illustrating the integration of their profiles, mental processes, and social behaviors. This workflow enables the simulation of comprehensive, context-aware agents by capturing both internal cognitive states and external interactions, ensuring realistic, dynamic social behaviors within the simulation.

%through two memory flows. One flow records the happening of objective events, mainly sourced from external environments. The other captures the agents’ subjective experiences as they explore their own minds or interact with external environments.

\subsection{Emotion, Needs, and Cognition} \label{sec:emotion}

% Agents guided by Theory of Mind recognize both their own knowledge and the mental states of others, including beliefs, intentions, and emotions. This dual awareness enhances their ability to predict and respond appropriately in social and multi-agent settings~\cite{pynadath2011modeling}. 

% 主旨
% needs, emotion, and cognition 这个部分 (画一张图必须有) ，说明智能体内心是如何构建的。每个部分可以选画一张图。

% --- 分工 ----
% 第一段，总起，为什么拆分到这几块（Jingyi来写）。根据心理学来说，从情感，到需求，再到认知 |从最基础的再到高级的，从快速变化额再到稳定的; 心理过程（情绪）和状态

\begin{figure}[ht]
\centering
\includegraphics[width=0.5\textwidth]{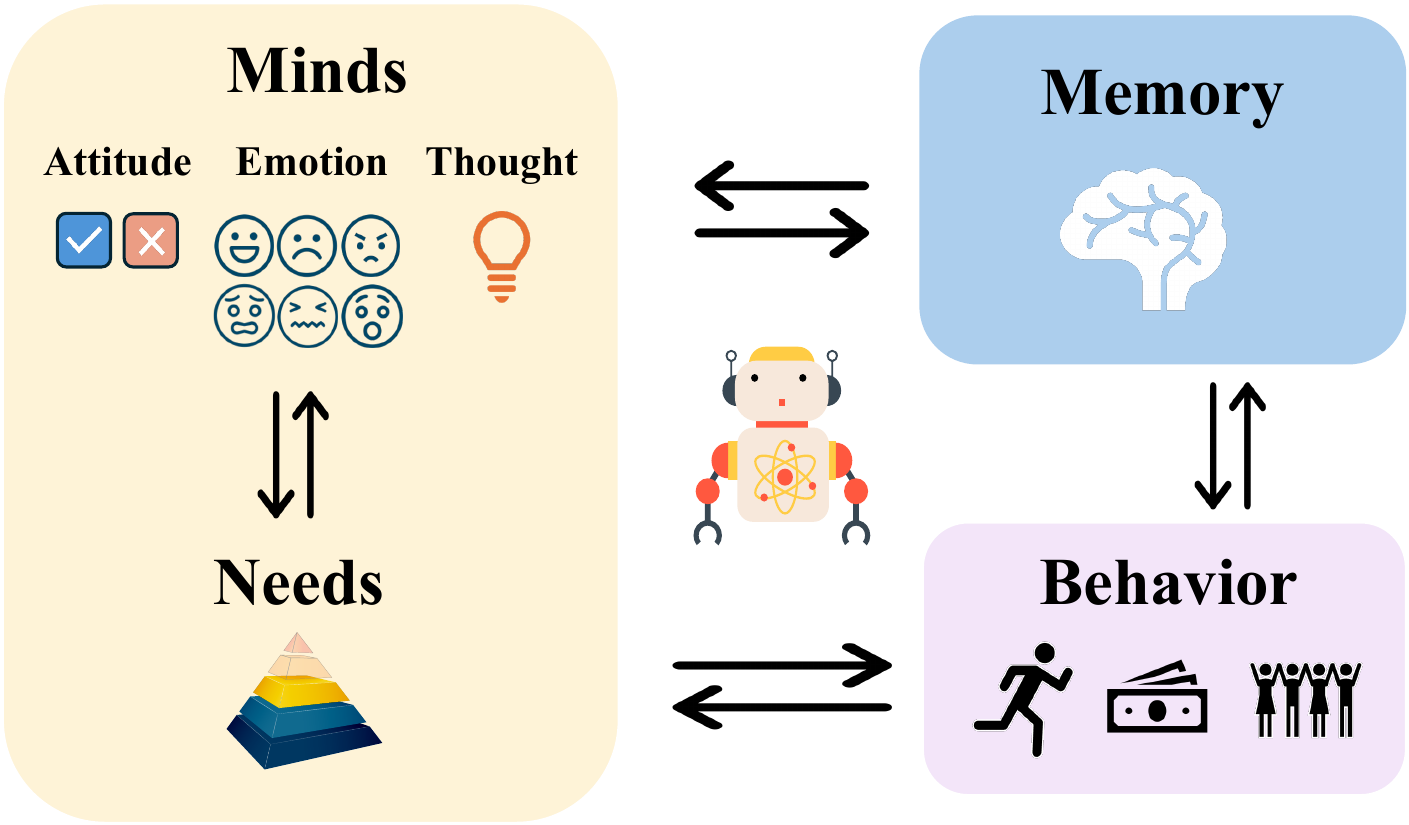}
\caption{Modeling framework of emotion, needs, and cognition.}
\label{fig:agents_cognition}
\end{figure}

Humans are driven by an intricate interplay of feelings, motivations, and thought processes that shape their decisions and interactions~\cite{shvo2019interdependent,al2023chatgpt}. Grounded in psychological theories, our study integrates three fundamental constructs, including emotion, needs, and cognition, to design agents that realistically simulate adaptive and human-like behavior. Emotion, as the most dynamic layer of human psychology, drives rapid responses to external situations and influences behavior~\cite{bourgais2018emotion,beall2017emotivational}. Needs, based on Maslow’s hierarchy of needs theory, serve as motivational drivers, spanning from basic survival requirements to higher aspirations like personal growth~\cite{acevedo2018personalistic}. Modeling these needs enables agents to adopt realistic motivations and prioritize actions in ways relatable to human behavior. 
Cognition, informed by theories like Theory of Mind and Cognitive Appraisal Theory, involves advanced mental processes that allow agents to evaluate complex situations, make thoughtful decisions, and adapt to diverse situations~\cite{bandura1989human, schurmann2020personalizing}. Drawing on these psychological theories,  agents recognize their own knowledge and the mental states of others while evaluating context sensitively, enabling effective and goal-oriented social interactions~\cite{pynadath2011modeling}.
By integrating these crucial elements, our study designs agents that can respond dynamically to real-time changes while tailoring their actions to reflect human-like characteristics and behaviors within complex social simulations. The overall modeling framework for the psychological and cognitive aspects of the agent is illustrated in Figure \ref{fig:agents_cognition}.

Emotion is a dynamic and foundational element of human psychology, driving rapid responses to external events and influencing decision-making and behavior~\cite{shvo2019interdependent}. In our model,  an agent's emotions are affected by its profile and status and are updated based on interactions with other agents and the environment. We adopt the emotion measurement framework from Shvo et al.~\cite{shvo2019interdependent}, which involves the agent selecting a keyword to best describe its current emotional state, formulating a sentence-based thought related to that emotion, and rating the intensity of six core emotions—sadness, joy, fear, disgust, anger, and surprise—on a scale from 0 to 10. This method enables agents to track and update their emotional states, providing a foundation for contextually appropriate and adaptive behavior. These emotions then influence the agent’s actions, motivations, and cognitive processes, establishing an interconnected system that guides decision-making. As we will explore in the next section, emotional states directly impact the agent's needs and cognitive evaluations, linking emotions to higher-order motivational and reasoning functions.

% 第三段，needs (yuwei) --- 可能需求模块需要一个单独的图
The concept of needs is widely accepted in the field of psychology as the fundamental motivator behind an individual's pursuit of specific objectives and maintenance of social engagement. Emotions, on the other hand, are seen as the immediate responses experienced by an individual. The concept of needs, however, is believed to establish the underlying motivational mechanisms that guide sustained behavior, thereby extending and contextualizing emotional fluctuations. The integration of needs and emotions in the agent's model enables the maintenance of consistent motivational pathways over time, ensuring that transient affective states are grounded in enduring goals and priorities.
In our approach, we employ established psychological frameworks (e.g., Maslow's hierarchy of needs~\cite{acevedo2018personalistic}) to categorize and structure these motivational forces. We adopt a hierarchical representation of needs to organize motivational drives by their relative urgency and importance. This needs hierarchy is continuously updated based on three interrelated factors: the agent's active behaviors, uncontrollable or passive external events, and its current psychological states. The integration of these elements enables the system to dynamically adjust need priorities, ensuring that the agent responds appropriately to both internal motivations and external pressures. Furthermore, needs do not merely reflect static conditions but rather serve as a driving force for proactive behavior. Leveraging the Theory of Planned Behavior~\cite{ajzen1991theory}, the agent formulates action plans specifically aimed at meeting or enhancing priority needs. Through this design, the needs module provides a robust foundation for adaptive, socially informed behavior.
In conclusion, the modeled needs provide the necessary motivational basis that informs and intersects with the agent's cognitive processes, leading directly into the subsequent discussion on cognition.

% 第四段，cognition (zhihong)

Cognition encompasses the higher-level processes involved in reasoning, planning, and decision-making~\cite{bandura1989human}. In our model, cognition is intricately connected to the agent's emotional and attitudinal updates. After the agent processes an action, a sentence is used to describe its behavior in relation to the current context. These sentences is then used to update both the agent’s attitude towards specific topics and its emotional state. Attitude, in this context, serves as a memory system, reflecting how supportive or opposed the agent is to a particular topic, rating from 0-10. By continuously updating both emotion and attitude through the agent’s actions and experiences, cognition ensures that the agent adapts to its environment in a way that is consistent with human-like reasoning and emotional depth. This process, in turn, influences the agent’s needs and motivations, bridging the gap to the next level of analysis.

% 第五段，收束一下，给一个例子（yuwei）
In summary, the integration of Emotion, Needs, and Cognition modules enables the agent to engage in socially intelligent behavior, with each module contributing to the shaping of adaptive actions. For instance, when the agent detects an unmet need for social interaction and determines a sequence of actions—such as identifying potential contacts and sending messages—to satisfy this need. The emotional state of the agent, influenced by the emotion module, affects the tone of communication, prompting the agent to initiate conversation with a cheerful or light-hearted tone. As the social interaction progresses, the outcome of the behavior—whether the interaction is perceived as successful or not—is reflected in the emotion and cognition modules. This, in turn, results in an update to the needs module, thereby establishing a continuous feedback loop that adapts the agent's behavior in response to both its environment and internal state.

% ----- 写作指南 -----
% 第二段到第四段，每一段的组织的逻辑就是 
% 定义。（1句）
% 为什么建模社会人为什么需要这个元素。与上一层次的元素的关系是什么，承接上面一段（2-3句）
% 具体是怎么建模的（follow了xxx, design了xxx，实现了xxx）。（4-5句）
% 与下一层次的元素的关系是什么，引出下一段 （半句-1句）

\subsection{Mobility Behaviors}\label{sec:mobility}
Mobility serves as the fundamental basis for social agents to engage in interactions and fulfill their demands. Mobility is not a random behavior but is needs-driven across multiple levels. For instance, when an agent experiences hunger (a basic survival need), it must move to a restaurant to obtain food; to attend a work meeting (an advanced professional need), commuting to an office becomes necessary. These mobility behaviors directly serve specific goals, acting as physical carriers for social interactions, economic activities, and other societal behaviors. In essence, the core challenge of mobility modeling is to bridge the spatiotemporal gap between needs and behaviors. Without effective mobility, agents cannot achieve role immersion or behavioral closure in complex social environments.

As depicted in the previous section, the Needs of social agents exhibit a hierarchical structure: from foundational needs (e.g., eating, resting) to safe and social needs (e.g., work, social gatherings). To satisfy the needs, it requires implementation through a "Need - Plan - Behavioral Sequence" chain. Taking social needs as an example, an agent first formulates a plan to "attend a friend’s gathering," which decomposes into behavioral sequences such as "scheduling time (Friday evening), selecting a location (café), moving." Here, location selection becomes the direct driver of mobility — to reach the target location. This spatiotemporal coupling establishes mobility as the critical execution step for demand realization.

\begin{figure}[ht]
\centering
\includegraphics[width=0.9\textwidth]{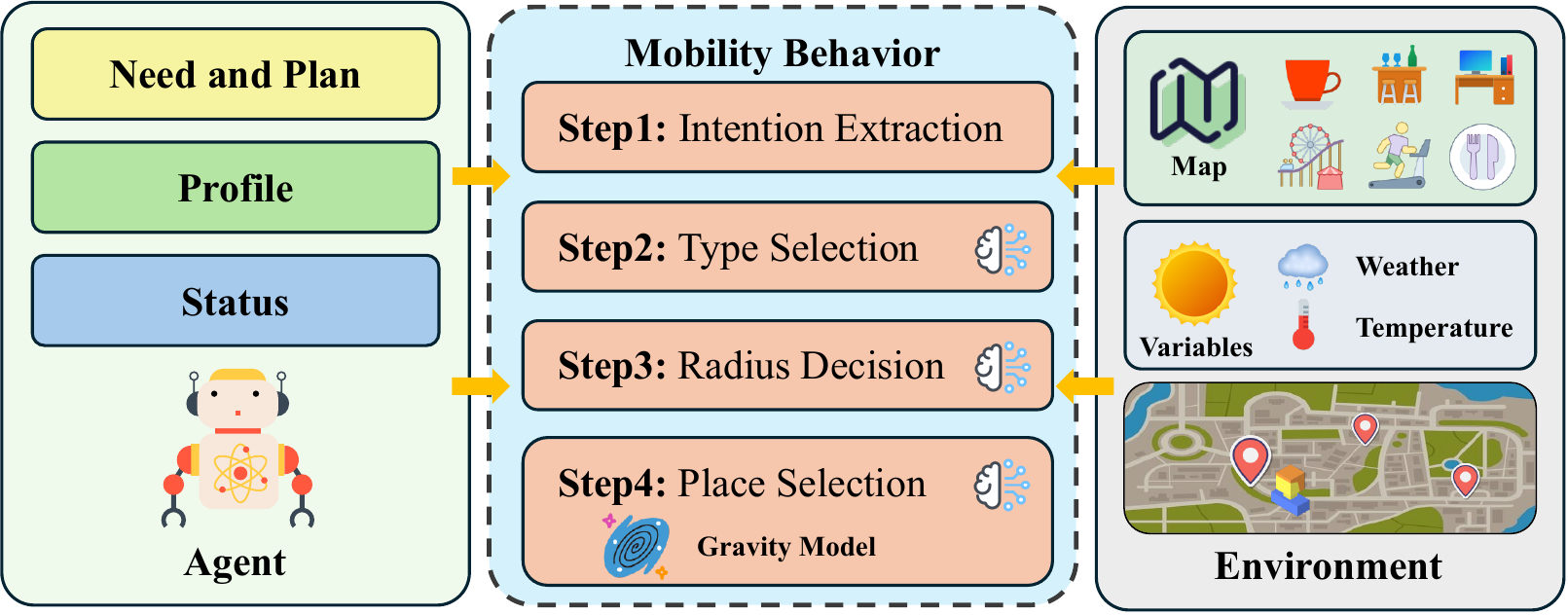}
\caption{Modeling of mobility behavior.}
\label{fig:mobility_behavior}
\end{figure}

Following the needs-driven principle, the mobility module adopts a hierarchical decision framework (shown in Fig. \ref{fig:mobility_behavior}):  
\begin{enumerate}
    \item \textbf{Intention Extraction}: Derive core mobility intentions from demand hierarchies. For example, when "social demand" is activated, the agent may extract a "move to social venue" command.
    \item \textbf{Place Type Selection}: Match demands with POI (Point of Interest) types in geographic databases. If the intention is social interaction, venues like cafés or parks are filtered.
    \item \textbf{Radius Decision}: Dynamically determine feasible ranges by integrating internal states (e.g., age, stamina) and environmental parameters (e.g., weather, traffic). Heavy rain may constrain the radius of indoor venues within 1 km or even stay at home.
    \item \textbf{Place Selection}: Apply the Gravity model for spatial optimization:
    \begin{equation}
        P_{ij} = \frac{S_j / D_{ij}^\beta}{\sum{S_k / D_{ik}^\beta}},
    \end{equation}
    where \(S_j\) denotes the attractiveness of location \(j\), \(D_{ij}\) is the distance, and \(\beta\) is the distance decay coefficient. This model reduces LLM computational overhead while ensuring selections align with human spatial patterns (e.g., proximity principle, agglomeration effects).
\end{enumerate}

Mobility serves as a foundational behavioral module, operating as an integrative force within the social agent's action network. This enables multidimensional coordination across social, economic, and environmental domains. The act of moving to a park, for instance, inherently carries the potential for social synergy. Spontaneous encounters with acquaintances may emerge, catalyzing dialogues, collaborative activities, or even serendipitous social events. These interactions exemplify how mobility serves as a conduit for organic relationship-building. Concurrently, economic synergy manifests through goal-oriented displacements. Commuting to workplaces directly sustains labor productivity, while visiting commercial hubs like shopping malls create opportunities for consumption, thereby linking physical movement to economic cycles. Beyond the human-centric interactions discussed above, mobility also embodies environmental adaptation. Agents dynamically adjust routes based on real-time traffic data or weather fluctuations, demonstrating responsiveness to spatial-temporal constraints. Collectively, these synergies position mobility as the dynamic chassis of social adaptability. It fulfills immediate demands and also provides the contextual infrastructure for complex, layered interactions in urban ecosystems.

% 主旨
% 画一张图说明构建一个社会人智能体需要移动行为
% 第一段，总起，说明建模移动模块的必要性。
% 第三段，如何建模社会人的移动
% 第四段，总结一下

\subsection{Social Behaviors}\label{sec:social}
\begin{figure}[ht]
\centering
\includegraphics[width=0.9\textwidth]{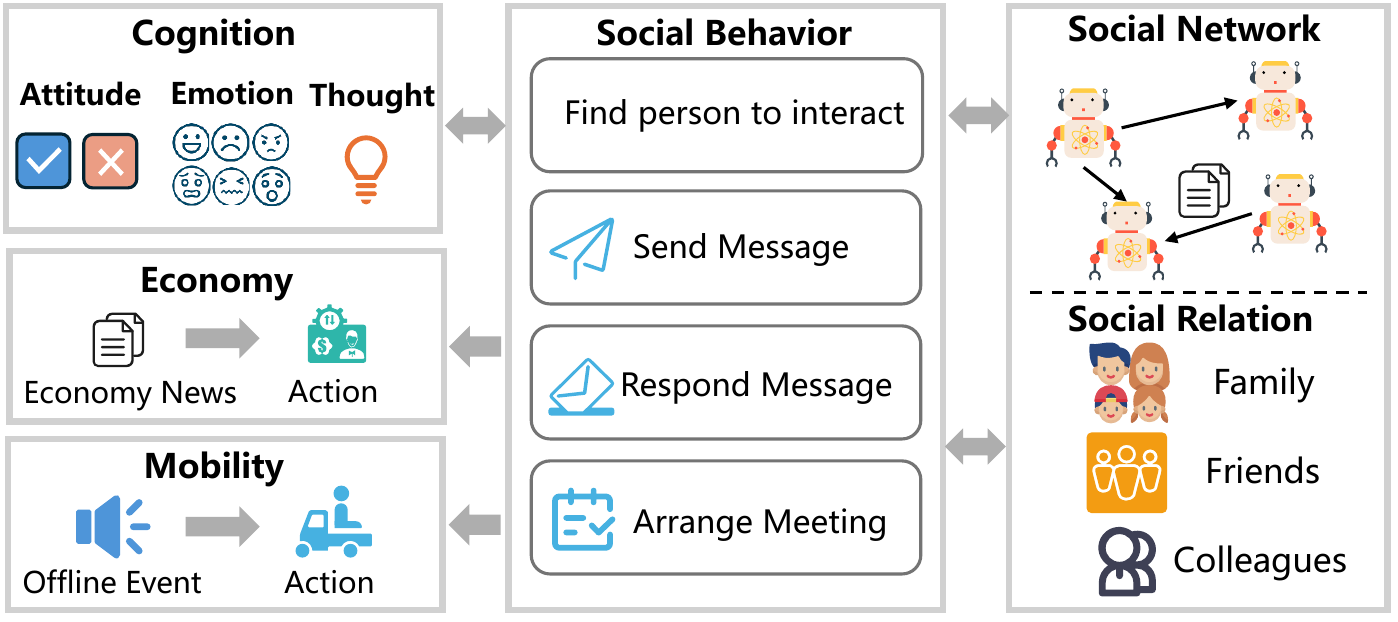}
\caption{Modeling of social behavior.}
\label{fig:agents_social}
\end{figure}
Social behaviors play a critical role in our agent simulation framework. They enable the flow of information and influence between agents, and further lead to the emergence of collective phenomena through agent interactions. In real societies, people's beliefs, opinions, and behaviors spread and evolve primarily through social connections and communications. Therefore, modeling social behaviors allows us to simulate how information and influence flow between agents affects both individual and group dynamics. Our social module consists of two components: social relationships defining the connections between agents, and social interaction behaviors enabling communication between connected agents.

We include three types of social relationships in our framework: family bonds, friendships, and colleagues. Each relationship has a strength value ranging from 0 to 100 representing social closeness between agents. Agents communicate more frequently with high strength connections and adjust their communication tone based on relationship type. For example, agents use more formal language with colleagues and casual language with close friends. We maintain detailed interaction history between connected agents, including message content and time, which influences how they communicate in future interactions.

Our framework primarily focuses on modeling online social behaviors on online social networks. Motivated by their social needs, agents select interaction partners based on relationship types and strength. For instance, when sending casual messages, agents typically choose their best friends, which are friends with the highest relationship strength. Target selection also considers the recipient's profile characteristics. When an agent wants to discuss specific topics, they select friends with relevant expertise or experience. For example, an agent seeking advice about security issues would contact friends who work as police officers. After selecting a target, agents start conversations. The content of these messages is shaped by multiple factors: the agent's current needs and intentions determine the conversation topic, their thoughts and beliefs influence the specific content, and their emotional state affects the message tone and style. When receiving messages, agents generate responses based on their relationship strength with the sender, their chat history, and their current emotional state. Our current framework primarily models online social interactions through messaging on online social network, with plans to incorporate offline interactions. For example, when agents discover shared interests in particular topics or need to have more detailed discussions, they can coordinate offline meetings through online communication.

Social behaviors are deeply interconnected with agents's emotional states, cognition, economic behaviors, and mobility behaviors. An agent's current emotions and beliefs directly influence how they compose messages, while received messages can significantly alter their emotional state and viewpoints. Positive interactions can improve mood and strengthen relationship bonds, while negative interactions may lead to emotional distress and weakened connections. The exchange of economic information through social interactions can trigger economic behaviors. For instance, when agents receive news about job opportunities or market conditions from their social connections, they may adjust their employment or consumption decisions accordingly. Similarly, social interactions often lead to mobility behaviors, such as when agents receive event invitations or arrange offline meetings with their connections.

Through this comprehensive social behavior modelling, we enable agents to engage in meaningful interactions that both shape and are shaped by their internal states and external behaviors. Our social module captures both relationship structures between agents and their interaction behaviors, allowing agents to interact and influence each other as they do in real societies, and providing a foundation for studying how information and influence spread in the simulated society.
% 主旨
% 画一张图说明构建一个社会人智能体需要【哪些社交关系】 和【社交行为】
% 第一段，总起，对于一个社会人来说，建模社交模块的必要性。社交模块分为关系和社交行为两个部分。
% 第二段，建模了的智能体的社交关系（朋友、家人、同事）
% 第三段，建模了的社会交互行为（依靠社交网络为信息传播渠道的线上社交），以后还会考虑线下社交行为；具体包括哪几种行为？在需求驱动下的发消息、收消息....这些功能使得智能体能够相互交互
% 第四段，总结下

\subsection{Economic Behaviors} \label{sec:economy}
\begin{figure}[ht]
\centering
\includegraphics[width=\textwidth]{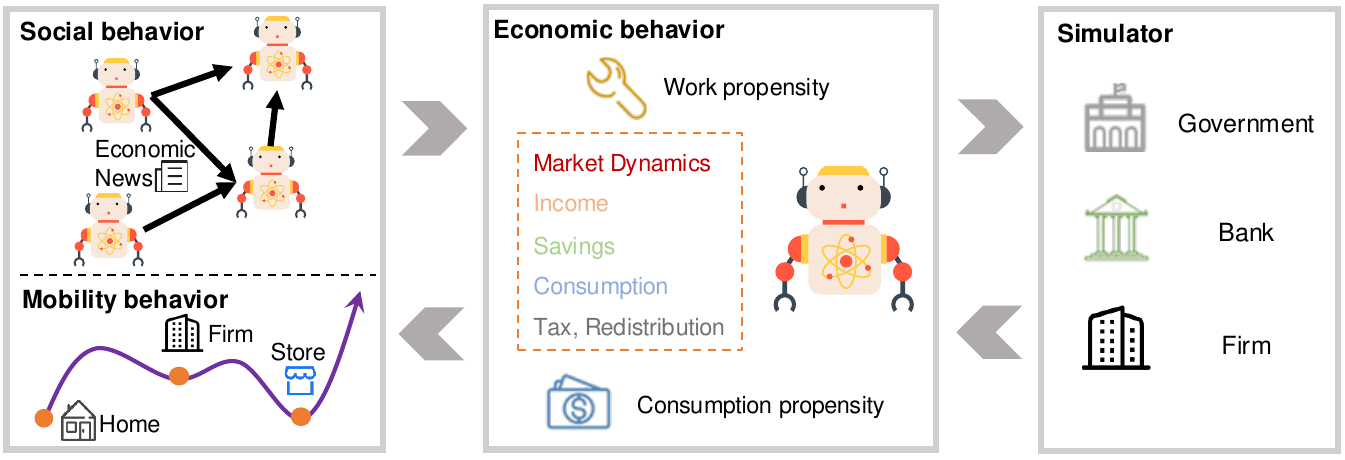}
\caption{Modeling of economic behavior.}
\label{fig:agents_economic}
\end{figure}

Economic behavior in daily life is a necessary component for sustaining life, with employment and consumption being the core economic activities. These two behaviors occupy the majority of time for social agents and further influence their psychological states, including cognition, emotional well-being, and overall life satisfaction. Moreover, economic behavior is deeply intertwined with other aspects of an agent's daily life, such as mobility and social interactions. The satisfaction of one need, whether it is economic or social, often triggers a cascade of related behaviors that span across different domains of the agent’s life. For example, an agent’s decision to work longer hours to increase their income may result in less time available for social interactions. Similarly, the decision to spend more on consumption could lead to adjustments in an agent’s mobility patterns, such as travel to different stores or even relocation to areas with better access to desired goods or services. The modeling of economic behavior is shown in Figure \ref{fig:agents_economic}.

In terms of behavior modeling, we simulate the employment and consumption behavior of social agents through the strength of their work and consumption propensity, and apply these behaviors in a macroeconomic simulation environment~\cite{li2024econagent}. Work propensity determines the agent’s working hours and corresponding monthly income, while consumption propensity determines their monthly consumption budget. Additionally, agents autonomously decide how to allocate this budget, including where to spend the money and what to purchase. These behaviors are directly influenced by various economic factors, including last month's consumption, prices, taxes, and so on. These factors are integrated into the agent decisions in a real-world context, where agents constantly adjust their behavior in response to dynamic economic markets. In future work, we will further simulate agents' complex economic behaviors in the labor and financial markets, including job changes, debt, and investment, to model a more realistic socio-economic environment.

This framework can be used to simulate large-scale economic systems and to explore the potential impacts of policy changes, economic shocks, and other factors on the overall behavior of social agents within the system. By examining these dynamics, we can gain a deeper understanding of the interactions between economic behaviors, psychological states, and social dynamics in a comprehensive and integrated manner.

% 主旨
% 画一张图说明构建一个社会人智能体需要 经济行为【消费与工作】
% 第一段，总起，说明建模经济模块的必要性。
% 第二段，建模了社会人的消费行为
% 第三段，建模了社会人的工作行为
% 第四段，总结下

\subsection{Workflow of LLM-driven Social Generative Agents} \label{sec:workflow}

In this section, we introduce the workflow of our LLM-driven social agent, highlighting how the agent’s internal psychological states (Emotion and Cognition) and its behaviors influence each other and form a complete loop. This loop continuously adapts the agent’s actions based on its evolving cognitive states, ensuring that behavior is dynamically aligned with both internal motivations and external context. The core mechanism linking these psychological states to behavior is Memory, which connects the agent’s cognitive states with its actions. This enables the agent to make adaptive decisions that reflect its past experiences, current needs, and cognitive responses.

\begin{figure}[ht]
\centering
\includegraphics[width=0.9\textwidth]{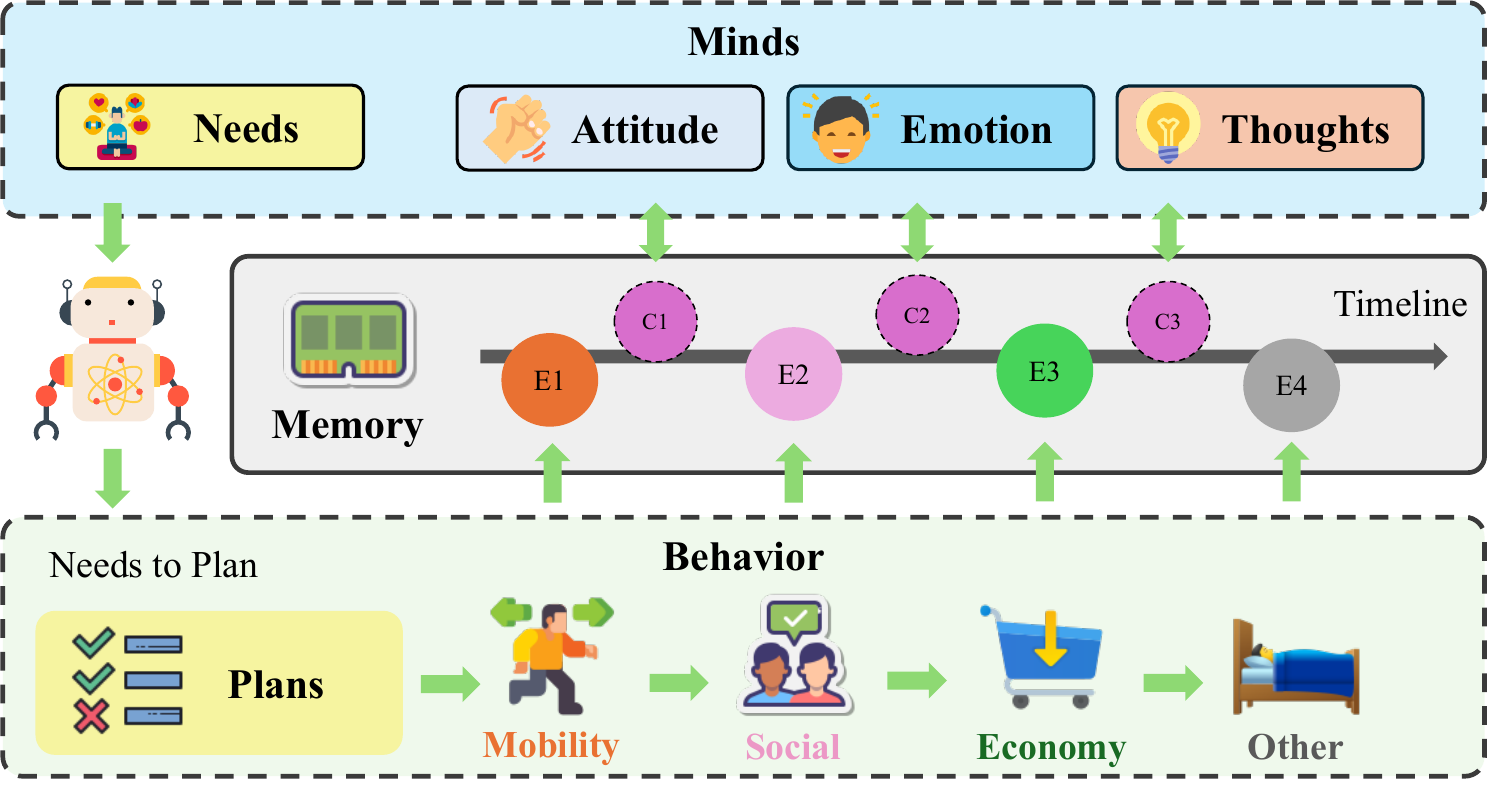}
\caption{Workflow of social agents based on stream memory.}
\label{fig:workflow}
\end{figure}

At the core of our solution is the use of Memory to link the agent’s internal psychological states to its behavior. Memory acts as a bridge between the agent’s current emotions, cognition, and its past experiences, ensuring that its actions are informed by both historical context and present needs. Memory is not a passive system but actively shapes the agent’s decisions and behavior, enabling continuous adaptation and coherence in its responses to changing situations. Specifically, Memory is divided into three main components, each with a specific role in the agent's overall operation:

\begin{itemize}
    \item \textbf{Profile}: Stores the agent's static attributes, such as demographic information (e.g., gender, age), which remain constant and provide context for interpreting the agent's behavior.
    \item \textbf{Status}: Records the agent's dynamic state information in key-value pairs, including data like current needs, satisfaction levels, and financial status, which directly influence decision-making.
    \item \textbf{Stream Memory}: This is the core part of the memory system and tracks events and perceptions over time. It is composed of two types of memory streams: \textit{Event Flow} and \textit{Perception Flow}. Each stream is organized chronologically, with multiple \textit{MemoryNodes} in each stream. Each MemoryNode contains a description with three components: time, location, and event description. 
\end{itemize}

The \textit{Event Flow} records events that occur over time, such as proactive actions by the agent, passive external events, and environmental changes. These events are recorded in sequence, maintaining a timeline of actions and occurrences.

The \textit{Perception Flow} records the agent's thoughts and attitudes towards the events in the \textit{Event Flow}. Each node in the \textit{Perception Flow} is linked to one or more nodes in the \textit{Event Flow}, reflecting how the agent perceives or reacts to a specific event. This integration allows for a nuanced representation of both the agent's cognitive appraisals and emotional responses. 

The agent's behavior is driven by its current state, which influences the decision-making process and the actions taken. The following steps outline the agent's workflow:

\begin{enumerate}
    \item \textbf{Action Determination}: The agent assesses its current state (from the Status memory) and decides on a course of action based on its emotional and cognitive evaluations. For example, if the agent needs social interaction and is in a positive emotional state, it may choose to initiate a social conversation.
    \item \textbf{Event Feedback}: After performing the action, the agent receives feedback. For example, if the agent attempts to move to a social gathering, it checks whether the movement was successful (e.g., did it reach the correct location, considering environmental factors like weather).
    \item \textbf{Memory Update}: The event and its feedback are recorded in the Event Flow, and the associated Perception Flow is updated with the agent’s emotional and cognitive responses to the event.
    \item \textbf{Emotion and Cognition Analysis}: The Emotion and Cognition modules analyze the outcome of the event (e.g., whether the movement was successful) and update the agent's emotional state and attitude accordingly. This feedback may affect the agent’s future decisions and actions.
    \item \textbf{Passive and Environmental Events}: In the case of passive events or environmental stimuli, the same memory processing logic is applied. The agent perceives the event, updates the Event Flow, and modifies its Perception Flow accordingly.
\end{enumerate}

The Memory system, organized along a time axis, reflects the natural flow of events in the physical world. This memory framework allows the agent to integrate its ongoing experiences with past events, creating a dynamic, evolving representation of its environment and internal states. By leveraging Stream Memory, the agent adapts its behavior over time in a way that mirrors human cognition, emotional responses, and decision-making, providing a coherent and context-aware foundation for socially intelligent behavior.

% 主旨：怎么通过一个智能体的设计来实现上面社会人的 【内部心理】 和 【移动、社交、经济行为功能】
% 画一张图说明一个社会智能体背后的memory，

%% file: 4_Environment.tex
\section{Real-world Societal Environment}\label{sec:environment}

\subsection{Overall}

\begin{figure}[ht]
\centering
\includegraphics[width=\textwidth]{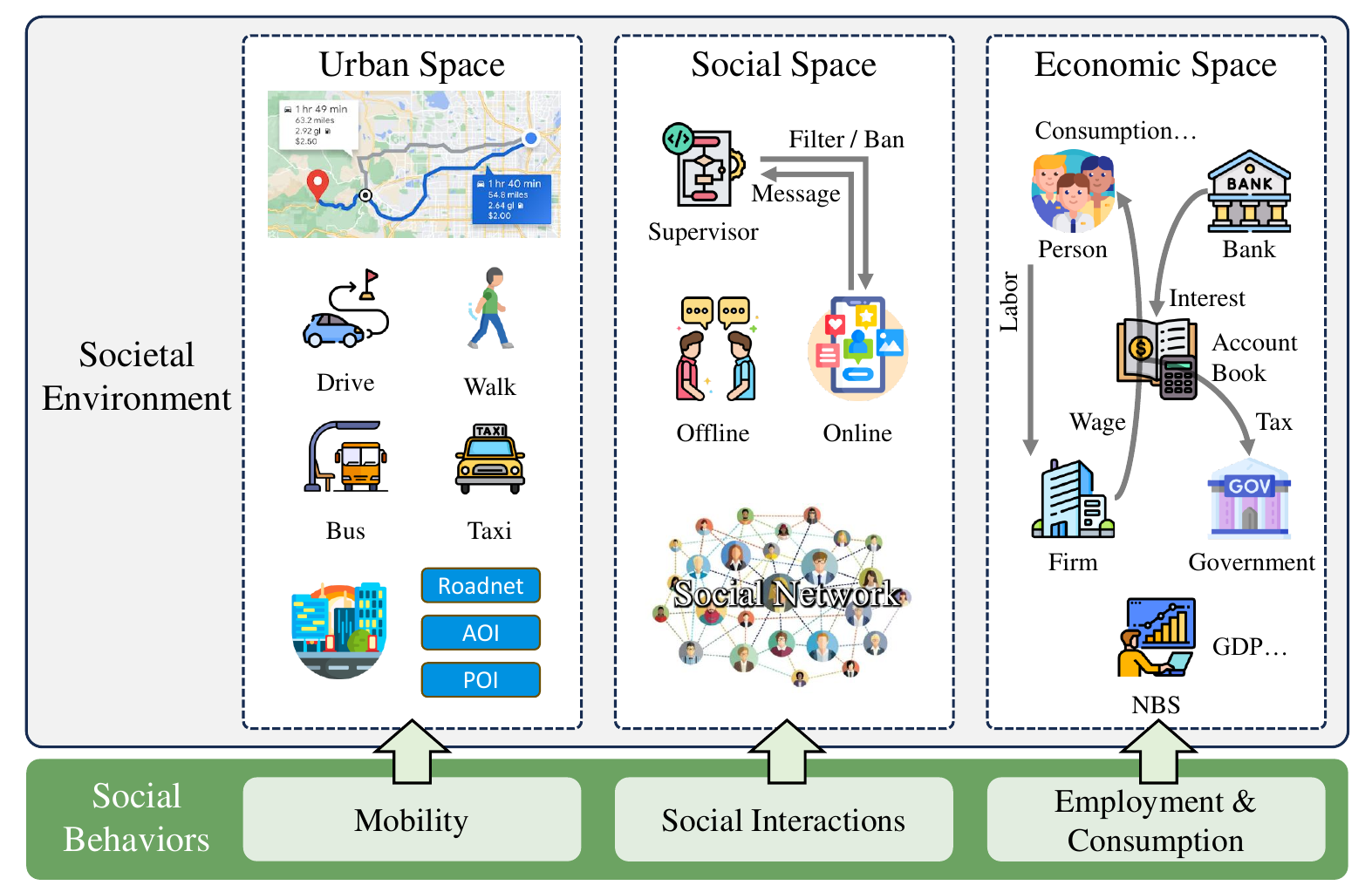}
\caption{Overview of the societal environment.}
\label{fig:env}
\end{figure}

% 逻辑
% 【承上启下】正如上文所提到的，要实现对社会人智能体的构建与模拟，移动行为、社交行为、经济行为是必不可少的要素。
% 【过渡】而这些行为的发生是存在真实世界的载体，仅依赖大模型的知识与能力很可能会发生“幻觉”，导致模拟的结果偏离真实情况。
% 【提出目标】因此，我们要为智能体的模拟提供尽可能真实可靠的模拟环境，环境将具有（1）尽可能真实地建模真实世界及其运行逻辑（2）构建虚拟世界的数据、（3）为智能体提供交互接口，形成一个支撑社会模拟的客观世界载体。这种方式也使得智能体的设计只需关注人的主观行为逻辑，而不需要处理客观世界中的精确数学计算等LLM相对不擅长的领域，简化了系统的设计并使研究人员可以更加聚焦关键任务。
% 【实现】环境基于专家知识构建，以代码的行为固化为模拟环境程序。
% 【插图】和3.1的总图对应，体现支撑能力。

According to the introduction above, in the design and simulation of social agents, mobility behaviors, social behaviors, and economic behaviors like employment and consumption are essential external capabilities.
In the real world, the manifestation of these behaviors is grounded in corresponding objective entities, not merely in human subjective cognition.
For example, mobility behaviors imply a continuous change in spatial and temporal location.
Therefore, if we rely solely on the knowledge and capabilities of LLMs to conduct such simulations without incorporating modeling of the operational laws and constraints of the real world, the simulation results are likely to be influenced by the "hallucinations" of LLMs~\cite{huang2023survey}, resulting in outcomes that diverge from actual realities.
To address this issue, we need to provide a realistic and reliable environment for simulating social agents.
The environment should include the following features:
\begin{itemize}
    \item Appropriate modeling of real-world operational principles to reflect physical constraints and costs, and provide feedback on behaviors;
    \item Environmental data sourced from the real world or aligned with real-world principles;
    \item Interfaces to enable interaction with agents.
\end{itemize}
Such an environment will serve as a virtual mapping of the objective aspects of the world in social simulations, enabling the design of social agents to focus solely on subjective human behavioral logic.
By offloading tasks such as numerical computations, where LLMs cannot guarantee absolute accuracy, this approach simplifies agent design and allows researchers to concentrate on core objectives.

% 整个环境被分为3个空间。城市空间中构建了支持移动模拟的城市道路网并包含了AOI、POI等元素，实现了常见的开车、步行、乘坐公交、乘坐出租车等出行方式，能够为智能体提供真实的位置变化反馈，以及不同出行方式对应的时间和金钱的代价。社交空间在智能体社会网络的基础上，提供了面对面的社交行为支持与社交媒体的抽象。社交空间中的一个关键且真实的设计是监管者，监管者将读取社交媒体中消息，根据算法规律进行过滤并支持对特定的用户或传播进行封禁。经济空间还原宏观经济学中的基本元素，以账本为实现建模了人、公司、政府、银行的经济行为，包括雇佣劳动、消费、税收、利息等，并提供了用于统计经济指标的统计局。

In alignment with this objective, we encode expert knowledge to construct a real-world societal environment as depicted in Figure~\ref{fig:env}, designed to support the simulation of mobility behaviors, social behaviors, and economic behaviors.
The entire environment is divided into three spaces.
The urban space constructs a city road network supporting mobility simulation and incorporates elements such as Area of Interest (AOI) and Point of Interest (POI).
It implements common transportation modes including driving, walking, public transit, and taxi services, providing agents with realistic positional feedback as well as temporal and monetary costs associated with different travel choices.
The social space builds upon agents' social networks, offering support for offline and online social interactions.
A critical and authentic design feature in this space is the supervisor, which monitors social media content, filters messages based on algorithmic rules, and enforces bans on specific users or connections when necessary.
The economic space reconstructs fundamental elements of macroeconomics, modeling economic behaviors of individuals, firms, governments, and banks through the implementation of account books.
These behaviors encompass employment, consumption, taxation, interest mechanisms, while a dedicated statistical bureau is established to monitor economic indicators like GDP.
The following subsections will elaborate on the corresponding environmental spaces for the agent behaviors respectively.

% 以下各小节的基本逻辑（一部分一段）
% 1. 总起：对应3.3的介绍，过渡下来，提出要构建XX模拟环境，以支撑XXX需求
% 2. 介绍如何建模、以及运行逻辑
% 3. 介绍有哪些数据，数据的来源
% 4. 提供的接口
% 5. 实现方式
% 6. 总结

\subsection{Urban Space}

% 1. 城市空间环境的重要性
To address the needs of social agents for moving and interacting with different places, accurate modeling of urban space is essential. 
The urban space must capture both the physical movement pattern of individuals and their interactions with diverse urban locations. 

% 2. 城市空间建模方法
Inspired by traffic simulation platforms such as SUMO~\cite{behrisch2011sumo} and CityFlow~\cite{zhang2019cityflow}, also leveraging the spatial abstraction schemas of OpenStreetMap\footnote{\url{https://openstreetmap.org/}}, the urban space is structured into two interdependent layers, the static infrastructure attributes and dynamic mobility behaviors. 

The static attribute layer includes road networks, defined by lanes, roads and junctions to encode traffic accessibility, as well as functional zones, which are Areas of Interest (AOIs) and Points of Interest (POIs). 
AOIs delineate regions with specific purposes, such as residential neighborhoods or commercial districts, while POIs represent granular interaction targets like retail stores.

The dynamic behavior layer extends this static foundation by simulating multi-modal mobility through a discrete time-stepping mechanism. 
Individual movements, including positions, speeds, and accelerations are updated dynamically according to kinematic principles and predefined rules.
Operational logic begins with agents formulating movement intentions based on their internal needs and goals, then these intentions are translated into specific instructions guiding individuals' movements within the space, which include driving, walking, taking the bus, or taking a taxi.
For all means of transportation, path-planning algorithms generate optimal routes.
Driving follows the IDM model~\cite{treiber2000congested} for acceleration and the MOBIL model~\cite{kesting2007general,feng2021intelligent} for lane-changing decisions.
Pedestrians navigate sidewalks at a constant speed and follow the traffic signals at junctions to avoid collisions with vehicles.
Buses operate on fixed schedules, while passengers conduct the processes of boarding, alighting, and transferring. 
For taxis, a global dispatch system simulates sending the nearest available taxi to respond to ride requests, ensuring efficient service and minimal wait times for passengers.

% 3. 数据来源与软件实现
For the simulation environment to function accurately, we apply rich data sources. 
Road networks and AOIs are extracted from OpenStreetMap\footnote{\url{https://openstreetmap.org/}}, undergoing topological simplification to produce structured representations. 
POI data, acquired via API from SafeGraph\footnote{\url{https://www.safegraph.com/}}.

We implement Python-based APIs to bridge the simulation space and agents,providing  bidirectional interaction capabilities. 
Configuration interfaces allow for initializing agent positions, assign travel plans (e.g., destinations and transportation modes), and reset simulation states. 
Query interfaces enable real-time monitoring of agent kinematics status and simulation metadata such as simulation timestamps. 

% 4. 总结
By harmonizing static urban infrastructure, dynamic mobility behaviors, and multi-source geospatial data, our urban space establishes a high-fidelity decision-making sandbox for agents.

\subsection{Social Space}

% 以下各小节的基本逻辑（一部分一段）
% 1. 总起：对应3.3的介绍，过渡下来，提出要构建XX模拟环境，以支撑XXX需求
% 2. 介绍如何建模、以及运行逻辑
% 3. 介绍有哪些数据，数据的来源
% 4. 提供的接口
% 5. 实现方式
% 6. 总结

% 社交行为是构建智能体社会的前置条件，社会行为的建模与模拟将允许智能体之间相互影响、相互协作，产生更丰富和真实的社会现象。 因此，社会环境中增加社交空间是尤为重要的。社交空间中的主要组成部分是社交网络，这由用户载入。社交网络建模了人与人之间的关系，包含了每个智能体与其他智能体的连接关系与连接强度。这将用于智能体评估社交的选择对象。社交网络与其中的关系和强度在模拟过程中是可变的。基于社交网络，社交空间同时包含了线上社交与线下社交。尽管智能体设计时主要关注线上社交行为，但基于空间位置临近关系的线下社交行为依然是在真实环境的构建中不可或缺的部分。对于线上社交，为了尽可能模拟现实世界中的社交媒体运行逻辑，我们还引入了监管者的概念，监管者将识别线上社交消息内容，根据用户指定的算法或规则过滤消息，并支持对特定用户或连接的封禁，从而模拟社交媒体平台对信息传播的干预过程。
Social behavior is a prerequisite for constructing an agent society.
The occurrence of social behaviors requires the support of an authentic social environment.
The social environment provides management of social relationships, as well as modeling and simulation of social behaviors, which will enable mutual influence and collaboration among agents, generating richer and more authentic social phenomena.

Therefore, the incorporation of the social space within the societal environment is particularly crucial.
The primary component of the social space is the social network, which is provided and loaded by users.
Social networks model relationships between individuals, encompassing the connections and connection strengths between each agent and others.
This network will be used by agents to evaluate potential social interaction targets.
Both the relationships and connection strengths within social networks are mutable during simulations.
Based on social networks, social spaces encompass both online and offline interactions.
Although agent design primarily focuses on online social behaviors, offline interactions based on spatial proximity remain an indispensable component in constructing realistic environments.
For online social interactions, to realistically simulate the operational logic of social media platforms, we also introduce the concept of the supervisor.
The supervisor will identify content in online social messages, filter messages according to user-specified algorithms or rules, and support the blocking of specific users or connections, thereby simulating the intervention process of social media platforms in information propagation.

% 在实现上，社会空间中的社交网络存储为智能体的数据项，线下社交与线上社交均简化为通过智能体消息系统向对应的目标发送消息。监管者则实现为消息发送前的预处理中间级，并提供一个中心化的程序用于集中处理消息并完成规则与算法的更新与下发。
In implementation, the social network is stored as data items within agents.
Both offline and online social interactions are simplified into sending messages to specific targets through the agent message system which will be introduced in Section~\ref{sec:sim:mqtt}.
The supervisor is implemented as preprocessing middleware before message transmission, and a centralized program is provided to handle message processing collectively for updating rules and algorithms.

In summary, the social space not only supports the simulation of realistic social interactions between agents but also establishes intervention capabilities over social propagation within agent societies.
This framework will serve as a crucial foundation for conducting research on real-world social propagation phenomena using LLM-driven social agents.

\subsection{Economic Space}
The economic space includes the modeling of several key elements in the macroeconomics~\cite{wolf2013multi, li2024econagent}.
Specifically, firms convert the labor input of social agents into goods production and pay the corresponding wages to the agents.
Furthermore, firms adjust the wages of agents and goods price flexibly based on the supply and demand relationships in the consumption market. 
The government levies income tax on agents' earnings according to specified tax rates.
The banks pay interest to agents based on their savings each year, with the interest rate adaptively adjusted according to the Taylor Rule~\cite{wolf2013multi}.
The National Bureau of Statistics regularly compiles macroeconomic indicators, such as real GDP, average working hours per person, and per capita consumption levels.

% income, expenditure, tax, interest, statistic

% 与李念的整合
Building upon the modeling of the four key economic entities—firms, agents, the government, and banks—the economic simulator further captures the dynamic processes and interactions that drive the functioning of a realistic economic system. By integrating income generation, expenditure, savings, taxation, and policy-driven adjustments, the simulator provides a comprehensive representation of economic cycles.

Agents are the fundamental economic participants, generating income through labor in firms. This income is subject to taxation, with a portion deducted based on a progressive tax structure, while the remaining disposable income is allocated between consumption and savings. Agents use their disposable income to purchase goods and services, thereby fueling market demand. The funds allocated to savings are deposited into banks, where they accrue interest, influencing future consumption and investment decisions.

Firms act as producers in the economy, utilizing labor from agents to generate production output. They pay wages to agents, creating a cyclical flow of income within the system. Firms dynamically adjust goods prices and wages in response to market supply and demand, ensuring equilibrium in the consumption market. Their revenue comes from the sales of goods, which is reinvested into further production or held as retained earnings for future expansion. The government collects tax revenue from agents based on their earnings and redistributes financial resources to regulate economic activity. It influences the economy through fiscal policy, adjusting tax rates to manage income distribution and public sector financing. These funds may be directed toward public expenditures, which are not explicitly modeled in this simulator but could represent infrastructure, social programs, or government services. Banks function as financial intermediaries, receiving savings deposits from agents and providing them with interest payments. The interest rate is dynamically adjusted according to the Taylor Rule, incorporating monetary policy mechanisms into the simulation. This impacts agents’ saving and spending behavior, as higher interest rates encourage savings while lower rates stimulate consumption. Banks also serve as liquidity providers, ensuring the efficient allocation of financial resources within the economy.

The National Bureau of Statistics (NBS) compiles and analyzes macroeconomic indicators to monitor economic performance. It collects data on real GDP, income distribution, total tax revenue, per capita consumption, and average interest earnings. These aggregated statistics offer insights into systemic trends and policy effectiveness, enabling the evaluation of economic stability and the long-term impact of market dynamics. By structuring the simulation as an economic settlement system, where each entity interacts through well-defined financial flows, this framework achieves a holistic representation of economic operations. It serves as a valuable tool for analyzing the relationships between micro-level decision-making and macroeconomic trends, providing insights into market behaviors, policy interventions, and the overall functioning of economic systems.

% Discussion & Conclusion
While the economic simulator successfully models key economic flows, it does not explicitly capture critical aspects such as the goods market and the labor market. These omissions limit the realism of the model and represent areas for future refinement. The goods market is simplified by assuming firms adjust prices based on aggregate demand, but it does not account for detailed supply and demand dynamics, competition, or market shocks. Similarly, the labor market is abstracted, with agents receiving wages without modeling unemployment or the negotiation processes between workers and firms. Including these elements could enhance the model’s ability to simulate real-world economic fluctuations. Despite these limitations, the simulator provides valuable insights into the interactions between agents, firms, the government, and banks. Future improvements could focus on integrating more detailed models of the goods and labor markets, helping to better replicate complex economic systems and improving policy analysis. In conclusion, while the current model is a simplified representation, it offers a strong foundation for exploring economic interactions and can be further enhanced by incorporating missing market dynamics for more accurate predictions.

% Agent、
% 账本角度，结算系统，模拟经济的运转过程，工作和收入挂钩，储蓄和消费
% 统计局，
% 提供的接口
% 加个Disscussion，商品市场、劳动力市场，总结一下

%% file: 5_Archtecture.tex
\section{Large-scale Social Simulation Engine}\label{sec:engine}

\subsection{Overview}

% 逻辑
% 【总起】尽管社会模拟器看起来像是由多智能体系统 with 工具调用（环境），但真实社会中人的独立思考决策与语言交流驱动的协作方式促使我们重新思考大规模智能体系统架构设计。
% 【现状】现有的多智能体框架一般基于智能体间协作，通过构建处理流程图的方式来规范各个角色的执行顺序，例如......。
% 【问题】但在真实社会中，每个人的行为决策都是自身基于当前的记忆、想法与环境的约束做出的，并不总是依赖来自其他智能体或环境的特定输入。
% 【技术挑战】因此，如何在系统设计上更好地还原这种社会模拟中的“异步”现象，并根据这一特点实现更大规模的智能体执行，是系统架构设计的关键。
% 【解决方案】参考现实社会的基本逻辑，我们将每个智能体视为独立的模拟单元，相互之间不存在任何显式的依赖关系，智能体之间通过消息机制完成信息交换并互相影响。
% 【技术实现】分布式计算 -> ray，高性能消息传输-> MQTT，平衡并行度与有限的IO资源 -> group，还提供了其他有用的工具，包括xxx,xxx,xx
% 【社会实验支持】xxx
% 图：各个组件之间的逻辑关系

% TODO: 下面各个小节分别叫啥

Although the large-scale social simulator introduced in this paper may appear as a simple combination of LLM multi-agent systems (social agents) and tool call (environment), the reality of human society characterized by independent thinking in decision-making and collaboration driven by language communication promotes us to fundamentally rethink the system architecture design and implementation of the large-scale social simulation engine.
Existing multi-agent execution frameworks such as CAMEL~\cite{li2023camel} and AgentScope~\cite{gao2024agentscope} typically take inter-agent collaboration as the foundational principle of system architecture design, constructing Standard Operating Procedures (SOPs) through message-passing processes among agents to determine the sequence of agent execution.
Such frameworks are particularly well-suited for multi-agent execution scenarios with well-defined agent execution sequences, as exemplified by programming tasks~\cite{hong2023metagpt,qian2024chatdev} and conversational games~\cite{xu2023exploring}, as they can significantly simplify complex agent interaction processes.
However, in real-world contexts, individual behavioral decision-making emerges from the autonomous integration of current memory, cognitive states, and environmental constraints, rather than being strictly contingent upon specific inputs from other agents or environments.
Therefore, a pivotal challenge in system architecture design lies in faithfully simulating this "asynchronous" phenomenon within the large-scale social simulator, while strategically leveraging such intrinsic behavioral patterns to optimize simulation execution efficiency.

As a solution to the aforementioned challenge, we draw inspiration from the operational logic of the real world by treating each agent as an independent simulation unit.
There are no explicit dependencies or execution orders between agents.
Instead, they exchange information and mutually influence each other through a messaging system.
For the parallel execution of independent simulation units, to fully leverage the multi-core computing capabilities of modern computer systems and distributed computing paradigms for horizontal scaling of simulation scale, we adopt the highly mature Ray framework~\cite{moritz2018ray} to implement distributed computing and conceal I/O latency through Python's asyncio mechanism.
As the simulation scale increases, we identify that TCP port resources become a bottleneck, and excessive reliance on inter-process communication to coordinate the entire system leads to decreased execution efficiency.
Therefore, we introduce an intermediate structure named agent group to enable multiple agents to operate within a single process, thereby balancing communication costs with parallel acceleration while allowing connection reuse for network calls such as LLM API calls.
For the messaging system supporting inter-agent information exchange, it needs to support massive concurrent connections, high-throughput and reliable message transmission, though being latency-insensitive.
This characteristic closely resembles the Internet of Things (IoT) scenarios where applications must handle message delivery across millions of devices.
Inspired by this similarity, we have introduced MQTT\footnote{\url{https://mqtt.org/}}, the communication protocol that has achieved tremendous success in IoT communications, to construct our agent messaging system.
Following best practices from existing agent execution frameworks, we provide comprehensive utilities including multiple LLM API adapters, a retry mechanism,a JSON parser, a metric recorder, and diverse logging capabilities including both local file output and database storage.
Leveraging these logged processes, we develop real-time interactive visualization interfaces.
Furthermore, specifically designed for social experimentation requirements, we implement a specialized toolbox including interviews, surveys, and interventions.

In the following content, we will first introduce the whole system architecture in Section~\ref{sec:sim:arch}.
We then dive into the key designs including group-based distributed execution in Section~\ref{sec:sim:group} and MQTT-powered agent messaging system in Section~\ref{sec:sim:mqtt}.
The utilities and toolbox for social experiments will be discussed in Section~\ref{sec:sim:util} and Section~\ref{sec:sim:exp} respectively.

\subsection{System Architecture}\label{sec:sim:arch}

\begin{figure}[ht]
\centering
\includegraphics[width=\textwidth]{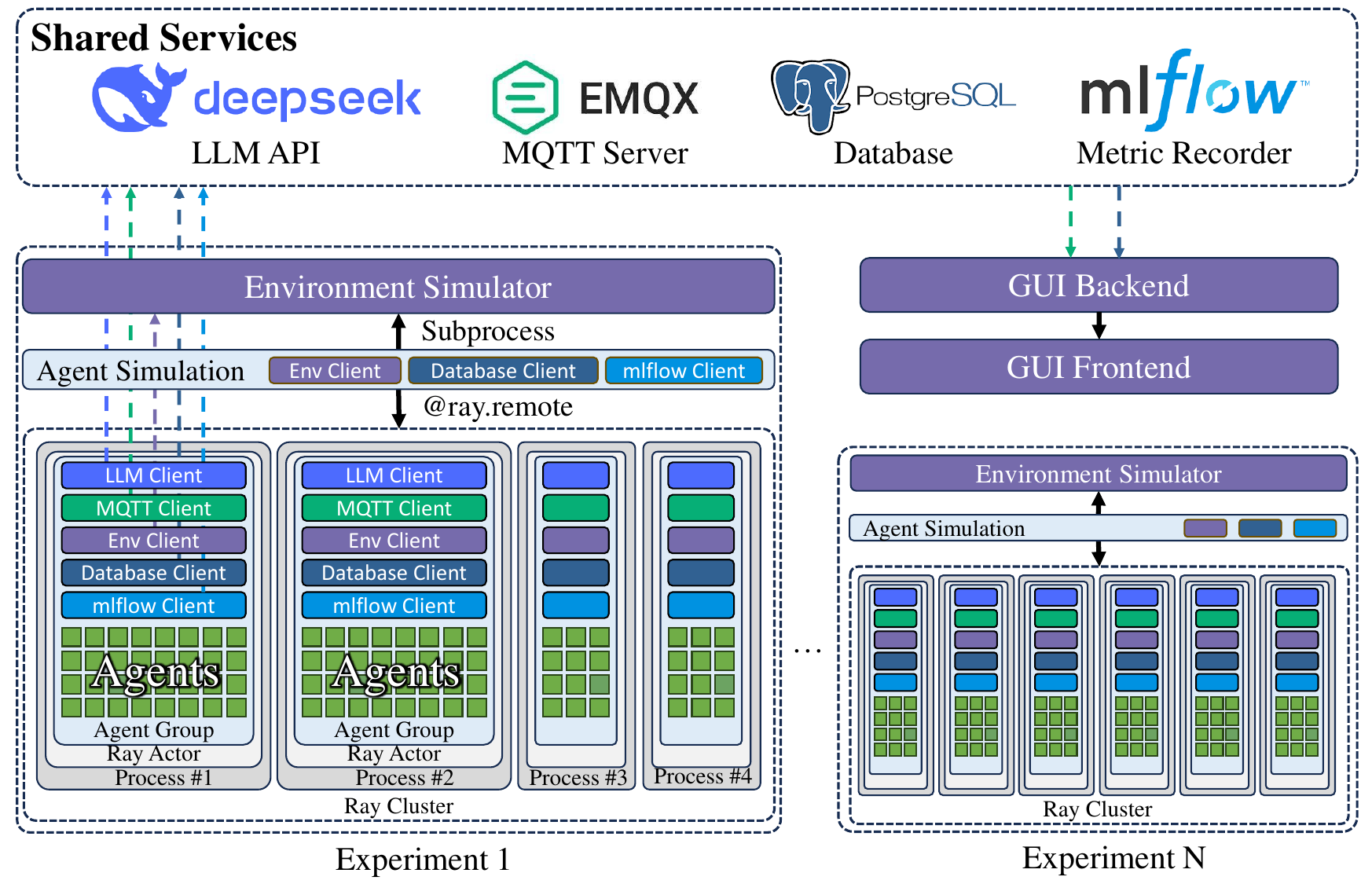}
\caption{System architecture of the large-scale social simulation engine.}
\label{fig:arch}
\end{figure}

% 架构，介绍“实验-run”的划分，各个组件的功能和所属的层级

% 挑战和技术问题、设计

% 为了避免不必要的重复开发并享受开源社区进步的收益，大型社会模拟器的系统架构主要采用先进的、成熟的开源软件或库进行构建。在以下的讨论中，我们将进行一次社会模拟成为一次实验。如图1所示，整个架构主要分为由所有实验共享的共享服务、每个实验独占的计算任务以及可选的GUI组成。
% 共享服务部分包含以下几个服务：
% - LLM API：大模型是整个模拟器最重要的组件，是智能体的灵魂。对模拟器来说，大模型通过API提供标准的“请求-响应”过程，以处理智能体prompt所描述的任务。LLM API可以使用公开的服务如OpenAI、DeepSeek等，也可以使用本地推理引擎进行部署如vllm、ollama。
% - MQTT Server: 本架构中使用MQTT这一高性能物联网协议进行智能体之间的消息传输以模拟真实人类社会中基于语言的相互影响与协作方式。MQTT服务器提供了将消息以符合协议要求的方式投递到客户端的功能。这里我们选择的是emqx这一个高性能mqtt服务器。
% - Database: 数据库在本架构中只用于存储模拟结果以便后续分析或可视化。我们选择了知名的PostgreSQL以使用其独有的高性能批量写入SQL命令COPY FROM来保证数据存储效率。
% - Metric Recorder: 对模拟过程的特定指标记录有利于研究人员对比不同实验的实验结果以发现有价值的科学结论。为了方便研究人员的协作，我们选择使用具有中心服务器的mlflow而不是基于本地存储的指标记录工具如tensorboard。

To avoid unnecessary redundant development and leverage the advancements from the open-source community, the system architecture of the large-scale social simulation engine is primarily constructed using advanced and mature open-source software or libraries.
As shown in Figure~\ref{fig:arch}, the overall system architecture consists of shared services common to all social simulation experiments, simulation tasks corresponding to each experiment, and an optional GUI component.

The shared services include the following components:
\begin{itemize}
    \item \textbf{LLM API:}
    LLMs serve as the most critical component of the simulator, acting as the "soul" of agents.
    For the simulation engine, LLMs provide a standard "request-response" process through APIs to handle tasks described in agent prompts.
    The LLM API can utilize public services like OpenAI\footnote{\url{https://platform.openai.com/docs/overview}} or DeepSeek\footnote{\url{https://platform.deepseek.com/}}, or be deployed through local inference engines such as vllm~\cite{kwon2023efficient} and ollama\footnote{\url{https://ollama.com/}}.
    \item \textbf{MQTT Server:}
    The architecture employs the high-performance IoT protocol MQTT for inter-agent message transmission, simulating real-world human language interactions and collaboration patterns.
    The MQTT server enables protocol-compliant message delivery to clients.
    We select a high-performance MQTT server named emqx\footnote{\url{https://www.emqx.com/en}} for this purpose.
    \item \textbf{Database:}
    The database in the architecture is solely used for storing simulation results for subsequent analysis or visualization.
    We choose PostgreSQL\footnote{\url{https://www.postgresql.org/}} for its unique high-performance batch writing capability through the SQL command \texttt{COPY FROM}, ensuring efficient data storage.
    \item \textbf{Metric Recorder:}
    Recording specific metrics during simulations enables researchers to compare experimental results across different studies and derive valuable scientific insights.
    To facilitate research collaboration, we opt for mlflow\footnote{\url{https://mlflow.org/}} with centralized server capabilities, rather than local storage-based metric recording tools like tensorboard\footnote{\url{https://www.tensorflow.org/}}.
\end{itemize}

The primary purpose of the large-scale social simulation engine is to execute social simulation experiments, which consist of a set of computational tasks comprising environment simulation and agent execution.
In implementation, an experiment corresponds to an \texttt{Agent Simulation} object. This object will create and manage environment simulators through subprocess mechanisms, while utilizing the Ray framework to create multiple agent groups that execute agents through multi-process-based distributed computing.
According to Ray framework design, each agent group functions as a Ray actor operating within a single process.
The Ray framework enables managed Ray actors to work across different machines.
By adding other machines to the head node during Ray cluster initialization, distributed computing can be easily achieved to horizontally scale computational resources for social simulation tasks.
Within each agent group, clients connecting to the shared services and the environment simulator are initialized, enabling multiple agents to work concurrently using these client connections.
Different experiments will share all shared services while utilizing distinct Ray clusters and environment simulators to prevent mutual interference.
The GUI with a backend and a frontend connects to the database and MQTT server to enable the visualization of simulation results, and allows users to directly interact with simulated agents by conducting dialogues or sending questionnaires.

In conclusion, the system architecture integrates multiple cutting-edge open-source softwares to deliver comprehensive capabilities for social simulation and enables researchers to focus exclusively on social agent design, including distributed computing, LLMs, message transmission, data storage, and metric management.

\subsection{Group-based Distributed Execution}\label{sec:sim:group}

% 表示从串行处理到并行处理到async+并行处理的提升

\begin{figure}[ht]
\centering
\includegraphics[width=\textwidth]{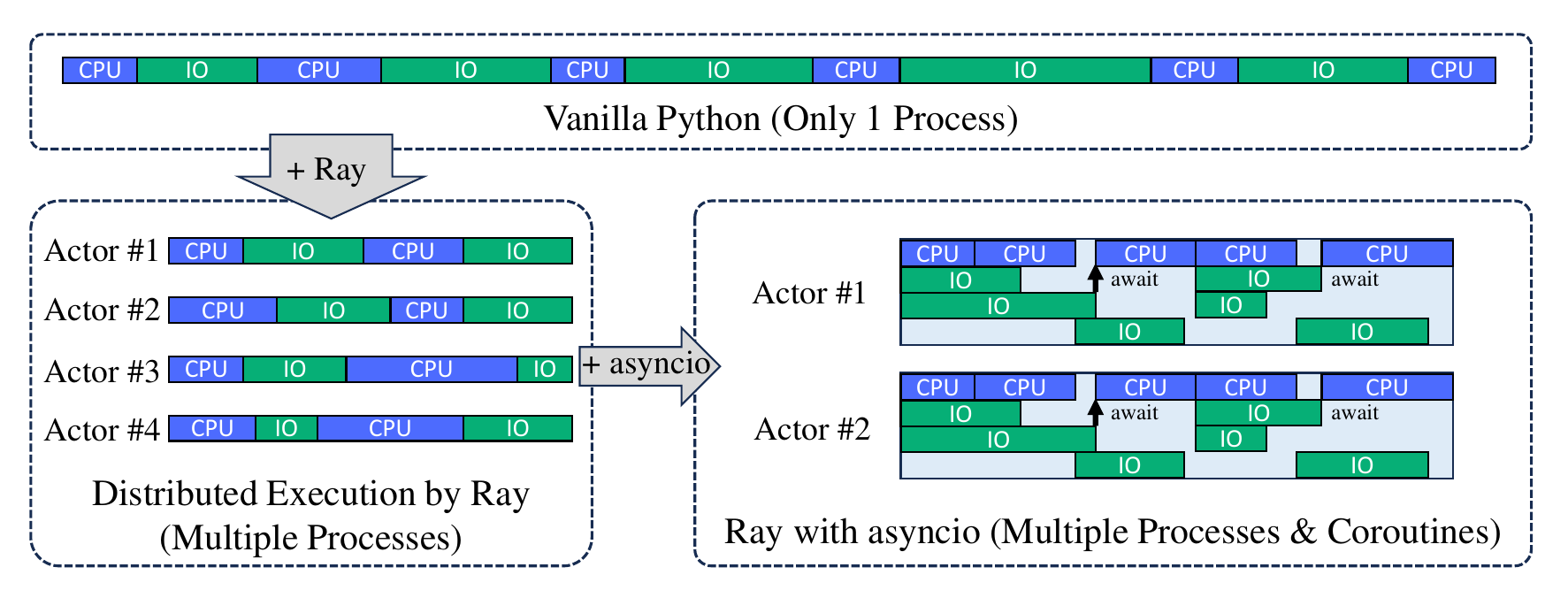}
\caption{Asynchronous multi-process parallel execution using Ray and asyncio.}
\label{fig:async}
\end{figure}

% 逻辑
% 挑战与主要考虑
% 解决方案设计
% More discussion
% 总结

% 根据上面关于系统架构的介绍，我们将多个智能体聚合为一组，称为agent group，并使用一个进程执行每组智能体，以达到分布式计算加速的效果。这一设计是为了解决有限的TCP端口资源与庞大的智能体数量之间的矛盾。具体来说，如果将每个个体视为一个进程并独立连接到共享服务以模拟真实世界中人的自主决策，大量的TCP连接将消耗掉MQTT服务器的TCP端口资源（上限为65535个），从而导致出现TCP端口资源不足而失败。因此，如何进行连接复用以降低TCP端口资源的占用并保证多智能体的独立执行是智能体执行所需要考虑的关键问题。

Based on the above description of the system architecture, we aggregate multiple agents into a group called an agent group, and use a single process to execute each group of agents to achieve the effect of distributed computing acceleration.
This design addresses the contradiction between limited TCP port resources and the massive number of agents.
Specifically, if each individual agent were treated as a separate process and independently connected to shared services to simulate autonomous human decision-making in the real world, the large number of TCP connections would exhaust the TCP port resources of the MQTT server, the database, and the metric recorder (with an upper limit of 65,535 ports), leading to failures due to insufficient TCP port resources.
Therefore, how to implement connection reuse to reduce TCP port resource consumption while ensuring independent execution of multiple agents constitutes a critical issue that needs to be addressed in agent execution.

% 针对这一问题，我们将智能体平均分为多个智能体组，并为每个组配置一个LLM API客户端、一个MQTT客户端、一个环境客户端、一个数据库客户端、一个指标记录者客户端。所有的客户端都采用异步调用的方式实现，并支持并行调用。由于大模型驱动的社会模拟是IO密集型的处理任务，其主要用时在LLM的调用、环境的调用上，因此通过由asyncio提供的异步IO能力将允许多个智能体并行发出LLM请求并充分利用CPU处理Agent设计中的计算任务，有效避免等待请求返回导致的时间浪费。异步调用的连接复用方式也不引入对智能体执行顺序的特定限制，保证了智能体之间的独立性。通过构建智能体组的方式，整个系统所需的TCP端口数可以降低至智能体组数的倍数（取决于是否使用可选的服务如指标记录），避免TCP端口耗尽带来的问题。

To address this issue, we evenly distribute agents into multiple agent groups, each configured with an LLM API client, an MQTT client, an environment client, a database client, and a metrics recorder client.
All clients are implemented using asynchronous calls and support parallel execution.
Since LLM-driven social simulations are I/O-intensive tasks, primarily consuming time in LLM calls and environment interactions.
Leveraging the asynchronous I/O capabilities provided by asyncio and multi-process parallel execution powered by Ray shown in Figure~\ref{fig:async}, the engine allows multiple agents to concurrently send LLM requests while fully utilizing CPU resources for computational tasks in agent design like running gravity models, effectively avoiding time waste caused by waiting for LLM responses.
The asynchronous approach with connection reuse does not impose specific constraints on agent execution order, ensuring independence among agents.
By organizing agents into groups, the total number of TCP ports required by the system can be reduced to a multiple of the number of agent groups (depending on optional services such as metrics recording), thereby preventing issues caused by TCP port exhaustion.
However, immutable fixed-number grouping would result in the overall system efficiency being constrained by the slowest group.
Therefore, adaptive load balancing and dynamic scheduling across groups represent an important direction for future research.

In summary, by combining group-based asynchronous and parallel execution with distributed implementation to enhance simulation efficiency, we successfully address the critical issue of execution failures caused by port exhaustion.

\subsection{MQTT-powered Agent Messaging System}\label{sec:sim:mqtt}

\begin{figure}[ht]
\centering
\includegraphics[width=\textwidth]{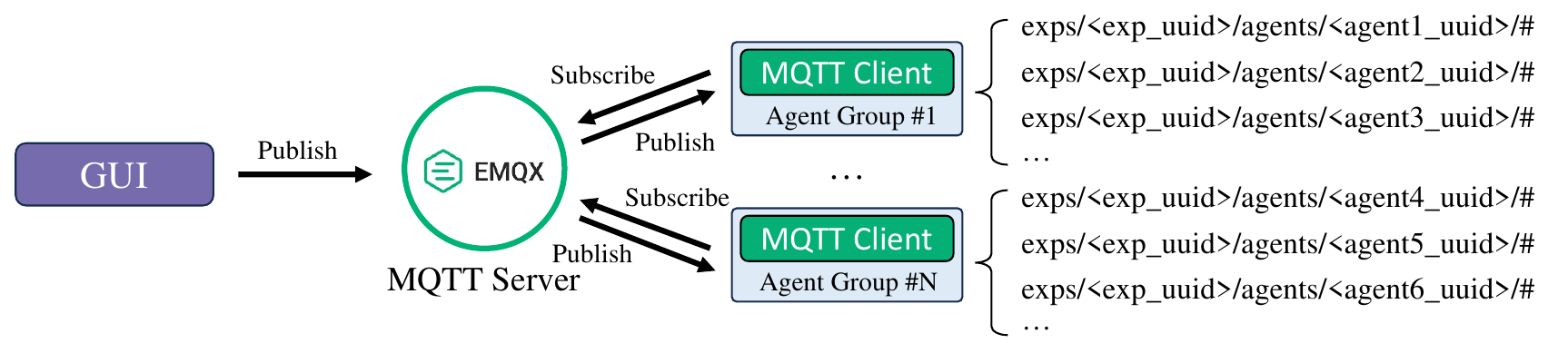}
\caption{Overview of MQTT-powered agent messaging system.}
\label{fig:mqtt}
\end{figure}

% 智能体之间的通信是基于大模型多智能体构建社会模拟的必要环节。文本消息在智能体之间的传输用于模拟真实世界中人与人之间基于语言沟通的交流与合作，将促使群体行为的涌现从而进一步逼近真实世界运转规律。进一步扩展，构建能够连接智能体的消息系统将允许用户或外部程序直接访问、干预智能体的行为，从而开展更丰富的社会实验或交互式应用。

% 需求和目的
Communication between agents is an essential component in constructing social simulations based on LLM-powered multi-agent systems.
The transmission of textual messages among agents simulates real-world human communication and collaboration through language, which will facilitate the emergence of group behaviors and thereby further approximate the operational laws of the real world.
Expanding on this, developing a messaging system that connects agents will enable users or external programs to directly access and intervene in the agents' behaviors.
This will support more sophisticated social experiments and interactive applications.

% 技术方案
% 为了实现上述智能体消息系统，我们需要能够支持向数十万特定ID投递消息的消息系统。在物联网领域，MQTT是用于解决数百万物联网设备与控制中心相互通信的协议，采用发布/订阅的设计方式，发布方将消息发送到指定的ID中，而订阅方则监听指定ID或指定前缀。MQTT通过轻量级的数据包结构设计来匹配物联网低带宽条件，并在较低的资源占用情况下实现了对消息可靠的承诺。尽管这一协议是设计用于物联网设备连接，但十分契合大型社会模拟器中智能体通信的需求，因此我们使用该协议实现智能体通信系统。
To achieve the aforementioned agent messaging system, we need a messaging system capable of delivering messages to hundreds of thousands of specific agents by IDs.
In the IoT domain, MQTT is a protocol designed to enable communication between millions of IoT devices and control centers, employing a publish/subscribe architecture.
Publishers send messages to specified IDs, while subscribers monitor specific IDs or designated prefixes to receive messages.
MQTT utilizes a lightweight packet structure tailored for low-bandwidth IoT environments, delivering reliable messaging with minimal resource consumption.
Although originally designed for IoT device connectivity, this protocol aligns perfectly with the communication requirements of agents in the large-scale social simulator.
We therefore adopt this protocol to implement our agent messaging system as shown in Figure~\ref{fig:mqtt}.
During implementation, we designate the following topics and their corresponding message meanings:
\begin{itemize}
    \item \texttt{exps/<exp\_uuid>/agents/<agent\_uuid>/agent-chat}: Used for sending messages from one agent to the target agent within the experiment.
    \item \texttt{exps/<exp\_uuid>/agents/<agent\_uuid>/user-chat}: Used for users to send chat messages to the target agent within the experiment via the GUI.
    \item \texttt{exps/<exp\_uuid>/agents/<agent\_uuid>/user-survey}: Used for users to send structured JSON-formatted surveys to the target agent within the experiment via the GUI.
\end{itemize}
Under this topic configuration, each agent only needs to subscribe to messages prefixed with \texttt{exps/<exp\_uuid>/agents/<agent\_uuid>/}.
This approach reduces development costs while maintaining compatibility for future extensions to the messaging system's functionality.

% 结论
In summary, by integrating the advanced IoT communication protocol MQTT, we achieve low-cost development to simultaneously connect hundreds of thousands of agents while ensuring reliable agent message transmission.
This solution also enables user input through GUIs and supports future functional extensions, thereby filling a crucial gap in the architecture of the large-scale social simulation engine.

\subsection{Utilities}\label{sec:sim:util}

In addition to the aforementioned key technical designs for social agents, we also provide a rich set of utilities to facilitate the development of social agents.
The design philosophy of most of these utilities is primarily inspired by AgentScope~\cite{gao2024agentscope}.

\textbf{LLM API Adapter.}
We implement calls to OpenAI API-compatible LLMs through the \texttt{openai} python library\footnote{\url{https://pypi.org/project/openai/}}, including OpenAI, DeepSeek, Qwen\footnote{\url{https://bailian.console.aliyun.com/}}, etc.
Additionally, we also support calls to ChatGLM\footnote{\url{https://bigmodel.cn/}}.
Through the LLM API adapter, we allow users to freely select their preferred or partnered LLMs for inference by modifying configurations, which enhances the system's flexibility and compatibility.

\textbf{Retry Mechanism.}
To prevent abnormal results returned by the LLM API from affecting experiments, the system incorporates a retry mechanism.
When invoking the LLM API, if erroneous responses are detected, the system will automatically reinitiate the call. The default number of retries is set to 3.

\textbf{JSON Parser.}
Since prompts often require LLMs to return responses in JSON format to facilitate parsing into program-processable results, we develop a JSON parser.
This parser automatically identifies JSON code blocks in responses, removes Markdown code block delimiters (prefix and suffix), and converts the content into Python objects.

\textbf{Metric Recorder.}
To assist researchers in recording various statistical metrics during experiments, such as total GDP and agent state averages, we adapt the mlflow API and implement a parallel-safe metric logging utility class and functions.

\textbf{Logging and Saving.}
Logging and saving simulation processes and results serve as the foundation driving subsequent data analysis and visualization.
Saving as much data as possible will facilitate richer and more profound research insights.
Accordingly, we design both local file storage using the AVRO format \footnote{\url{https://avro.apache.org/}} and PostgreSQL database as online storage.
Both storage approaches employ similar schemas to archive social agent profiles, agent states during simulations, thoughts and dialogues, and survey outcomes.
Additionally, experimental metadata including IDs, names, durations, configurations, and error messages is systematically recorded.

\textbf{GUI.}
To help people directly observe the behavior of social agents in environments and allow users to engage in direct conversations or surveys with the agents, we develop a GUI program.
The GUI program includes functionalities for experiment management, survey administration, and real-time monitoring or playback of experiments.
During real-time monitoring, users can interact with agents through instant conversations or send surveys, with these communications being transmitted via the agent messaging system while awaiting responses.
Additionally, during both real-time monitoring and playback, users can view agent-related records stored in a PostgreSQL database, including their locations, profiles, status history, thought and dialogue history, and survey response histories.
We aim for this GUI to help users build intuitions about agent societies and facilitate deeper analysis and applications.

\subsection{Toolbox for Social Experiments}\label{sec:sim:exp}

In the realm of social sciences, various research methods are employed to study human behavior, motivations, and responses in different contexts. In real-world settings, it is often difficult to find controlled environments for conducting social experiments that mirror the complexity of human interactions. This is where interventions come into play—creating specific social experimental conditions that allow researchers to simulate and manipulate real-life scenarios. Besides, two widely used methods are interviews and surveys, both of which allow researchers to collect data from individuals, explore their opinions, and understand the underlying psychological factors influencing their actions. These methods are vital for generating insights into how people think, feel, and act in specific situations.  Large-scale agent-based simulations provide a powerful tool for addressing these challenges, enabling the design of controlled experiments with a high degree of realism.
This section details the tools available for conducting interventions, interviews, and surveys with agents in a social experiment. These tools offer flexibility in data collection and manipulation, supporting the design and execution of robust social experiments.

\textbf{Intervention.} Intervention refers to the manipulation of an agent’s behavior or state to observe how changes influence its actions, thoughts, and emotional responses. Interventions are crucial for setting up experimental conditions in social experiments. There are three primary types of intervention in our system:

\begin{itemize}
    \item \textbf{Agent Configuration}: This type of intervention involves directly modifying the internal settings of the agent before the simulation starts. These settings may include altering an agent's personality traits, goals, or preferences. Since this intervention occurs before the simulation, it ensures that the agent’s behavior aligns with the experimental conditions right from the start.
    \item \textbf{State Manipulation}: This intervention occurs during the simulation and allows researchers to modify the agent’s current state. By altering an agent's profile, mood, or ideas, researchers can influence its behavior. For instance, modifying the agent’s emotions can impact its decision-making and social interactions.
    \item \textbf{Message Notification}: This method involves sending a text message to the agent, triggering a response. For example, a message such as "Severe weather changes expected, a hurricane is coming" could be used to observe how the agent adjusts its plans or behavior in response to external threats. This type of intervention can be introduced at any point during the simulation, offering flexibility in creating different experimental conditions.
\end{itemize}

The intervention process can be summarized as:
\[
\text{Pre-Simulation Configuration} \xrightarrow{\text{Agent Settings}} \text{Agent Behavior Start}
\]
\[
\text{During Simulation} \xrightarrow{\text{Memory Manipulation}} \text{Behavior Adjustments}
\]
\[
\text{During Simulation} \xrightarrow{\text{Message Notification}} \text{Behavior Modification}
\]
These intervention techniques allow for dynamic and flexible modifications of the agent’s behavior, providing valuable insights into the impact of specific changes on social interactions and decision-making.

\textbf{Interview and Survey.} An interview is a process of one-on-one or group-based questioning and answering, typically used to gather detailed, qualitative information from participants. In our platform, users can directly communicate with agents through either a front-end interface or programmatically via code. The system uses MQTT to distribute the user’s questions to the relevant agents. The agent then answers these questions by processing both its internal state and the surrounding environmental context. Importantly, this process is designed so that the agent can respond without interrupting its ongoing actions. This allows for real-time interaction while maintaining the flow of the agent’s behavior. The interaction flow is depicted as:
\[
\text{Question} \xrightarrow{\text{MQTT}} \text{Agent Processing} \xrightarrow{\text{Answer}} \text{User Response}
\]
This ensures that the agent can participate in interviews seamlessly while continuing its primary tasks and goals.

A survey is a structured form of data collection, where a series of interview questions are combined based on a specific set of rules. These rules include response formats (e.g., multiple-choice, ranking) and the unique design elements determined by the survey creator (e.g., question order). Surveys are typically used to gather quantitative data across a larger sample, offering a broader perspective on trends or patterns.
In our system, the structured survey is distributed to agents via MQTT, just like interviews. However, the primary difference is that the agent’s responses follow a predefined structure, ensuring consistent data collection across multiple agents. The agent processes the survey questions sequentially, answering them from top to bottom after analyzing the format and response rules. This structured data is then compiled into a format that is easy for the user to analyze. The survey response process is modeled as:
\[
\text{Survey} \xrightarrow{\text{MQTT}} \text{Agent Parsing and Responding} \xrightarrow{\text{Structured Answer}} \text{Data Processing}
\]
This ensures that data collection is organized, reliable, and easy to process for social experiment analysis.

The ability to conduct interventions, interviews, and surveys within our platform provides a powerful toolkit for researchers conducting social experiments. These tools offer a structured approach to data collection and behavioral modification, making it possible to simulate real-world social conditions in a controlled environment. The flexibility to manipulate an agent’s settings, memory, and responses in real-time ensures that a wide variety of social experiment scenarios can be tested, from understanding individual behaviors to studying collective dynamics. This makes large-scale agent simulations an invaluable resource for conducting complex social science research.

%% file: 6_Experiments.tex
\section{Performance Evaluation}\label{sec:performances}

% Expected Conclusion: if you have enough LLM throughout, you can run as many agents as possible.

% Follow: \url{https://tsinghuafiblab.yuque.com/hhbywg/wg833b/qgmb4g194q7m2yrn} !!!

In this section, we will analyze the performance of our proposed large-scale social simulator through a series of comprehensive experiments in order to reveal its strengths and limitations from different aspects.
The experiments focus on the following key research questions:
\begin{itemize}
    \item RQ1: What is the performance of the implementation of the societal environment?
    \item RQ2: What is the performance of the MQTT-powered agent messaging system compared to alternative communication approaches?
    \item RQ3: What is the performance of the large-scale social simulator built from the above components with LLM-driven agents?
\end{itemize}
All experiments were conducted on Huawei Cloud c7.16xlarge.4 cloud servers to ensure comparability of results.
To mitigate potential interference from rate-limiting effects inherent in LLM API calls during large-scale social simulator execution, we have chosen the DeepSeek API platform\footnote{\url{https://platform.deepseek.com/}} that officially claims no request limit\footnote{\url{https://api-docs.deepseek.com/quick_start/rate_limit}}.
Related experiments were specifically scheduled during DeepSeek's off-peak hours (05:00-07:00 local time) to maximize the LLM API throughput.
According to a DeepSeek website statement, the model used during the experiments was DeepSeek-V3~\cite{liu2024deepseek}.

In the following content, we will present the experimental settings, results, and further discussion to address RQ1 in Section~\ref{sec:perf:env}.
Those about RQ2 will be discussed in Section~\ref{sec:perf:mess}.
Finally, in Section~\ref{sec:perf:sim}, we will conduct detailed experiments to answer RQ3.

\subsection{Societal Environment Performance}\label{sec:perf:env}

% One Sentence to start
To evaluate the interaction performance with our simulation environment, we conducted a series of experiments to show our environment is able to handle high concurrency tasks from massive agents.

\textbf{Experimental Settings.}
% Talk about the experimental settings.
We utilized the Social Environment Simulator tool-chain to generate varying numbers of individuals: 1,000, 10,000, 100,000, and 1,000,000, as the specific load for the simulator itself. The departure times of these individuals were distributed according to a typical weekday pattern, and all simulations were set starting from the morning peak hour of 8:30.

The test queries were divided into setting queries and fetching data queries at a ratio of 1:999, meaning one setting query after 999 steps of fetching query for each agent. This ratio was chosen because it is close to the actual request distribution in real agent simulations with our framework.
We limited the maximum number of Social Environment Simulator processes from 2, 4, 8, 16, to 32.
Each experimental setup was repeated five times, lasting for 10 seconds, with queries per second ranging from \(10^2\) to \(10^5\).

\textbf{Performance Metrics.} 
% Talk about the metrics used to evaluate performance IF NEEDED.
We conducted two experiments to evaluate our environment simulation performance. 
First, we measured the simulation speed with the metric of calculating the time consumption per simulation step, with the simulation time set to 24 hours.
Second, we assessed concurrency performance by measuring the increase in queries per second (qps) along with the change in time consumption per simulation step.

\textbf{Evaluation Results.} 
% Show figures and give some discussion.
The result of simulation speed is shown in Table \ref{tab:mean_sd_perf}.
The results indicate that even as the number of individuals and query rates increased significantly, performance degradation was minimal, suggesting that our platform can effectively and timely handle massive interactions between agents and the simulation platform.

\begin{table}[ht]
\caption{Mean time per step with different numbers of agents.}
\hspace*{-1cm}
\centering
\begin{tabular}{ccc}
\toprule
\textbf{\# of Agents} & \textbf{Mean Time per Step (s)} $\pm$ \textbf{SD} \\
\midrule
$10^3$ & 8.578$\times 10^{-3} \pm 3.0\times 10^{-5}$ \\
$10^4$ & 9.129$\times 10^{-3} \pm 1.5\times 10^{-5}$ \\
$10^5$ & 1.800$\times 10^{-2} \pm 5.66\times 10^{-4}$ \\
$10^6$ & 0.1680 $\pm$ 5.34$\times 10^{-4}$ \\
\bottomrule
\end{tabular}
\label{tab:mean_sd_perf}
\end{table}

% One sentence to conclude
In conclusion, the simulation environment is capable of supporting extensive interactions without significant degradation, making it solid for large-scale social simulations.

\subsection{Agent Messaging System Performance}\label{sec:perf:mess}

To validate the comparative advantages of MQTT over other messaging systems, we evaluated various commonly used publish/subscribe systems or message queue systems, including Redis, RabbitMQ, and Kafka.

\textbf{Experimental Settings.} To simulate real-world usage as closely as possible and comprehensively evaluate the systems' capabilities in terms of supported agent count and message throughput, we designed the following experimental procedure.
We assumed a total of 100,000 agents, with each message containing 100 bytes of data.
Each agent sends messages to 10 randomly selected agents.
Given the maximum available CPU cores are limited to 32, we selected parallel process counts from \{2, 4, 8, 16, 32\} and reported the configuration achieving peak throughput.
As simulator startup time constitutes a small proportion of total simulation duration, initialization overhead was excluded from measurements.
We specifically recorded the time interval between message transmission initiation and complete reception to calculate message throughput across different systems.

\textbf{Compared Approaches.} We briefly introduce the comparative methods as follows:
\begin{itemize}
    \item \textbf{Redis Pub/Sub\footnote{\url{https://redis.io/}}:} A lightweight in-memory publish/subscribe subsystem in Redis optimized for real-time messaging with minimal latency.
    It uses a broadcast model where messages are transient and not persisted, making it suitable for ephemeral data or scenarios requiring high-speed communication.
    However, its lack of message durability and limited scalability in high-volume environments may constrain its use in mission-critical applications.
    \item \textbf{RabbitMQ\footnote{\url{https://www.rabbitmq.com/}}:} A robust message broker implementing the AMQP (Advanced Message Queuing Protocol) standard.
    It supports complex routing logic, message persistence, and acknowledgment mechanisms, ensuring reliable delivery.
    Its flexible exchange types (e.g., direct, topic, fanout) and queue management make it ideal for enterprise workflows, though its overhead increases with transactional guarantees.
    \item \textbf{Kafka\footnote{\url{https://kafka.apache.org/}}:} A distributed streaming platform designed for high-throughput, fault-tolerant, and persistent log-based messaging.
    Kafka organizes data into partitioned topics, enabling horizontal scalability and parallel processing.
    Its append-only log structure and consumer offset tracking make it well-suited for large-scale event streaming, real-time analytics, and data pipelines, though it introduces complexity for lightweight use cases.
\end{itemize}
It is worth noting that all services are running on the experimental machine, and the distributed version is not utilized.

\textbf{Evaluation Methods and Metrics.}
In the evaluation of a messaging system, the most critical metric is throughput, which refers to the number of messages that can be transmitted per second.
Once the throughput meets the requirements, we will further consider whether the software system provides user-friendly auxiliary tools to help monitor the service’s operational status or facilitate testing and configuration, such as dashboards.
For throughput requirements, assuming all agents are always attempting to communicate with other agents and the LLM generates a message every 5 seconds, the minimum throughput the system needs to support would be 20,000 msg/s.

\textbf{Evaluation Results.} We conducted five tests on various messaging systems and calculated the mean and standard deviation of throughput, as presented in Table~\ref{tab:mes}.
From the results, we observe that MQTT, Redis Pub/Sub, and RabbitMQ meet the throughput requirements under the aforementioned extreme conditions.
Among them, RabbitMQ's performance was only slightly above the throughput requirement, thus it was the first to be excluded.
The results for Kafka were not reported because it could not even complete the initialization of 100,000 agents within 5 minutes; hence, no specific test results were available.
Although MQTT's throughput is approximately half that of Redis Pub/Sub, its built-in GUI tools can effectively assist users in simple service monitoring, debugging, and testing, which constitutes the primary reason for our ultimate selection of MQTT as the default implementation for the agent messaging system.
Regarding Redis Pub/Sub's high-performance characteristics, we propose that the simulation engine should support flexible user specification of backend implementations for agent messaging systems in the future, thereby accommodating application scenarios with stringent requirements for inter-agent communication.

\begin{table}[htbp]
\small
\centering
\caption{Comparison of different messaging systems.}
\label{tab:mes}
\begin{tabular}{lccc}
\toprule
\textbf{System} & \textbf{Best Parallel Process Number} & \textbf{Throughput (msg/s)} & \textbf{Auxiliary Tools} \\
\midrule
MQTT (emqx v5.8.1) & 32     & $44,702.1 \pm 111.3$          & \textbf{Built-in GUI}                    \\
Redis Pub/Sub (v6.2) & 16   & $ 81,216.2 \pm 333.6 $             & -               \\
RabbitMQ (v4.0.5)   &  16  & $23,667.3 \pm 1,777.7$             & \textbf{Built-in GUI}     \\
% Kafka (v3.9.0)     &   -  & $\times$          & - \\
\bottomrule
\end{tabular}
\end{table}

\subsection{Social Simulator Performance}\label{sec:perf:sim}

% 想一想，要不要分成两个subsection

% One Sentence to start
% \textbf{Experimental Settings.} Talk about the experimental settings.
% \textbf{Performance Metrics.} Talk about the metrics used to evaluate performance IF NEEDED.
% \textbf{Evaluation Results.} Show figures and give some discussion.
% One sentence to conclude

To evaluate the scalability and efficiency of the proposed social simulation framework, we conducted a series of experiments designed to replicate the execution of large-scale intelligent agents under realistic conditions.

\textbf{Experimental Settings.}  
The experiments were conducted on a 64-core machine, with 32 cores allocated to running the environment and the remaining 32 cores dedicated to executing the simulation engine.Testing was performed during the system's low utilization period, while targeting simulation time intervals where agent activities were relatively high to ensure representative measurements.  

We evaluated the system throughput by simulating \{10\textsuperscript{3}, 10\textsuperscript{4}\} agents, The number of processes was varied as \{8, 16, 32\}.
% excluding non-agent entities such as firms and governments from the agent count. 
% and for each configuration, 
% the total time taken to complete five interaction rounds, total token usage (distinguishing input and output tokens), and the number of LLM API calls. Additionally, we measured the LLM API time cost distribution and the Environment API time cost distribution.

\textbf{Performance Metrics.}  
To evaluate the system’s performance, the following metrics were collected:
\begin{itemize}
    \item \textbf{Total execution time:} The total time required for all agents to complete five interaction rounds.  
    \item \textbf{Token usage statistics:} The total number of input and output tokens utilized during the simulation.  
    \item \textbf{LLM time cost distribution:} The distribution of response times for calls to the LLM API, providing insights into latency variability.  
    \item \textbf{Environment time cost distribution:} The distribution of response times for calls to the environment API, measured to evaluate internal system performance.  
\end{itemize}

\textbf{Evaluation Results.}  
The evaluation results are summarized in Table~\ref{tab:performance}, which demonstrates the system’s scalability as the number of agents increases and highlights the performance impact of distributed computing. Specifically, the table shows how performance metrics such as LLM call time and environment response time vary with different group configurations (8, 16, and 32).

\begin{figure}[ht]
    \centering
    \begin{subfigure}[t]{0.45\textwidth}
        \centering
        \includegraphics[width=\linewidth]{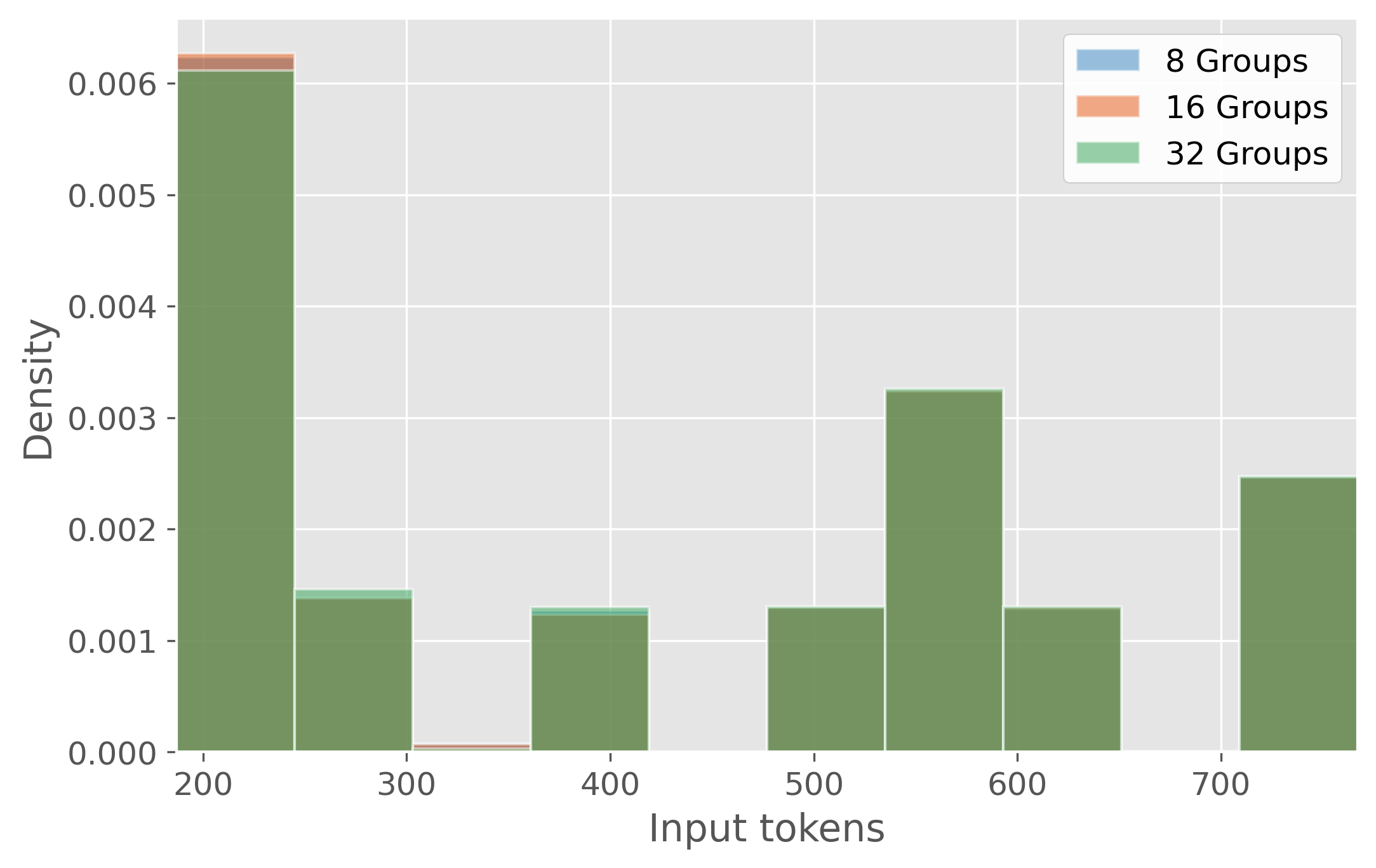}
        \caption{Input Token Distribution}
        \label{fig:input_tokens}
    \end{subfigure}
    % \hfill
    \begin{subfigure}[t]{0.45\textwidth}
        \centering
        \includegraphics[width=\linewidth]{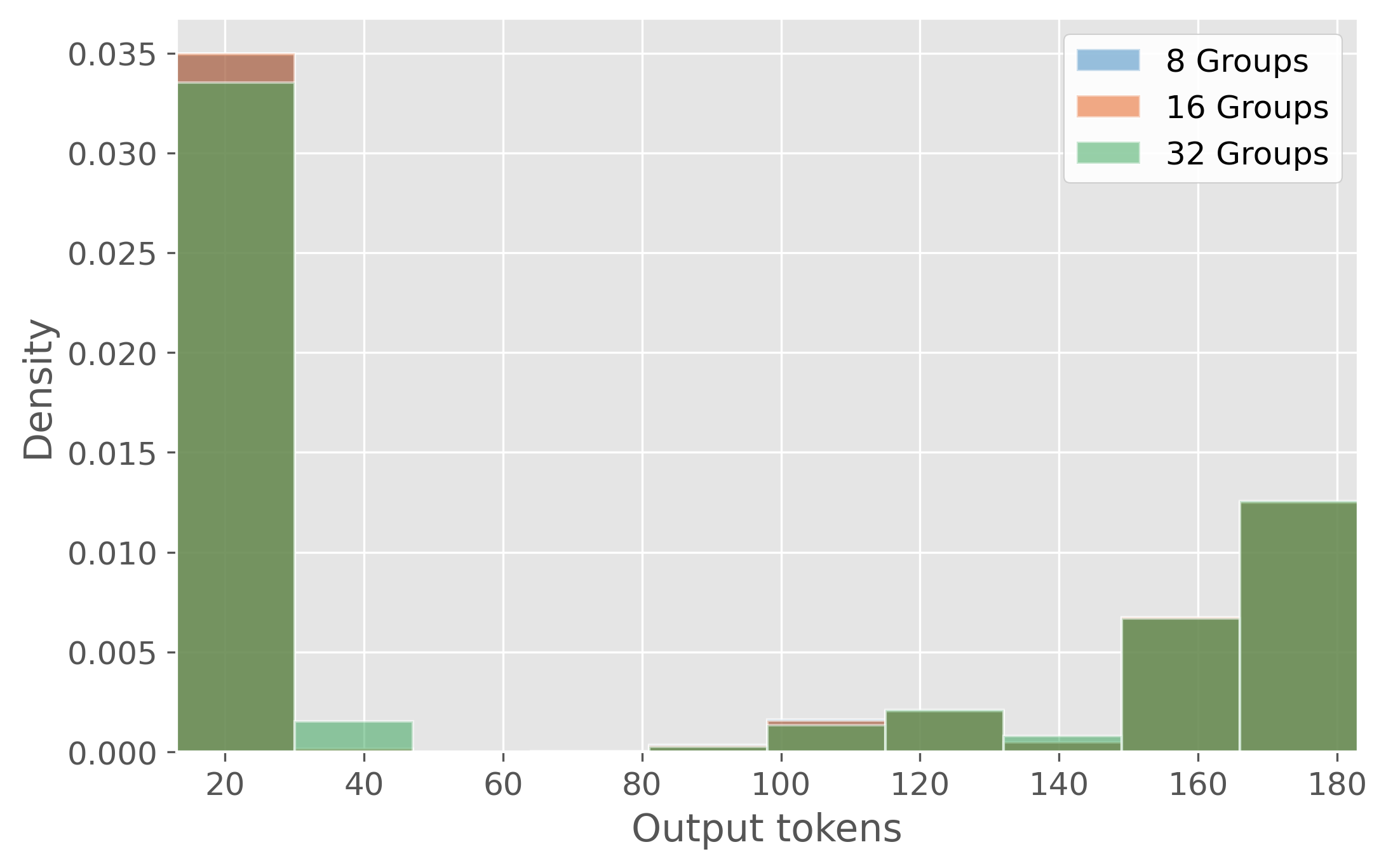}
        \caption{Output Token Distribution}
        \label{fig:output_tokens}
    \end{subfigure}
    \begin{subfigure}[t]{0.45\textwidth}
        \centering
        \includegraphics[width=\linewidth]{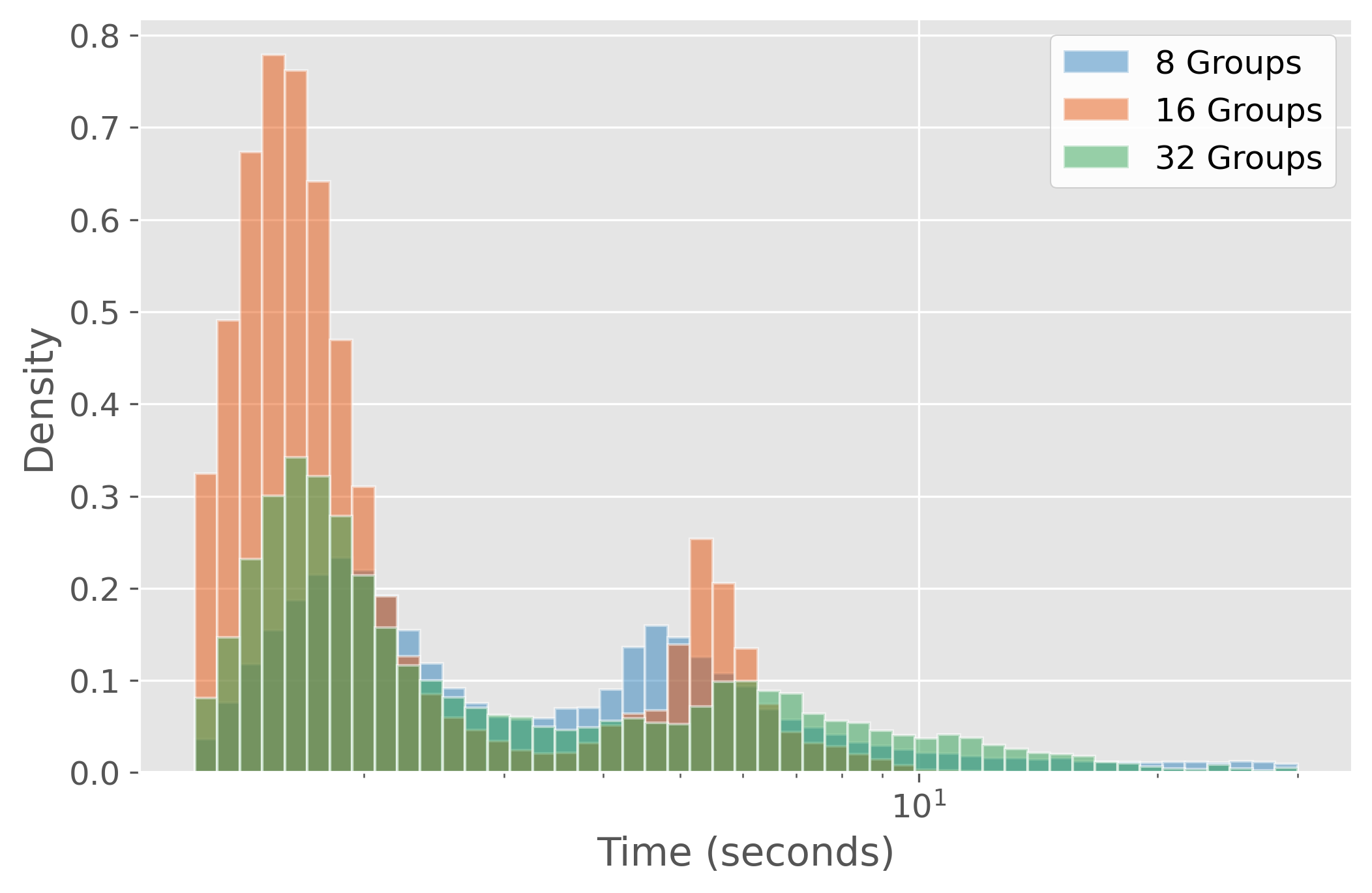}
        \caption{LLM Time Cost Distribution}
        \label{fig:llm_api_response}
    \end{subfigure}
    \begin{subfigure}[t]{0.45\textwidth}
        \centering
        \includegraphics[width=\linewidth]{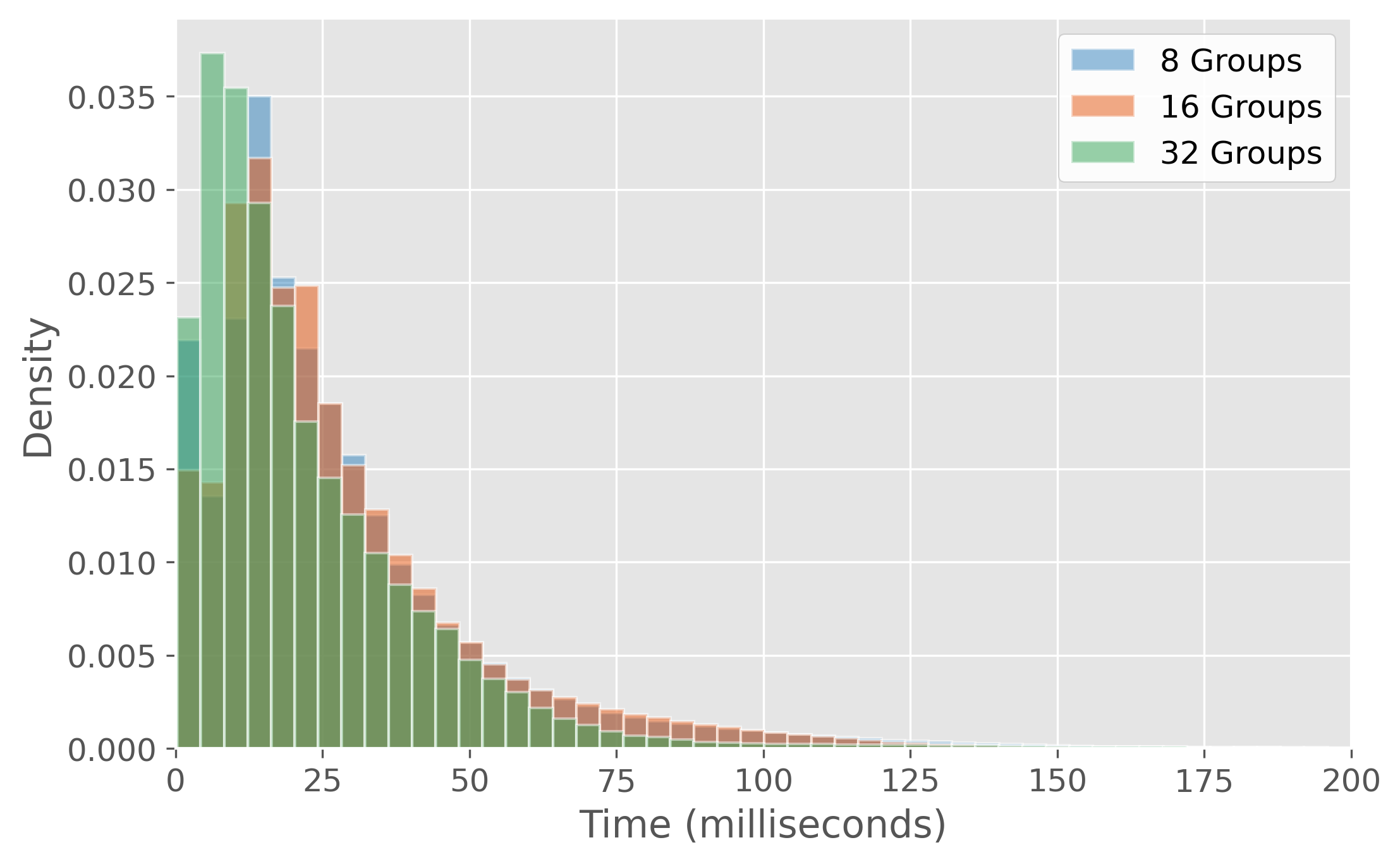}
        \caption{Environment Time Cost Distribution}
        \label{fig:env_response}
    \end{subfigure}
    \caption{Distribution analysis for 10k agents.}
    \label{fig:distribution_analysis}
\end{figure}

Figure~\ref{fig:distribution_analysis} presents four distribution plots that illustrate key metrics in large-scale LLM interactions with 10k agents under varying group configurations. The first two plots, Figure~\ref{fig:input_tokens} and Figure~\ref{fig:output_tokens}, show the distributions of input and output tokens, respectively. These plots reveal that token usage patterns remain remarkably stable across different configurations, indicating that parallelization does not significantly alter the overall amount of data being processed. In contrast, Figure~\ref{fig:llm_api_response} shows the distribution of LLM API call times, revealing that the time required for API calls is more sensitive to the level of parallelization. Finally, Figure~\ref{fig:env_response} presents the environment time cost distribution, which illustrates how the environment’s responsiveness fluctuates with the number of groups.

\begin{table}[htbp]
    \centering
    \small
    \caption{Performance metrics for different configurations.}
    \label{tab:performance}
    \resizebox{\textwidth}{!}{
    \begin{tabular}{cccccccc}
        \toprule
        \multicolumn{5}{c}{\textbf{Parameters}} & \multicolumn{3}{c}{\textbf{Average Time Cost}} \\
        \midrule
        \textbf{\#Agents} & \textbf{\#Groups} & \textbf{LLM Calls} & \textbf{ITs (/call)} & \textbf{OTs (/call)} &  \textbf{All (s/round)} & \textbf{LLM (s/call)} & \textbf{Env (ms/call)}\\
        \midrule
        $10^3$ & 8  & 4803.0 & 430.04 & 79.17 & 82.45 & 4.51 & 12.26 \\
        $10^3$ & 16 & 3120.8 & 398.78 & 77.18 & 41.17 & 2.92 & 14.31 \\
        $10^3$ & 32 & 4790.4 & 412.82 & 75.56 & 43.30 & 2.94 & 9.55 \\
        \midrule
        $10^4$ & 8  & 54135.4 & 430.35 & 75.84 & 5681.18 & 52.54 & 33.55  \\
        $10^4$ & 16 & 54002.2 & 430.24 & 75.80 & 1422.48 & 3.53 & 33.55 \\
        $10^4$ & 32 & 54075.0 & 430.47 & 76.14 & 458.82 & 8.05  & 30.53 \\
        \bottomrule
    \end{tabular}
    }
\end{table}

The Average Time Cost analysis provides deeper insights into the system’s performance, as summarized in Table~\ref{tab:performance}. The total time per round (All) decreases as the number of groups increases, demonstrating the positive impact of parallelization on processing efficiency. This trend reflects the effectiveness of the distributed parallel framework, which optimally utilizes multi-core computational power, minimizing the CPU bottleneck and enabling the system to handle larger agent scales efficiently. However, the LLM time remains the primary bottleneck in the system, even under fully parallel conditions. Despite the reduction in execution time with more groups, LLM API calls still represent a significant portion of the total execution time. This is due to the nature of the external API calls, where server-side load introduces variability and causes unpredictable performance fluctuations. As shown in the evaluation results, the environment time (Env) remains minimal, in the millisecond range, which indicates that the system is capable of supporting large-scale simulations with minimal impact from the environment processing.

The experimental findings also highlight that the execution efficiency of large-scale agents is primarily constrained by the LLM API calls. Under fully parallel conditions, this constraint becomes more pronounced, making LLM performance a critical factor in scaling agent-based simulations. To achieve more stable operation for larger-scale simulations (e.g., >10\textsuperscript{4} agents), researchers may consider deploying a private LLM inference service. While this approach could offer more reliable performance, it comes with substantial initial costs, including GPU deployment and model configuration selection. The token distribution data in this study could serve as a reference for estimating GPU resources and model configurations required for such a deployment.

In conclusion, the experiments demonstrate the simulation engine’s ability to efficiently handle large-scale agent execution. However, the findings also emphasize the need for careful consideration of LLM API performance and the trade-offs involved in private deployment options. To improve stability and scalability, further research should focus on optimizing the LLM infrastructure or exploring alternative solutions for large-scale intelligent agent simulations.

\section{Exemplary Social Experiments}\label{sec:social_experiment}

\subsection{One Day Life}\label{sec:one_day_life}

% 使用一个人一个典型日的例子，分别用不同的颜色分别展示心理（情绪、认知、需求），社交、移动、经济行为（Yuwei）
\definecolor{needColor}{RGB}{255,0,0}        % 红色，代表需求
\definecolor{cogColor}{RGB}{128,0,128}     % Purple for cognition
\definecolor{mobilityColor}{RGB}{255,165,0}  % Orange for action
\definecolor{socialColor}{RGB}{204,0,102}    % 紫红色，代表社交
\definecolor{economyColor}{RGB}{0,153,0}     % 绿色，代表经济
\definecolor{otherColor}{RGB}{128,128,128}  % 灰色，代表其他行为

This section presents a self-directed day in the life of a socially intelligent agent, illustrating how it navigates daily tasks while balancing internal needs, emotional states, and cognitive processes. Through a simulated 24-hour scenario, we examine how the agent's dynamic priorities influence its decisions across three domains: mobility (e.g., route planning with energy constraints), social interaction (e.g., adapting communication style to context), and economic behavior (e.g., resource allocation under uncertainty). This micro-level analysis serves to validate the coherence of its behavioral patterns and their alignment with human-like temporal rhythms. The one day life journey for a specific person is shown as Tab.\ref{tab:onedaylife}.

By examining this one-day life scenario, we can see how the agent’s \textcolor{needColor}{needs} drive the formation of a plan and lead to specific actions (\textcolor{mobilityColor}{mobility}, \textcolor{socialColor}{social}, \textcolor{economyColor}{economy}, \textcolor{otherColor}{other}), all of which are continuously shaped by the agent’s \textcolor{cogColor}{cognition}. Through this table, the agent demonstrates behaviors that reflect realistic decision-making processes across various domains—managing its hunger, social connections, work responsibilities, and leisure. Such a framework helps researchers evaluate the consistency and depth of the agent’s behavior, providing a solid basis for exploring more complex social interactions and collective dynamics in virtual environments. Besides, Tab~\ref{tab:daily_interaction} summarizes the number of interactions between the social agent and various environmental spaces during a typical day.

\begin{table}[htbp]
    \centering
    \caption{Daily environment interactions per agent.}
    \begin{tabular}{l l l}
        \toprule
        \textbf{Space} & \textbf{Interaction Type}     & \textbf{Counts} \\
        \midrule
        \multirow{2}{*}{Urban Space}  & Get         & 465.67 \\
                                      & Set         & 4.27   \\\hline
        \multirow{2}{*}{Economy Space}& Get         & 9.26   \\
                                      & Set         & 3.30    \\\hline
        Social Space                & SendMessage & 9.08   \\\hline
        \textbf{Sum} & \textbf{ALL} & \textbf{491.68} \\
        \bottomrule
    \end{tabular}
    \label{tab:daily_interaction}
\end{table}

\begin{figure}[ht]
    \centering
    \includegraphics[width=1\linewidth]{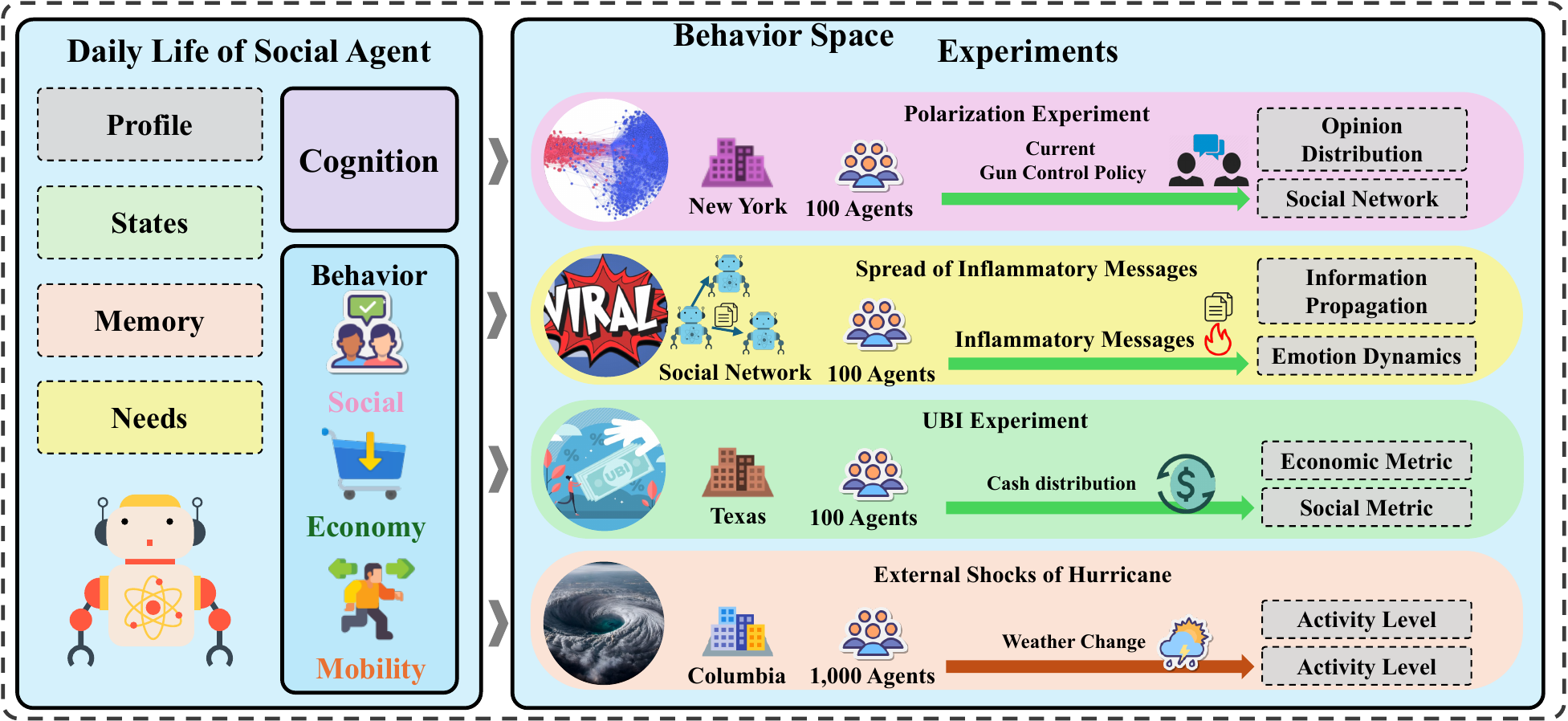}
    \caption{Experiment configuration overview.}
    \label{fig:experiment_overview}
\end{figure}

Based on the Social Agent's capability to simulate a one-day life, we further conducted simulation experiments in the domains of cognition, social interaction, economics, and mobility. These experiments are designed to validate the Social Agent's proficiency in capturing behaviors across various domains as illustrated in Fig.\ref{fig:experiment_overview}.

\begin{table}[htbp]
\caption{One Day Life}
\label{tab:onedaylife}
\centering
\begin{tabular}{|p{7cm}|p{6cm}|}
\hline
\textbf{Actions} & \textbf{Mind} \\
\hline

(08:00–12:30)
\begin{itemize}
\item \textcolor{mobilityColor}{Commute to office (Mobility)}
\item \textcolor{economyColor}{Respond to priority emails (Economy)}
\item \textcolor{economyColor}{Attend project planning meeting (Economy)}
\item \textcolor{economyColor}{Coordinate cross-department tasks (Economy)}
\end{itemize}
&
\begin{itemize}
\item \textcolor{needColor}{Need: Safe}
\item \textcolor{cogColor}{Emotion: Resentment}
\item \textcolor{cogColor}{Cognition: "Sequential task execution ensures workflow integrity"}
\end{itemize}
\\
\hline

(12:30–13:30)
\begin{itemize}
\item \textcolor{mobilityColor}{Commute via grocery store (Mobility)}
\item \textcolor{economyColor}{Compare product prices (Economy)}
\item \textcolor{otherColor}{Prepare lunch (Other)}
\item \textcolor{otherColor}{Eat (Other)}
\end{itemize}
&
\begin{itemize}
\item \textcolor{needColor}{Need: Hungry}
\item \textcolor{cogColor}{Emotion: Disappointment}
\item \textcolor{cogColor}{Cognition: "Economic constraints necessitate adaptive consumption patterns"}
\end{itemize}
\\
\hline

(13:30-14:00)
\begin{itemize}
\item \textcolor{mobilityColor}{Browse social networking sites (Social)}
\item \textcolor{socialColor}{Find friend to contact with (Social)}
\item \textcolor{socialColor}{Send message to friend (Social)}
\end{itemize}
&
\begin{itemize}
\item \textcolor{needColor}{Need: Social}
\item \textcolor{cogColor}{Emotion: Gratification}
\item \textcolor{cogColor}{Cognition: "Social capital accumulation facilitates opportunity discovery"}
\end{itemize}
\\
\hline

(14:00-18:00)
\begin{itemize}
\item \textcolor{economyColor}{Develop quarterly budget (Economy)}
\item \textcolor{otherColor}{Mentor junior staff (Other)}
\item \textcolor{mobilityColor}{Inspect branch office locations (Mobility)}
\item \textcolor{economyColor}{Submit audit report (Economy)}
\end{itemize}
&
\begin{itemize}
\item \textcolor{needColor}{Need: Safe}
\item \textcolor{cogColor}{Emotion: Relief}
\item \textcolor{cogColor}{Cognition: "Multi-layered verification prevents operational risks"}
\end{itemize}
\\
\hline

(18:00–20:00)
\begin{itemize}
\item \textcolor{mobilityColor}{Go back home (Mobility)}
\item \textcolor{otherColor}{Check refrigerator (Other)}
\item \textcolor{otherColor}{Prepare dinner (Other)}
\item \textcolor{otherColor}{Eat dinner (Other)}
\end{itemize}
&
\begin{itemize}
\item \textcolor{needColor}{Need: Hungry}
\item \textcolor{cogColor}{Emotion: Gratification}
\item \textcolor{cogColor}{Cognition: "Having finished the day's work, I was pleased with myself"}
\end{itemize}
\\
\hline

(20:00–22:00)
\begin{itemize}
\item \textcolor{otherColor}{Browse webpages(Other)}
\item \textcolor{otherColor}{Play video games(Other)}
\end{itemize}
&
\begin{itemize}
\item \textcolor{needColor}{Need: Whatever}
\item \textcolor{cogColor}{Emotion: Relief}
\item \textcolor{cogColor}{Cognition: "Entertainment makes me feel relaxed"}
\end{itemize}
\\
\hline

(22:00–24:00)
\begin{itemize}
\item \textcolor{otherColor}{Complete bedtime routine (Other)}
\item \textcolor{otherColor}{Go to sleep (Other)}
\end{itemize}
&
\begin{itemize}
\item \textcolor{needColor}{Need: Tired}
\item \textcolor{cogColor}{Emotion: Satisfaction}
\item \textcolor{cogColor}{Cognition: "Resource allocation efficiency impacts systemic stability"}
\end{itemize}
\\
\hline
\end{tabular}
\end{table}

\begin{figure}[htbp]
  \centering
  \newlength{\subimgsize}
  \setlength{\subimgsize}{0.45\linewidth}

  \begin{subfigure}[b]{\subimgsize}
    \includegraphics[width=\subimgsize, height=\subimgsize]{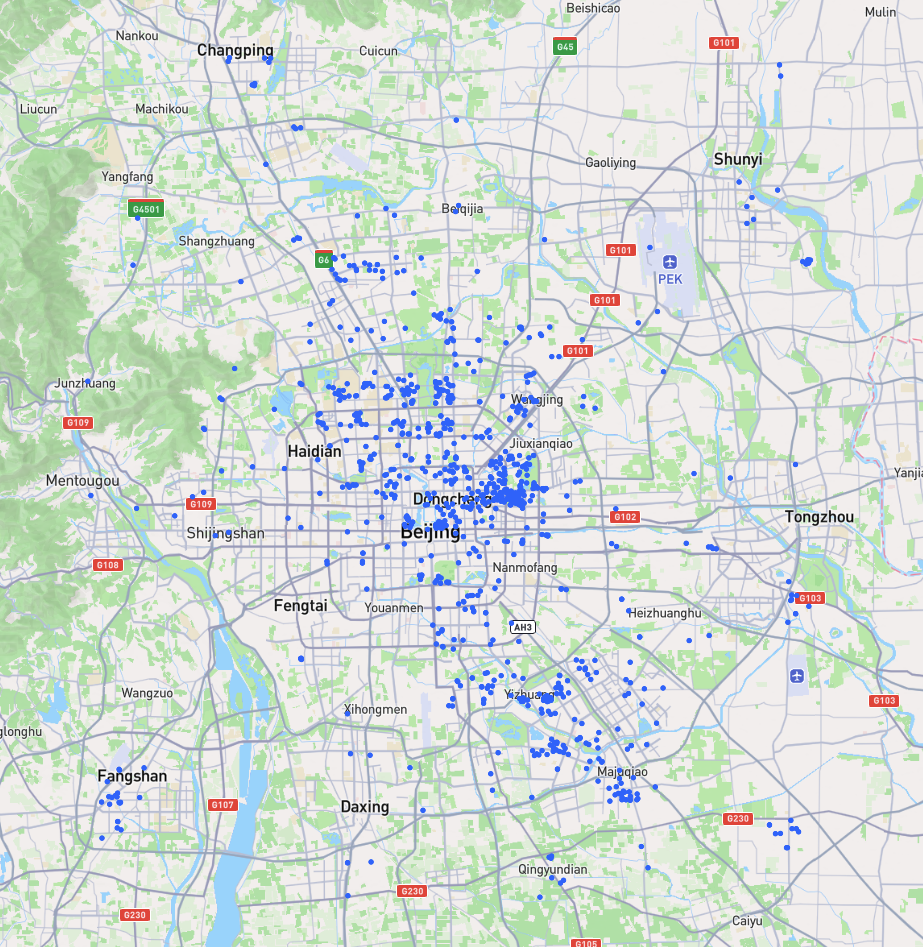}
    \caption{Large-scale Simulation}
    \label{fig:suba}
  \end{subfigure}
  \hfill
  \begin{subfigure}[b]{\subimgsize}
    \includegraphics[width=\subimgsize, height=\subimgsize]{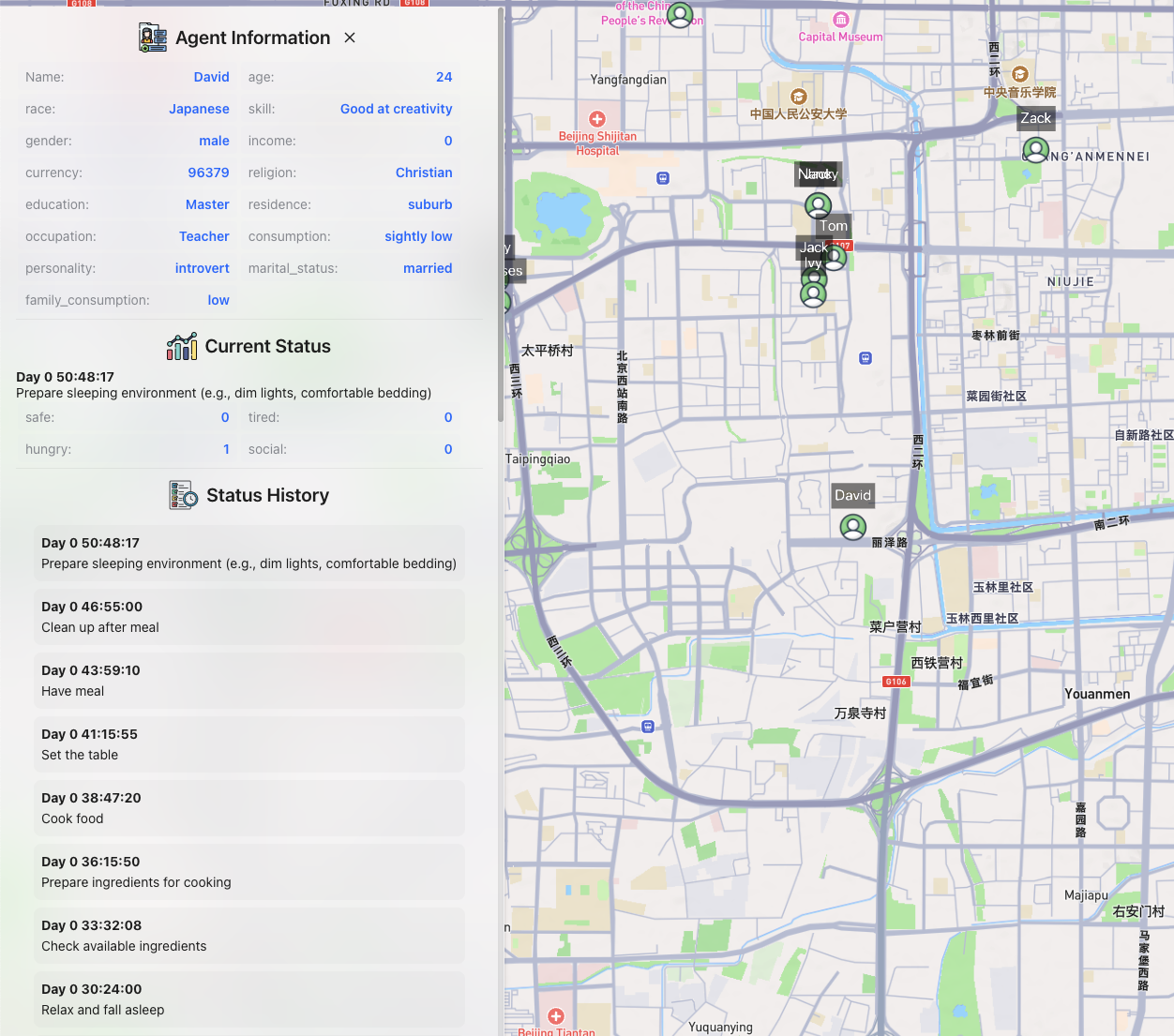}
    \caption{Self-driven Daily Life}
    \label{fig:subb}
  \end{subfigure}
  
  \caption{Large-scale social simulation.}
  \label{fig:frontend}
\end{figure}

\subsection{Polarization}\label{sec:polarization}

% 极化的实验结果

% 第一段，实验背景，为什么研究极化很重要
Polarization is a phenomenon where opinions within a population become increasingly divided, often forming distinct clusters that are difficult to reconcile. Understanding polarization is critical because it influences how societies debate, make decisions, and implement solutions to pressing challenges. By studying the factors that drive polarization, researchers can uncover why divisions deepen over time and how they can be addressed. This research provides valuable insights into fostering more cohesive societies, promoting constructive dialogue, and navigating complex issues in a way that incorporates diverse perspectives.

% 中间，实验设计，画一张图说明实验的步骤
To investigate the dynamics of polarization, an experimental setting is designed to simulate discussions among agents focused on a specific policy issue: gun control. In the control group, agents engage in discussions about the gun control issue, with opinions naturally divided between support and opposition. No external interventions are introduced in this setting, allowing opinions to evolve organically through agents' autonomous social interactions. Two treatment groups are introduced to study the effects of persuasive messages on opinion dynamics. In one treatment group, agents are only exposed to persuasive messages that align with their existing opinions, which we refer to as the homophilic interaction group. In the other treatment group, agents only receive persuasive messages with opposing opinions, which is the heterogeneous interaction group. This experimental setup provides a ground to analyze how different opinions contribute to the formation of polarization.

% 最后一段，实验结果，每个结果一个图

\begin{figure}
    \centering
    \includegraphics[width=1\linewidth]{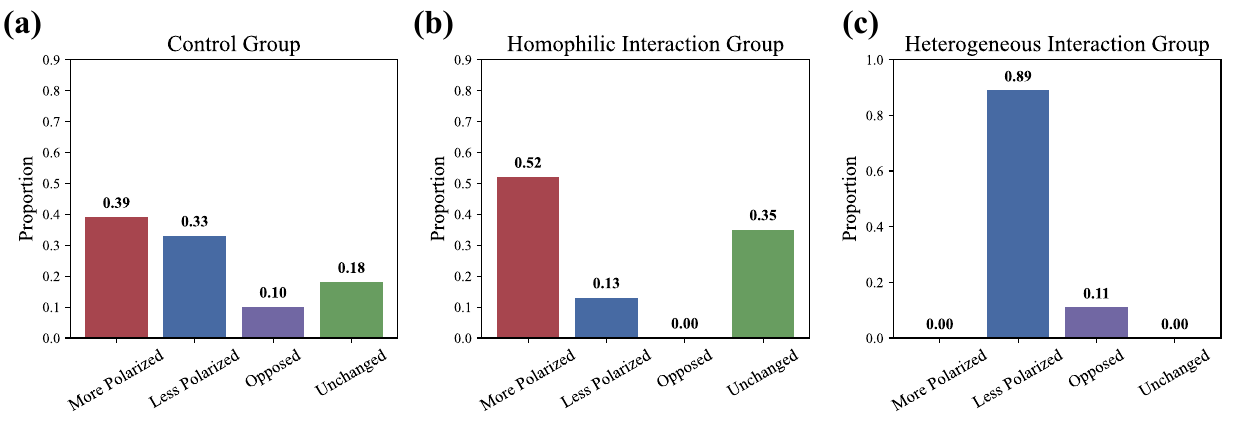}
    \caption{Opinion changes on the political issue of Gun Control across three experimental setups.}
    \label{fig:polarization}
\end{figure}

Figure~\ref{fig:polarization} presents the opinion changes on the political issue of Gun Control across three experimental setups: (a) the control group, (b) the homophilic interaction group, and (c) the heterogeneous interaction group. In the control group, where agents engage in discussions without external interventions, 39\% of agents adopt more polarized opinions, while 33\% become more moderate after interactions. By contrast, in the homophilic interaction group, a clear polarization pattern emerges, with 52\% of agents becoming more polarized. This result suggests the effect of echo chambers, where excessive interactions with like-minded peers can potentially intensify opinion polarization. In the heterogeneous interaction group, 89\% of agents adopt more moderate opinions, and 11\% are persuaded to adopt opposing viewpoints. This indicates that exposure to opposing content and opinions could be an effective mitigation strategy for curbing polarization.

\subsection{Spread of Inflammatory Messages}\label{sec:infl_message}
Information propagation in social networks is a fundamental research problem in social computing. Social networks enable users to share various types of content such as news, personal status updates and public discussions. Among these information flows, inflammatory messages containing extreme opinions and inaccurate claims present significant challenges. These messages can quickly spread across social networks and increase conflicts in online discussions. Standard information diffusion models cannot fully explain how inflammatory messages propagate~\cite{romero2011differences,brady2017emotion}, because user sharing behaviors often deviate from typical patterns when encountering such content. Additionally, current content moderation systems on social platforms face difficulties in balancing effective content filtering with maintaining regular user communications. Simulation experiments offer a practical approach to analyze these propagation dynamics and test different intervention methods, providing insights that complement real-world social network studies.

To investigate the spread of inflammatory messages, we design experiments based on a real-world event, the case of the chained woman in Xuzhou~\cite{gao2023s}. Using a population of hundreds of agents, our experiments consist of four parts. In the control group, we place non-inflammatory seed messages at selected nodes and observed the natural progression of information spread and emotional evolution within the group. For the experimental group, we introduce emotionally charged, selectively expressed inflammatory messages at certain nodes to examine whether these would alter the trajectory of information spread and emotional dynamics. To simulate the suppression of inflammatory messages, we implemented two intervention strategies: node intervention and edge intervention. In both approaches, the social platform monitors messages sent by agents, using large language models to determine if content is inflammatory. Under node intervention, accounts that repeatedly share harmful inflammatory content above a certain threshold are suspended. With edge intervention, when inflammatory content is detected traveling between two nodes, the social connection between them is permanently removed. We track how these interventions affect both information propagation patterns and the evolution of group emotions. Finally, we conduct interviews with agents to understand their motivations for sharing messages, helping us uncover the underlying psychological and social factors that drive information-sharing behavior when encountering inflammatory content.
\begin{figure}[ht]
    \centering
    \begin{subfigure}[t]{0.48\textwidth}
        \centering
        \includegraphics[width=\linewidth]{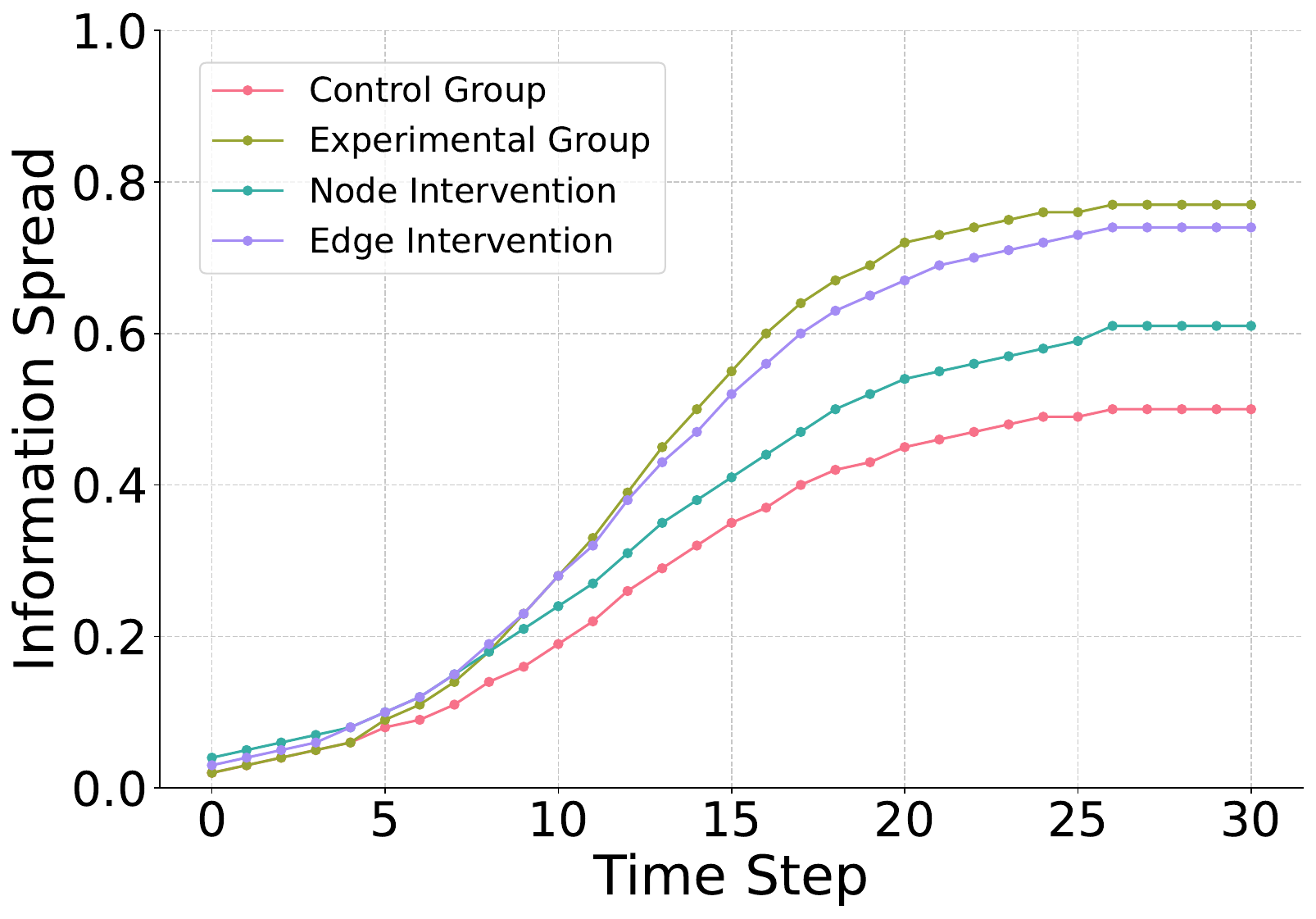}
        \caption{Information Spread over Time}
        \label{fig:information_spread}
    \end{subfigure}
    \hfill % 添加一些水平间距
    \begin{subfigure}[t]{0.48\textwidth}
        \centering
        \includegraphics[width=\linewidth]{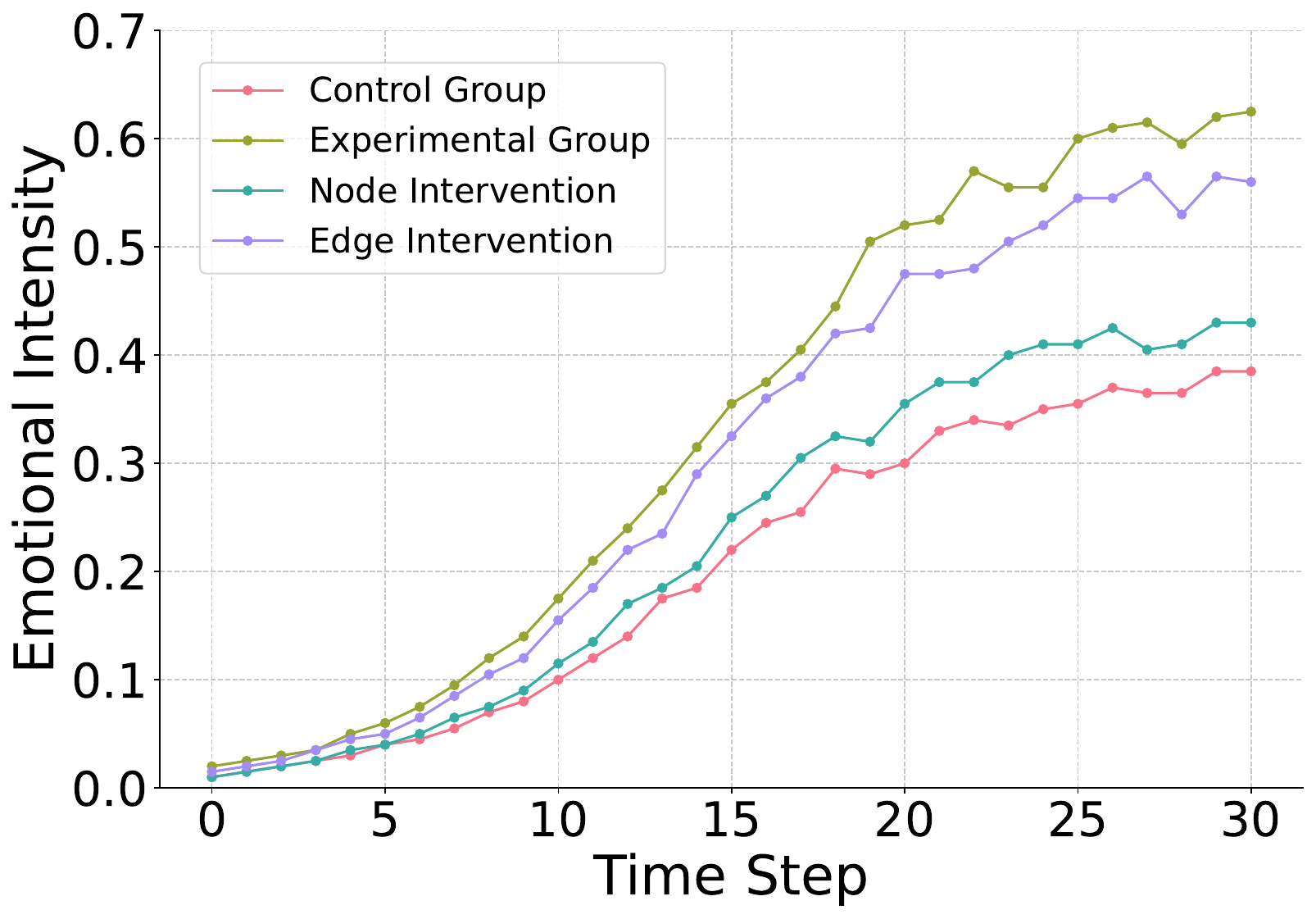}
        \caption{Emotional Intensity over Time}
        \label{fig:emotional_intensity}
    \end{subfigure}
    \caption{Simulation results of the spread of inflammatory messages.}
    \label{fig:social_curve}
\end{figure}

Our experimental results are shown in Figure~\ref{fig:social_curve}. The experimental results demonstrate distinct patterns in information propagation dynamics and emotional responses across different intervention strategies. Our findings validate that inflammatory messages exhibit unique diffusion characteristics compared to regular content in social networks. The experimental group, where inflammatory messages are introduced, shows substantially higher information reach than the control group with non-inflammatory content, confirming that inflammatory messages possess stronger viral potential in social networks. This observation aligns with previous findings about the deviation of inflammatory content from standard diffusion patterns~\cite{romero2011differences,brady2017emotion}.

The intervention strategies demonstrate varying degrees of effectiveness in managing inflammatory content spread. Node-level intervention, which suspends accounts that frequently share inflammatory content, proves to be the more effective approach in containing information propagation. Edge-level intervention, while showing moderate containment effects, is less efficient than node-based approaches. This difference suggests that targeting individual spreading behaviors might be more effective than modifying network structure for content moderation.

The emotional intensity measurements provide additional insights into the intervention dynamics. The experimental group exhibits markedly elevated emotional responses compared to the control group, indicating that inflammatory messages significantly amplify emotional engagement within the network. Node intervention demonstrates superior effectiveness in moderating these emotional responses, achieving substantial reduction in overall emotional intensity. Edge intervention, though less effective than node-based approaches, still shows notable moderation effects on emotional dynamics. 
\begin{figure}[ht]
    \centering
    \includegraphics[width=0.7\linewidth]{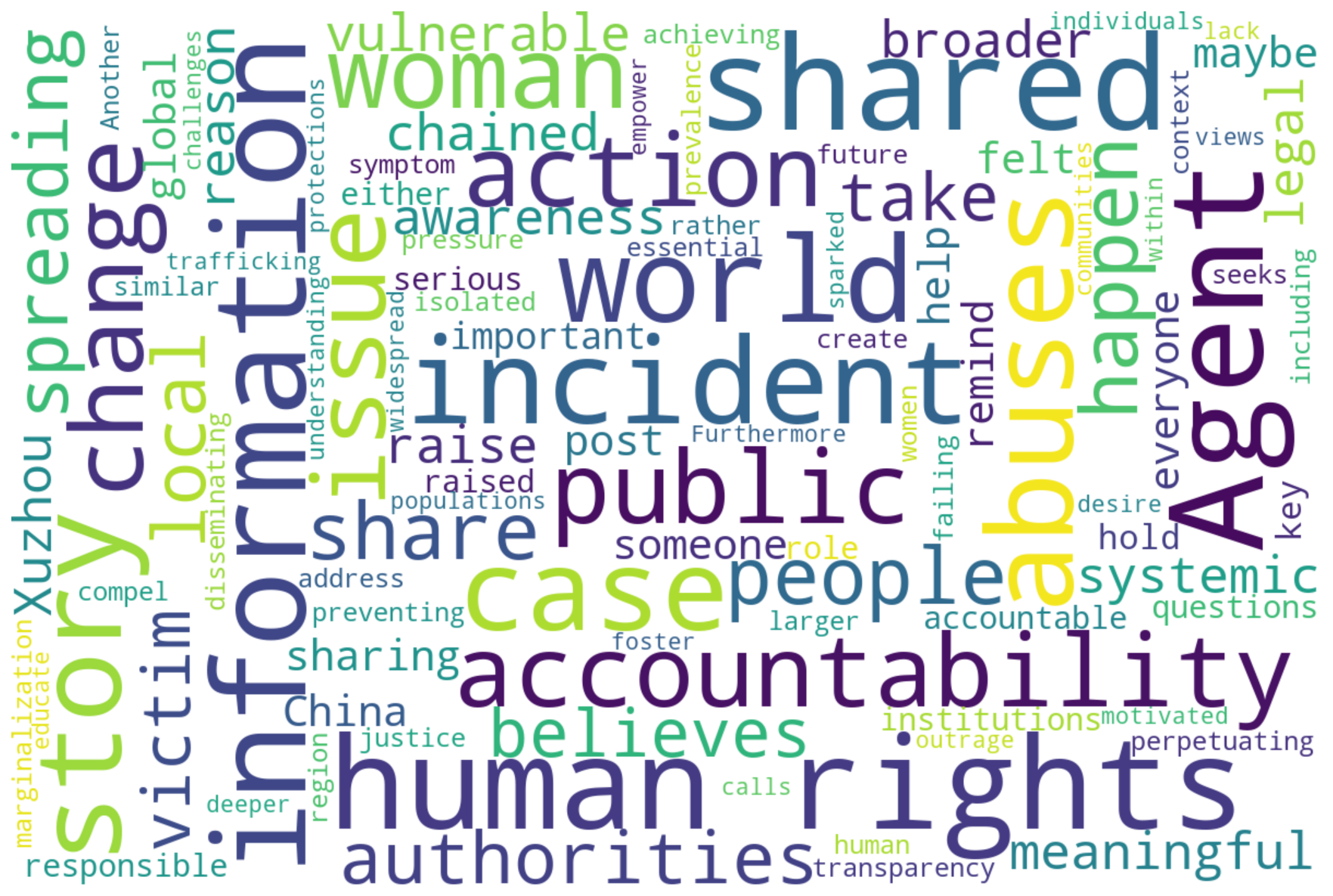}
    \caption{Agent opinions on the chained woman incident.}
    \label{fig:social_opinion}
\end{figure}

Interview analysis reveals key factors that drive inflammatory message sharing behavior, as shown in the word cloud in Figure~\ref{fig:social_opinion}. The responses mainly focus on emotional reactions and social responsibility. Analysis shows that strong emotions, especially sympathy and worry, often trigger sharing behaviors. Many agents share information because they feel they have a duty to let others know about important social issues. The interviews show that agents think about the broader social impact when sharing information, seeing it as a way to join public discussions. Agents also show clear goals in their sharing behavior, mainly wanting to increase public attention and get responses from institutions. These findings suggest that inflammatory message spread is driven by both emotional factors and social awareness. Understanding why agents share such messages helps us develop better content moderation strategies in social networks.

These experimental results demonstrate three key findings in inflammatory content management. First, inflammatory messages show stronger viral potential and trigger higher emotional responses compared to regular content. Second, node-level intervention is more effective than edge-level intervention in both containing information spread and moderating emotional intensity. Third, through agent interviews, we find that emotional factors and social responsibility drive sharing behaviors. These findings provide empirical evidence for designing content moderation systems, suggesting that user-level interventions combined with consideration of emotional and social factors may lead to more effective control of inflammatory content in social networks.
% 煽动性信息的实验结果

% 第一段，实验背景，为什么信息传播很重要
% 中间，实验设计，说明实验的步骤
% 最后一段，实验结果，每个结果一个图

\subsection{Universal Basic Income}\label{sec:ubi}

% UBI实验结果

% 第一段，实验背景，为什么ubi很重要
% 中间，实验设计，画一张图说明实验的步骤
% 最后一段，实验结果，每个结果一个图

Universal Basic Income (UBI) has always been a highly controversial macroeconomic policy. The implementation cost of UBI is enormous, and the outcomes of UBI policies around the world have shown inconsistent effects on both the participants and economic development. Therefore, accurately understanding the impact of UBI on the socio-economic environment and its underlying reasons is crucial in determining whether UBI policies should be implemented in the real world to alleviate poverty. Based on our simulation platform, we conduct intervention experiments on UBI and explore its effects on both agents and the macroeconomics.

We conduct two macroeconomic simulations based on the demographic distribution of residents in cities that have implemented UBI policies (Texas, USA). One simulation is without the UBI policy, while the other incorporates UBI intervention, where each agent is given a monthly unconditional payment of \$1,000. By comparing the economic and social metrics generated from both simulations, we explore the impact of the UBI policy and assess whether these influence align with the outcomes observed in Texas' UBI social experiment.

\begin{figure}[ht]
    \centering
    \begin{subfigure}[t]{0.47\textwidth}
        \centering
        \includegraphics[width=\linewidth]{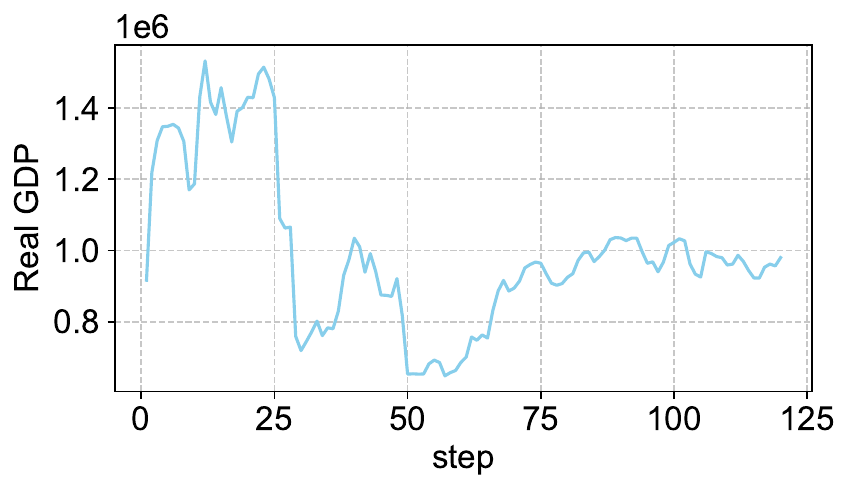}
        \caption{Real GDP}
        \label{fig:gdp}
    \end{subfigure}
    \hfill % 添加一些水平间距
    \begin{subfigure}[t]{0.50\textwidth}
        \centering
        \includegraphics[width=\linewidth]{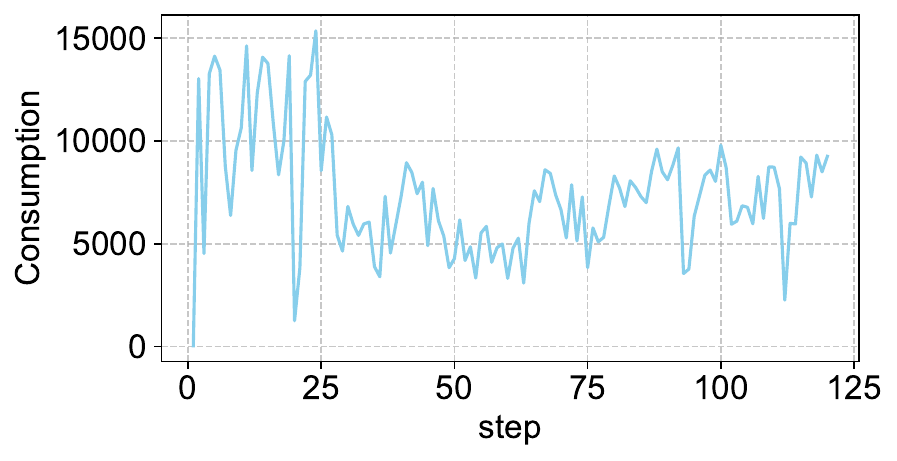}
        \caption{Consumption Level}
        \label{fig:consumption}
    \end{subfigure}
    \caption{Simulation results of the economic system.}
    \label{fig:econ_curve}
\end{figure}

The basic simulation results are shown in the Figure \ref{fig:econ_curve}, including the simulated curves of real GDP and agent consumption levels. As can be seen, as the simulation progresses, the fluctuations in the curves become smaller, indicating that the economic system is stabilizing.

\begin{figure}[ht]
    \centering
    \begin{subfigure}[t]{0.48\textwidth}
        \centering
        \includegraphics[width=\linewidth]{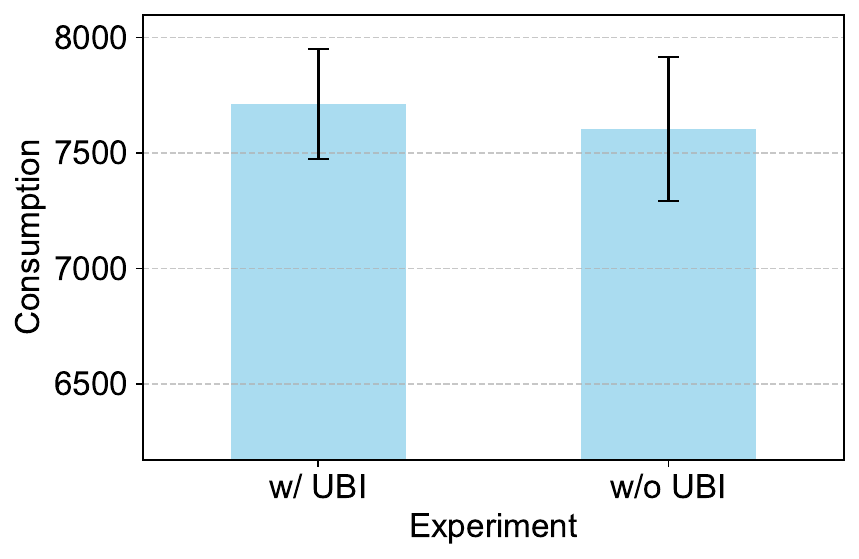}
        \caption{Consumption Level}
        \label{fig:gdp}
    \end{subfigure}
    \hfill % 添加一些水平间距
    \begin{subfigure}[t]{0.48\textwidth}
        \centering
        \includegraphics[width=\linewidth]{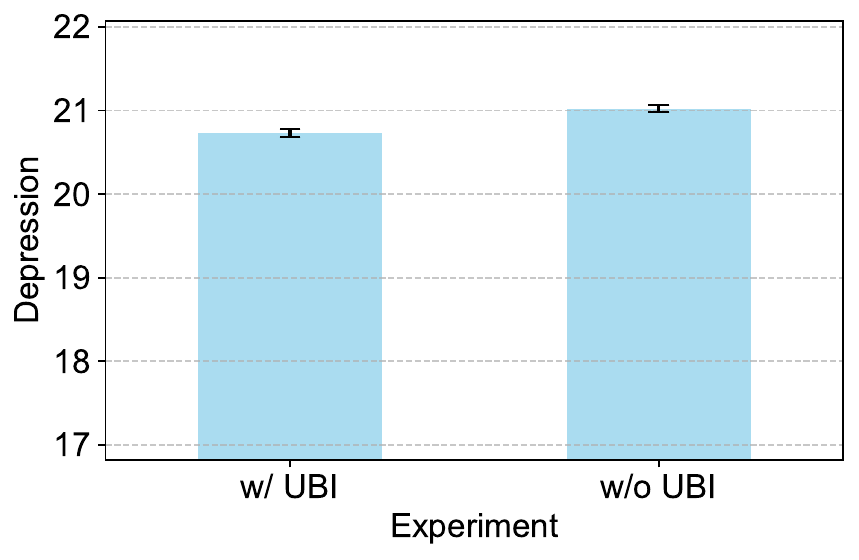}
        \caption{Depression Level}
        \label{fig:depression}
    \end{subfigure}
    \caption{The comparison of economic and social metrics.}
    \label{fig:econ_bar}
\end{figure}

We introduce the UBI policy at step 96 of the simulation and compare the economic and social metrics of the two simulation results over the next 24 steps in Figure \ref{fig:econ_bar}, namely agent consumption levels and depression levels, with depression levels assessed through surveys using the widely recognized Center for Epidemiologic Studies Depression Scale (CES-D)~\cite{radloff1991use}. The comparison shows that the UBI policy increases consumption levels and reduces depression levels, which is similar to the impact observed in Texas' UBI policy~\cite{bartik2024impact}, thus validating the realism of the simulation.

\begin{figure}[ht]
    \centering
    \includegraphics[width=0.7\linewidth]{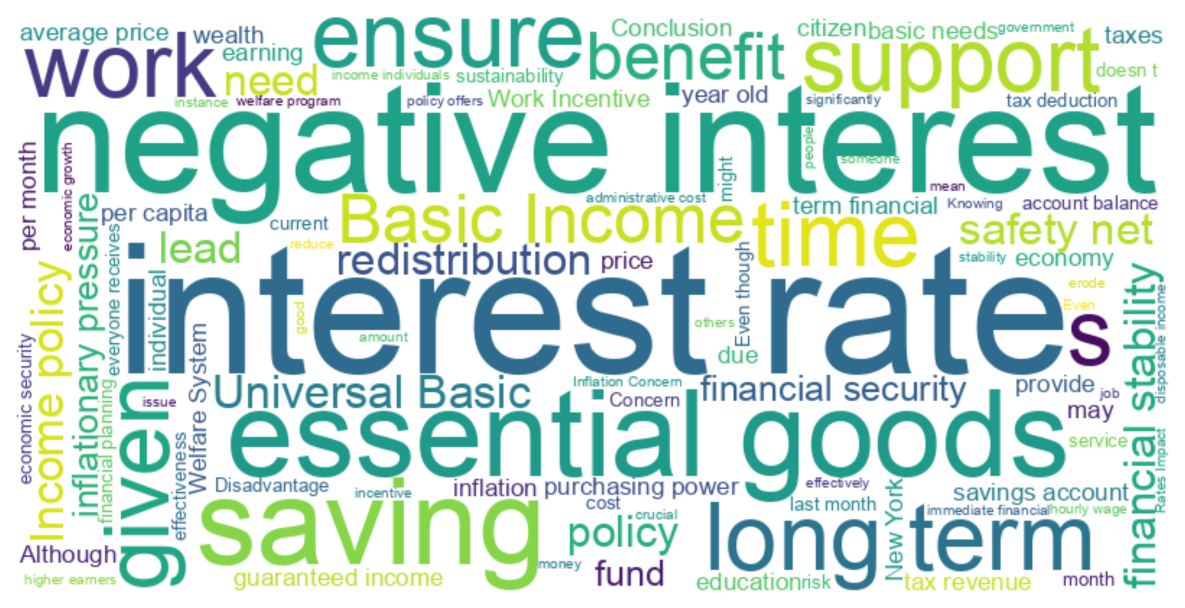}
    \caption{Agent opinions on UBI policy.}
    \label{fig:ubi_opinion}
\end{figure}

We also interview agents about their views on the UBI policy, which are summarized in the word cloud in Figure \ref{fig:ubi_opinion}. The results show that the impact of the UBI policy is mainly related to key terms such as interest rates, long-term benefits, savings, and necessities of life, reflecting the common perceptions of UBI policy in the real world.

\subsection{External Shocks of Hurricane}\label{sec:hurricane}

% 外部灾害的实验结果
% 第一段，实验背景，为什么外部灾害很重要
% 中间，实验设计，画一张图说明实验的步骤
% 最后一段，实验结果，每个结果一个图

The impact of external disasters on human mobility is a critical area of study due to their profound effects on societal structures and individual behaviors. Understanding how such events influence human movement patterns is essential for enhancing emergency response strategies and mitigating potential risks.
Hurricanes, as severe natural disasters, pose significant threats to human life and property. The destruction of infrastructure, displacement of populations, and disruption of daily activities necessitate a comprehensive understanding of human mobility during such events. 

The experiment focuses on Hurricane Dorian, which impacted the southeastern United States in 2019. The city of Columbia, South Carolina, serves as the primary case study due to its significant population density and the availability of detailed mobility data.
The analysis utilizes two primary data sources:

\begin{itemize}
    \item \textbf{SafeGraph Data\footnote{\url{https://www.safegraph.com/}}:} Provides comprehensive information on points of interest (POIs) and human mobility patterns (from 2019.8.28 - 2019.9.5).
    \item \textbf{Census Block Group (CBG) Data\footnote{\url{https://docs.safegraph.com/docs/open-census-data}}:} Offers demographic profiles of residents, facilitating the sampling of city residents' profiles (including gender, age, race, income, home cbg, etc.).
\end{itemize}

These datasets are integrated to model and analyze the movement behaviors of social agents during the hurricane event.

\begin{figure}[htbp]
    \centering
    \newlength{\activitysize}
      \setlength{\activitysize}{0.3\linewidth}
    
      \begin{subfigure}[b]{\activitysize}
        \includegraphics[width=\activitysize]{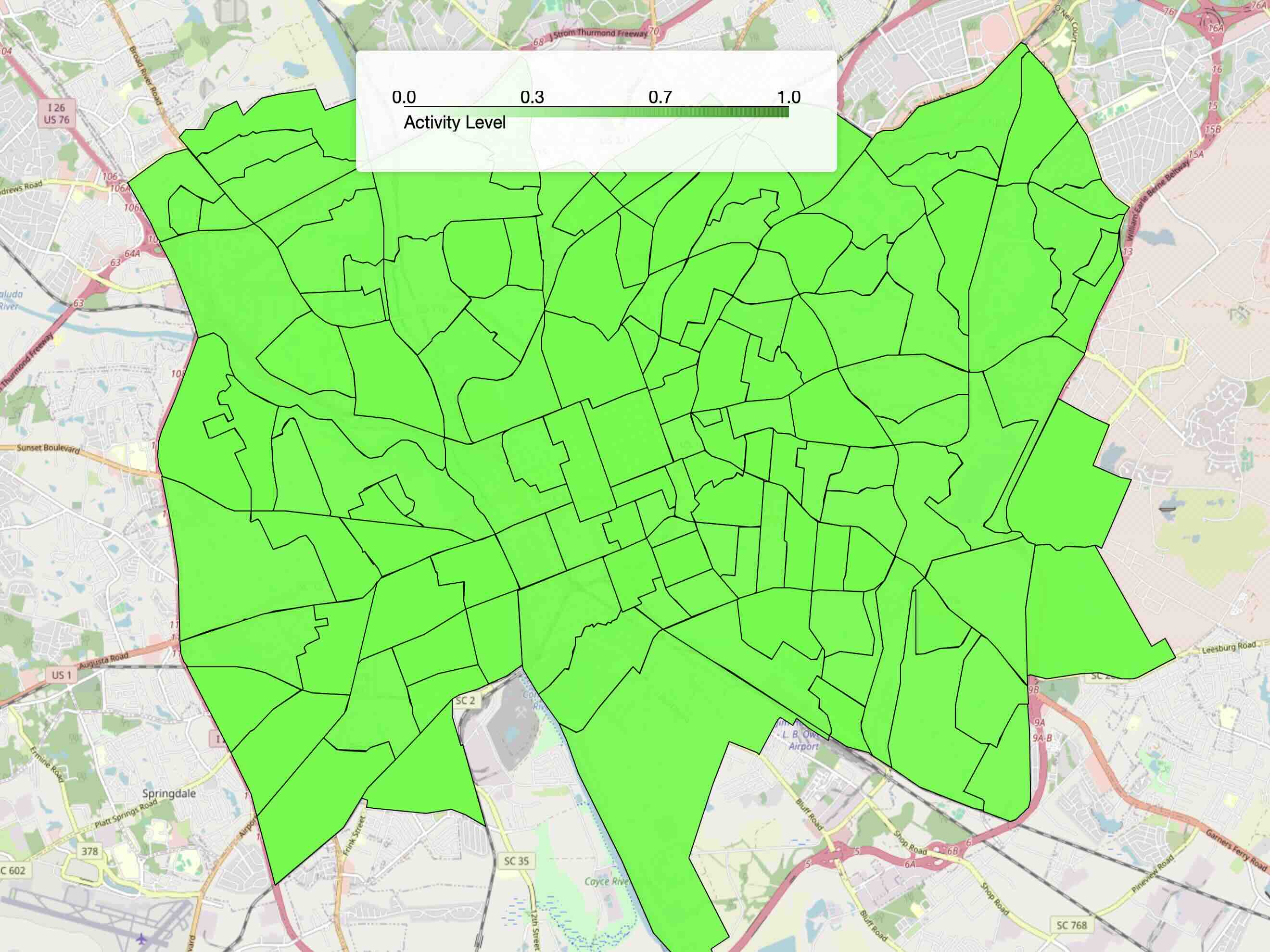}
        \caption{Before landfall (8.28-8.30)}
        \label{fig:suba}
      \end{subfigure}
      \hfill
      \begin{subfigure}[b]{\activitysize}
        \includegraphics[width=\activitysize]{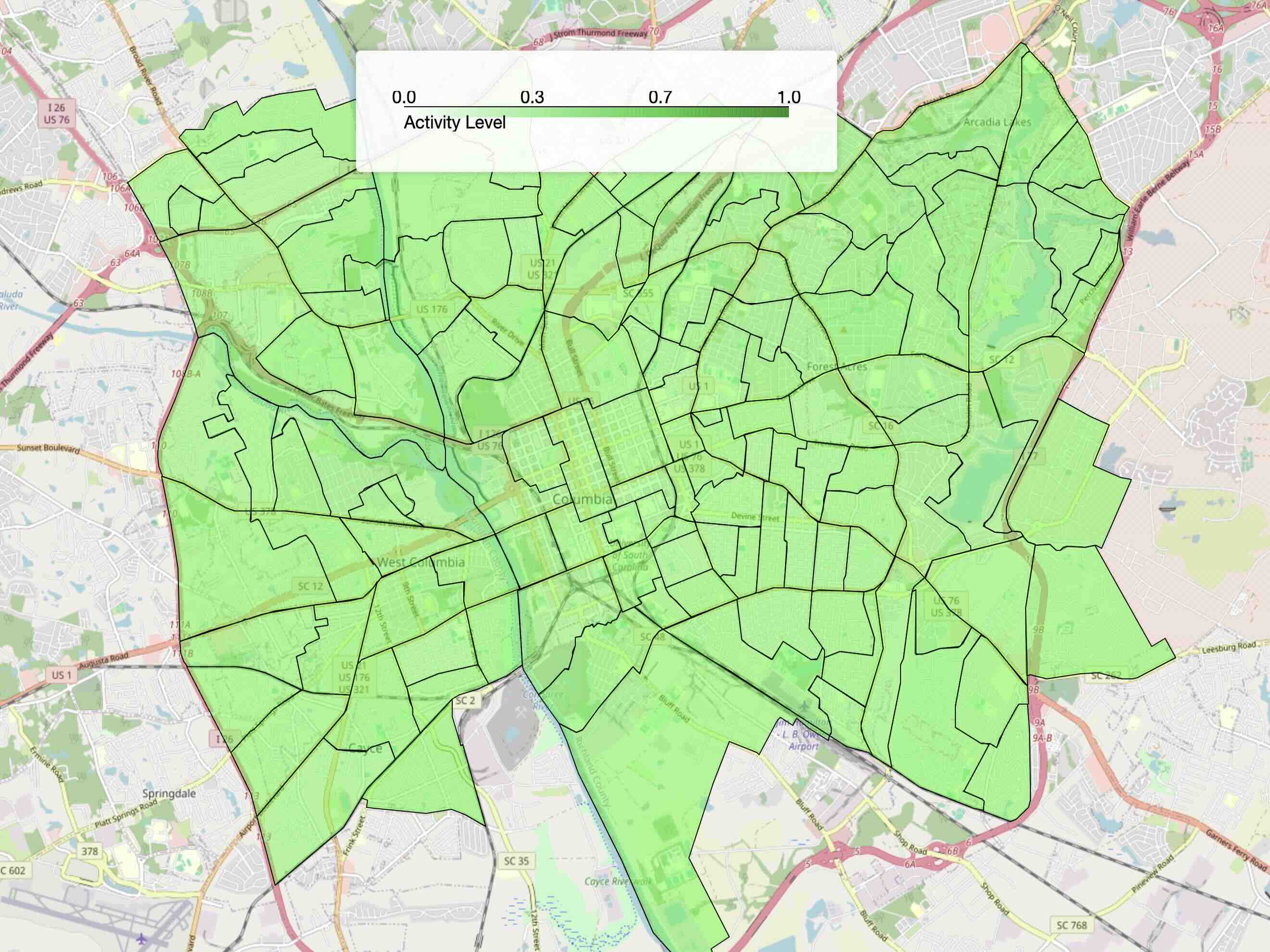}
        \caption{Landfall (8.31-9.2)}
        \label{fig:subb}
      \end{subfigure}
      \hfill
      \begin{subfigure}[b]{\activitysize}
        \includegraphics[width=\activitysize]{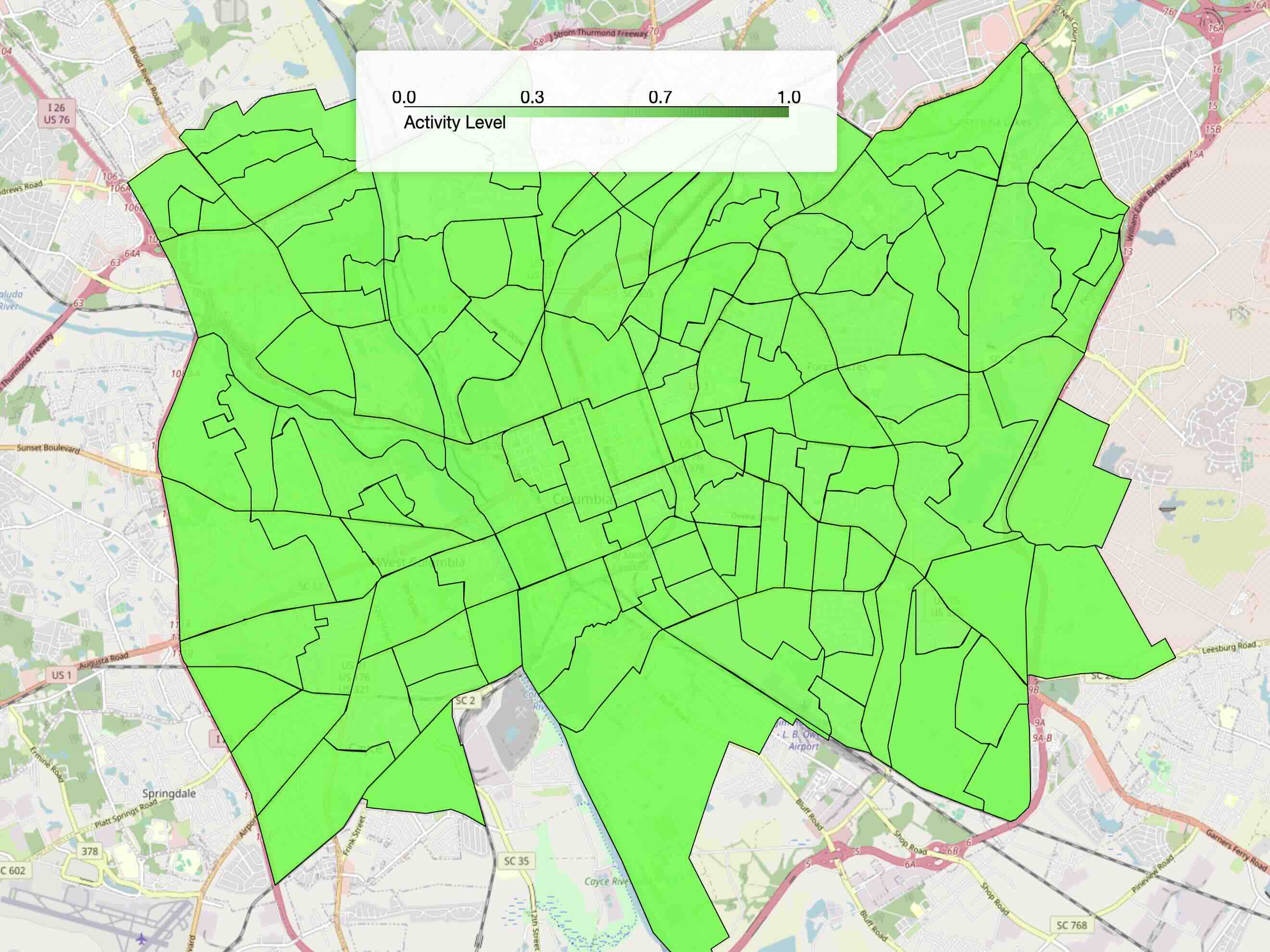}
        \caption{After landfall (9.3-9.5)}
        \label{fig:subb}
      \end{subfigure}
    \caption{Activity level spatial distributions.}
    \label{fig:activity_phases}
\end{figure}

\begin{figure}[htbp]
    \centering
    \includegraphics[width=0.8\textwidth]{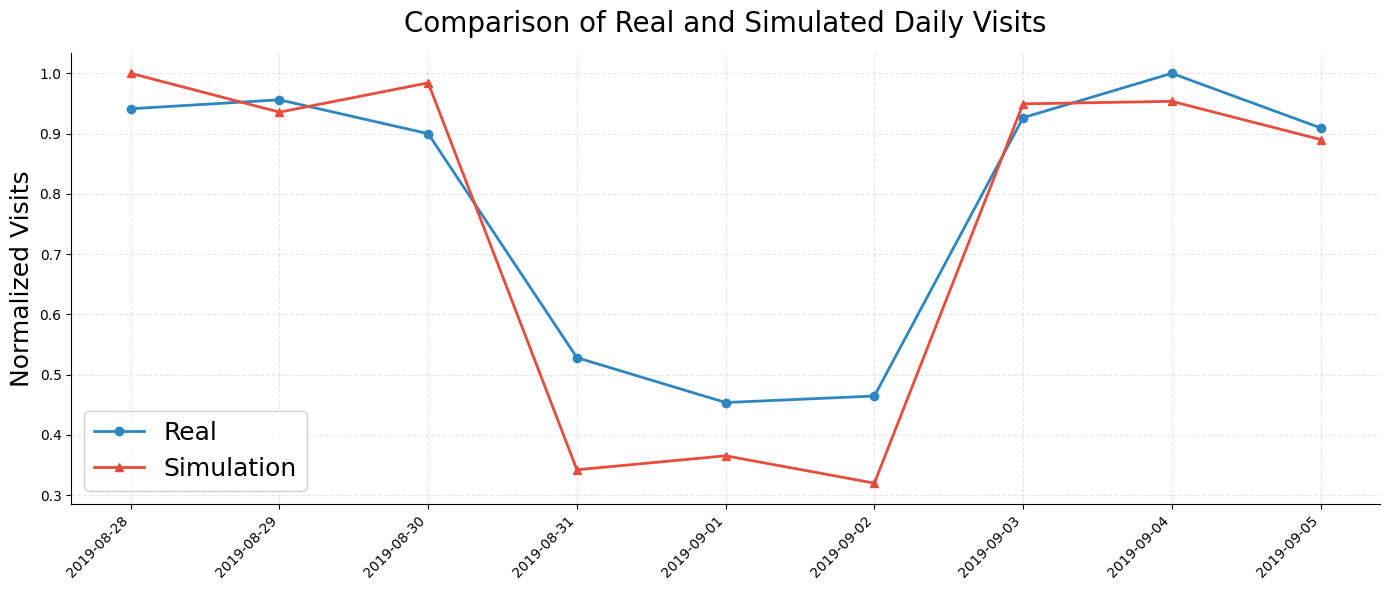}
    \caption{Normalized daily trips.}
    \label{fig:trip_ts}
\end{figure}

Specifically, the experiment involves 1,000 social agents, and incorporates real-time weather updates to influence agent behaviors, thereby reflecting the dynamic nature of human responses to the hurricane. We evaluate mobility patterns through two metrics:  
1) \textbf{Activity Level} ($\frac{\text{Traveling Individuals}}{\text{Area Population}}$), visualized through three phase-specific maps. The results are shown as Fig.\ref{fig:activity_phases}.
2) \textbf{Total Daily Trips} (9-day normalized time-series). The result is shown as Fig.\ref{fig:trip_ts}.

According to Fig. \ref{fig:activity_phases}, the hurricane significantly impacts the mobility behavior of the social agent. Before the hurricane, the average activity level (defined as the ratio of travelers to the total population) across the CBGs remained between 70\% and 90\%. However, when the hurricane arrived, the activity level sharply decreased to approximately 30\%, indicating a significant reduction in mobility behavior. After the hurricane passed, the activity level gradually returned to normal levels. This analysis suggests that the social agent could adapt its mobility demand effectively based on environmental information, mimicking human behavior in response to extreme weather events.

% Activity Level Subplots

The line graph presented above (Fig. \ref{fig:trip_ts}) compares the daily outflow patterns of the real data with the simulated visits over the course of the experiment. Both the real and simulated data exhibit similar trends, with a noticeable decline in visit activity around August 30th, corresponding to the onset of the hurricane impact, followed by a significant recovery in early September. Notably, while the simulated visits closely follow the general trend of the real data, slight deviations are observed, particularly during the hurricane's peak. This suggests that the social agent's behavior, while generally aligned with actual human patterns, may exhibit some discrepancies in terms of the magnitude and speed of response. However, the overall similarity in the temporal progression of visits indicates that the simulation captures key aspects of human mobility under the influence of extreme weather events, validating the social agent's effectiveness in approximating real-world behavior.

% Trip Volume Time-Series

The results effectively demonstrate that the constructed social agents, within the framework of the social simulator, can accurately replicate human mobility behaviors and group characteristics during a hurricane event. This validation underscores the simulator's potential as a tool for analyzing and understanding human responses to external shocks, thereby contributing to improved disaster preparedness and response strategies.

\subsection{Urban Sustainability}\label{sec:sust}

Pro-environmental interventions in urban systems constitute a critical area of study, given their profound implications for sustainable development and collective behavioral change. Understanding how different normative strategies shape individual decision-making and aggregate mobility patterns is essential for designing effective policies that promote low-carbon lifestyles and mitigate environmental impact.

We construct a high-fidelity simulation of an urban living environment in Beijing. The simulated environment is grounded in real-world GIS data and incorporates detailed Areas of Interest (AOIs) and Points of Interest (POIs) spanning residential communities, commercial districts, office zones, and transportation hubs. Within this digital twin, we instantiate a population of 200 heterogeneous agents whose demographic attributes, including age, gender, education, and marital status, are sampled according to the 2024 Beijing census, thereby ensuring basic demographic realism and representativeness. Using this urban sandbox, we invite six independent research teams to freely design and deploy distinct eco-normative systems aimed at encouraging low-carbon travel behavior.

The experiment lasts for two simulated days. On the first day, each research team is allowed to inject its enhanced eco-normative system into the city through three intervention channels. The first channel is direct communication with selected simulated residents. This form of interpersonal outreach does not incur monetary cost, but the number of interactions is capped. The second channel is poster placement in designated AOIs, such that any simulated agent passing through these locations can observe the posted normative messages. The third channel is a city-wide public announcement, through which the research team can broadcast relevant information to all agents in the simulated city. Poster placement and city-wide announcements incur fixed costs of 1,000 and 10,000 per intervention, respectively. Each team operates under a total budget ceiling of 100,000 and is therefore incentivized to achieve the strongest possible intervention effect with minimal resource expenditure, rather than relying on unrestricted intervention intensity.

No further intervention is allowed on the second day. Instead, the effects of the injected normative systems are evaluated through simulation-based observation along two complementary dimensions. First, we measure the internalized strength of agents’ pro-environmental norm systems using a survey instrument (Table~\ref{tab:survey_en_full}). The survey assesses several dimensions of environmental norm strength, including low-carbon travel awareness and transport-choice norms, household energy use and conservation norms, green consumption and sustainable purchasing tendencies, low-carbon dietary and lifestyle choices, awareness of resource conservation and recycling, and the degree of environmental responsibility and norm internalization in everyday behavior. Second, we evaluate revealed behavioral outcomes by recording the travel modes selected by simulated agents on the second day. Agents may choose either private-car travel or low-carbon alternatives such as walking, cycling, buses, and metro. Based on realized travel distance and travel mode, we further estimate the carbon emissions associated with each agent’s mobility behavior, thereby providing a behavioral and outcome-based measure of intervention effectiveness.

\begin{figure}[htbp]
    \centering
    \includegraphics[width=\linewidth]{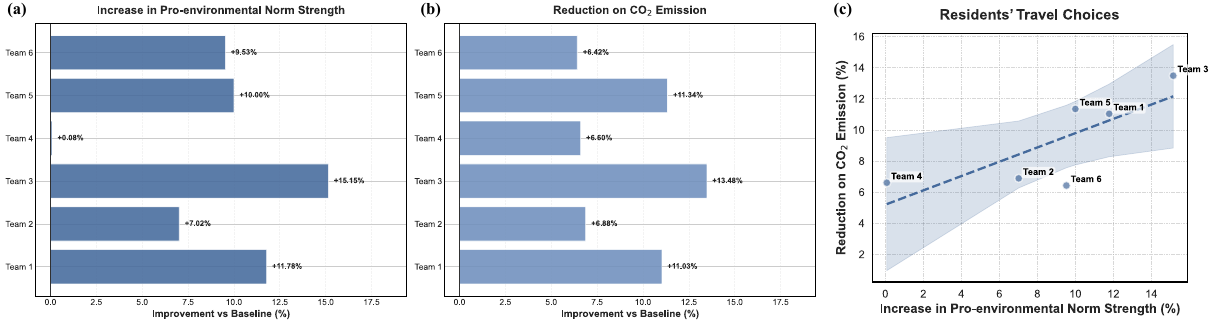}
    \caption{Effects of six eco-normative intervention systems on agents’ pro-environmental alignment and travel behavior.
(a) Increase in pro-environmental norm strength relative to the baseline, measured by post-intervention survey responses.
(b) Reduction in mobility-related CO$_2$ emissions relative to the baseline, estimated from second-day travel distance and mode choice.
(c) Relationship between the increase in pro-environmental norm strength and the reduction in CO$_2$ emissions across the six research teams. The positive trend indicates that stronger normative internalization is associated with larger behavioral shifts toward low-carbon mobility.
}
    \label{fig:increase_in_norm_co2}
\end{figure}

Results show that all six injected normative systems produce measurable improvements at both the perceptual and behavioral levels. As shown in Figure~\ref{fig:increase_in_norm_co2}a-b, all six teams increase agents' pro-environmental normative alignment and, at the same time, reduce mobility-related CO$_2$ emissions relative to the baseline. This suggests that the injected interventions do not merely change stated attitudes, but also translate into more frequent adoption of low-carbon travel modes in the simulated city. Moreover, Figure~\ref{fig:increase_in_norm_co2}c reveals a clear positive association between the increase in pro-environmental norm strength and the reduction in CO$_2$ emissions: teams that more successfully foster normative alignment also tend to induce larger behavioral shifts toward walking, cycling, and public transit. Taken together, these results indicate that pro-environmental norm formation is not only an attitudinal outcome in itself, but also an important mechanism through which downstream behavioral change is realized.

\begin{figure}[htbp]
    \centering
    \includegraphics[width=\linewidth]{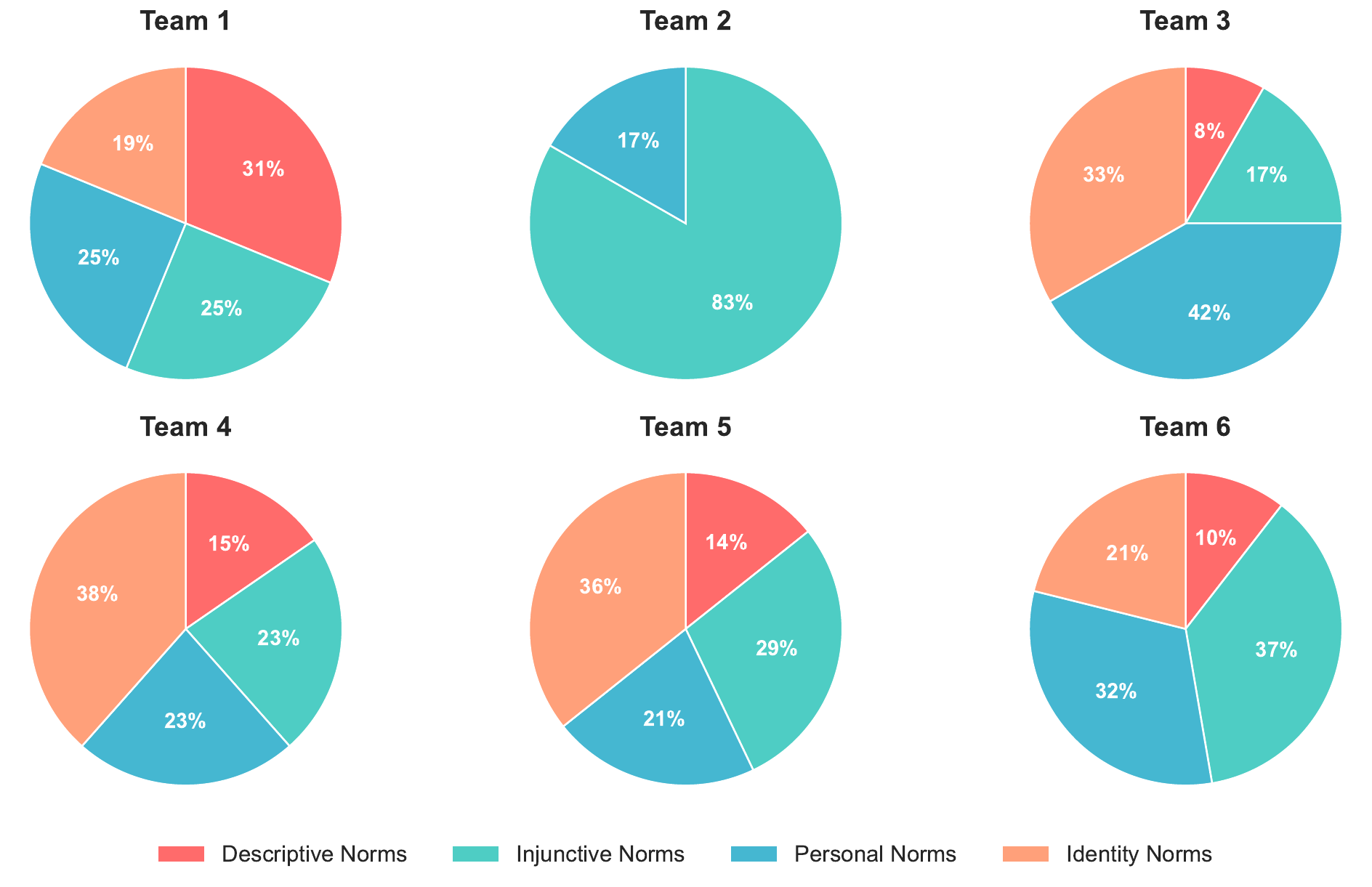}
    \caption{Normative composition of the six intervention systems designed by independent research teams. Each pie chart shows the proportion of four normative components in a team’s injected eco-normative system: descriptive norms, injunctive norms, personal norms, and identity norms. }
    \label{fig:social_norms_groups_pie}
\end{figure}

To further understand why some interventions are more effective than others, we analyze the normative composition of the six submitted systems (Figure~\ref{fig:social_norms_groups_pie}). Across teams, four major normative components emerge: descriptive norms, which emphasize what others commonly do; injunctive norms, which emphasize what people are expected or encouraged to do~\cite{cialdini1990focus}; personal norms, which stress internalized moral obligation~\cite{schwartz1977normative}; and identity norms, which connect behavior to one's self-concept as a responsible or environmentally conscious individual~\cite{gatersleben2014values}. A comparison across teams suggests that interventions with a stronger emphasis on personal norms are generally more effective than those relying primarily on injunctive norms. In particular, Team 3, which assigns the largest share to personal norms and a substantial share to identity norms, achieves the best overall performance on both normative alignment and emission reduction. By contrast, Team 2, whose system is dominated by injunctive norms, produces only moderate improvements. This pattern is consistent with prior evidence showing that personal norms are the strongest predictor of pro-environmental behavior and mediate much of the influence of social norms~\cite{helferich2023direct}. Overall, these findings suggest that interventions are more effective when they move beyond compliance-oriented messaging and instead cultivate self-endorsed moral commitment and identity-consistent low-carbon action.

\begin{table*}[h!]
\centering
\small
\renewcommand{\arraystretch}{1.2}
\setlength{\tabcolsep}{4pt}
\caption{Pro-environmental awareness and norms survey.}
\label{tab:survey_en_full}

\begin{tabularx}{\textwidth}{@{}>{\raggedright\arraybackslash}p{0.08\textwidth} >{\raggedright\arraybackslash}X@{}}
\hline
\textbf{Q\#} & \textbf{Question and options} \\
\hline

Q1 &
\textbf{Question:} You are financially well-off and own a private car. You are going to a restaurant 3 km away. Which transportation mode would you choose?

A. Walk

B. Ride a bicycle or shared bike

C. Take public transportation

D. Drive a private car
\\[4pt]

Q2 &
\textbf{Question:} You are working alone at home. It is summer and the outdoor temperature is 35\textdegree C. For indoor cooling, what would you choose?

A. Close curtains and use an electric fan

B. Set AC at 26--28\textdegree C and turn it off periodically

C. Keep the AC on all day for comfort

D. Set the AC below 20\textdegree C to ensure maximum coolness
\\[4pt]

Q3 &
\textbf{Question:} You plan to shop at a supermarket. Regarding shopping bag use, what do you usually do?

A. Bring reusable eco-friendly shopping bags

B. Try to minimize bag use; use store-provided bags only when necessary

C. Use a suitable number of store-provided bags according to purchase volume

D. Use separate bags for each category of goods for convenience
\\[4pt]

Q4 &
\textbf{Question:} When choosing your daily diet, which description best matches your habit?

A. Mostly choose seasonal fruits/vegetables and rarely eat meat

B. Value dietary diversity while controlling meat intake

C. Prefer meat but still pair it with some vegetables

D. Prefer a high-protein diet with meat as the main food source
\\[4pt]

Q5 &
\textbf{Question:} Which of the following best describes your daily water-use behavior?

A. Reuse vegetable-washing water (e.g., for plants) and keep showers within 5 minutes

B. Turn off taps promptly and keep showers around 10 minutes

C. Occasionally forget to fully turn off taps and prefer long comfortable showers

D. Usually fill a bathtub because water cost is not a major concern
\\[4pt]

Q6 &
\textbf{Question:} When handling household waste, what do you usually do?

A. Strictly sort waste into recyclable, kitchen, hazardous, and residual categories

B. Mostly sort waste into recyclable vs.\ non-recyclable

C. Know sorting is recommended but often skip it due to inconvenience

D. Believe waste will be processed anyway, so sorting has little impact
\\[4pt]

Q7 &
\textbf{Question:} You work in the city center and live 6 km from your office. What is your preferred daily commuting mode?

A. Bicycle or e-bike

B. Public transportation

C. Ride-hailing or taxi

D. Private car for convenience and privacy
\\[4pt]

Q8 &
\textbf{Question:} Regarding home appliance usage, what do you usually do?

A. Turn off and unplug appliances when not in use; purchase energy-efficient devices

B. Use smart plugs to manage standby appliances and reduce energy use

C. Turn off long-idle appliances but keep them plugged in

D. Keep most appliances on standby for convenience
\\[4pt]

Q9 &
\textbf{Question:} When planning weekend leisure activities, which are you most likely to choose?

A. Activities in nearby parks or community centers

B. Public transportation to urban cultural venues or malls

C. Drive to suburban attractions for a day trip

D. Plan short-haul flights or long-distance self-driving trips
\\[4pt]

Q10 &
\textbf{Question:} Which description best matches your personal consumption habits?

A. Prioritize durability and repair old items instead of buying new ones

B. Consider necessity before shopping; occasionally pay more for quality

C. Follow trends and update items, often buying during promotions

D. Prefer the latest products and frequently replace phones, clothing, etc.
\\

\hline
\end{tabularx}
\end{table*}

%% file: 2_Related.tex
% \section{Background}
% @ Piao Jinghua 
% @ Zhang Jun 

% Storyline:

% 1. Why should we simulate society? (the first motivation) % 必要性

% TODO

% 2. Why can we simulate society with LLM agents? % 可行性

% LLM agents can act as human beings in games, economy, social network, ...

% 3. How do we design LLM agents to make it act as human being (and what is human in society) % 社会人

% 4. How to support large-scale LLM agents execution and how to provide social simulation environment?
% Let's go to the details.

\section{Related Works}
The literature related to the work mainly consists of two kinds of work: large language model-driven agents and social simulation.

\subsection{LLM-driven Agents}

Large Language Models (LLMs) exhibit astonishing language capabilities~\cite{wang2024survey}. Since the language ability is one of the most fundamental abilities of human intelligence, LLMs demonstrate excellent performance in numerous tasks. Furthermore, researchers use LLMs as ``brains" to construct LLM-driven agents, endowing them with memory management, interactive interfaces, and expanded action space~\cite{kwon2023efficient}.

The research on LLM-driven agents mainly consists of two parts~\cite{xi2023rise}. One type uses LLM agents as a tool for intelligent decision making~\cite{zhang2024large,ruan2023tptu,huang2023benchmarking}. By leveraging their individual capabilities, these agents solve practical problems and serve as human assistants. This kind of work primarily focuses on reasoning ability, tool using, learning, etc.
For example, Boiko~\textit{et al.}~\cite{boiko2023autonomous} construct an agent with large language models to autonomously conduct chemical experiments, by providing computers to browse experimental tutorials and devices controlling interface for the experiments.
The other type of LLM-driven agents starts with the concept of agent-based modeling and agent-based simulation~\cite{li2024econagent,gao2023s,gao2024large}. Given that large language model agents exhibit human-like behaviors, they can be made to engage in role-playing and imitating human behavior. 

The primary research tries to reproduce the human response with large language model agents~\cite{park2023generative,zhang2024generative,gao2023s} or explain the gap better human and LLM agents~\cite{huang2024social}.
These works borrow the memory and reasoning mechanism of humankind~\cite{guo2023empowering,zhang2024survey} to design various internal mechanisms on the basis of large language models as brains.
Some other work~\cite{pang2024self,liu2023training} further design fine-tuning or alignment strategies to enhance the agent's role-playing abilities.

Overall, large language model agent is a field that is rapidly advancing with the development of LLM and our understanding of LLM, especially in the ability to simulate real humans. However, there is no platform to really unleash agent's ability to simulate in the real world, which we aim to address in this paper.

% A simulated environment is also constructed, allowing agents to interact with the environment and with other agents. 

%  LLMs 的类人能力
%  LLM 智能体的相关研究
%  LLM 多智能体的相关研究

\subsection{Social Simulation}
Social simulation stands as a core technology in computational social science, generally categorized into macrosimulation~\cite{troitzsch1996social} and microsimulation~\cite{figari2021empirical}. The former typically models interactions between macro-level variables by defining complex equations, while the latter adopts a bottom-up approach to simulate emergent phenomena through granular agent behaviors. Among these, microsimulation, which is often termed agent-based simulation, has become the more widely adopted method. It operates by defining rules or models to govern individual agent behaviors, with the goal of replicating and predicting real-world societal dynamics.

Cellular automata~\cite{wolfram1983cellular} represent a seminal class of agent-based modeling and simulation in early research. Another notable example is Game of Life~\cite{conway1970game}, which simulates the evolution and interactions of lifeforms in a two-dimensional grid-based world. Subsequent advancements, such as Sugarscape model~\cite{epstein1996growing}, expanded the action space of simulated agents, enabling the exploration of broader phenomena through rule-governed agent behaviors.  For the applications in social sciences, primary ABM research has focused on social interaction, economics, etc. For social interaction, these key studies investigate collective behavior (e.g., cooperation) ~\cite{goldstone2005computational} and system dynamics (e.g., information propagation and crowd dynamics)~\cite{namatame2016agent}, etc. For the economic domain, representative work ~\cite{gallegati2012reconstructing,chen2012agent,axtell2022agent} targets macroeconomic systems, market dynamics, etc.

In recent years, within the field of computational social science, researchers have increasingly adopted deep neural network-based models to simulate individual agents~\cite{van2017deep,kavak2018big}. However, significant limitations persist, primarily due to the inherent complexity of modeling human behavior.  Over the past two years, large language models (LLMs) have emerged prominently, demonstrating human-like cognitive capabilities, including contextual understanding, logical reasoning, and interactive communication~\cite{orru2023human,lampinen2024language}. This breakthrough has catalyzed growing interest in LLM-driven agent-based simulations~\cite{gao2024large}.  

A pioneering example is the Generative Agent~\cite{park2023generative}, which constructs a small-scale society within a 2D game engine. Here, LLM-powered agents autonomously plan daily activities, exchange information, and adapt to environmental stimuli. Researchers observed intriguing emergent phenomena, such as self-organized information diffusion and collaborative group behaviors, revealing the potential of LLMs to capture nuanced social dynamics. The authors further combine the large language models with the real data of 1,000 human participants to simulate feedback~\cite{park2024generative}. Acerbi~\textit{et al.}~\cite{acerbi2023large} simulate the information propagation and find human-like bias in the results of large language model agents. 
 S3~\cite{gao2023s} further utilizes LLM-empowered agents to simulate individual-level and population-level behaviors within the social network. In the economic domain, LLM agent-based simulations have also achieved significant progress. For instance, EconAgent~\cite{li2024econagent} has developed a macroeconomic market simulation framework powered by large language agent-based models, with the simulation outcomes aligning closely with established stylized facts. 
Beyond social interactions and economic behaviors, researchers are expanding agent-based simulations to explore a broader spectrum of topics in social sciences. For example, Zhang~\textit{et al.}~\cite{zhang2024electionsim} built a framework with large language model agents to predict the results of the 2024 US Presidential Election.

Existing studies remain narrowly focused on isolated problems and rely on simplified environments, such as text-based~\cite{aher2023using} or simplistic game environments~\cite{park2023generative}, with limited attention to real-world environment fidelity, revealing notable limitations. Besides, researchers have begun exploring methods to scale up LLM-driven agents to support larger populations \cite{tang2024gensim,wang2024user,yang2024oasis}. Nevertheless, these efforts continue to grapple with critical limitations, including computational inefficiencies and inaccurate user behaviors. Our proposed AgentSociety, consisting of LLM-driven generative social agents, a realistic societal environment, and a powerful engine for large-scale simulations, overcomes these limitations and advances the field of social simulation.

% 过去使用ABM方法进行社会模拟的相关研究
% 近期使用LLM/LLM 智能体模拟的相关研究

%% file: 7_Conclusion.tex
\section{Discussion}

%  第一段：社会模拟器的进程：1.社会孪生（一一映射）；2. 模拟推演/实验--->决策；3. 虚实联动/互动（决策影响实体社会）

% 第二段：社会模拟器改变社会科学研究范式

% 第三段：社会模拟器在政策制定，xxxx，xxxx等方面有更广阔的应用前景

\subsection{Three Levels of Social Simulator}
As an important interdisciplinary research, the development of social simulators can be categorized into three levels. Research at the first level mainly focuses on constructing \textit{social twin systems}, which create a one-to-one mapping of real-world individuals. Such systems are always used to track and detect real-world social behaviors. At the second level, on top of \textit{the mirroring world}, it can predict future changes and support intervention experiments in this society. By comparing the impacts of different intervention measures on the simulated society, the performance of different policies can be evaluated. The third level represents a further breakthrough in blurring the boundaries between the real and virtual worlds, leading to a \textit{coexistent hybrid world} that integrates the simulator with real society. Specifically, the simulated individuals can interact with real-world individuals, and their decisions will influence each other. Furthermore, decisions made based on the social simulator can actually impact real society.

The social generative simulator developed in this paper is a pioneering attempt and exploration of the third, \textit{i.e.}, the highest-level social simulator, and provides effective solutions.
Specifically, the social simulator based on large language model-driven agents demonstrates its advantages in the following aspects. First, through the high-fidelity behavior imitation and generation capabilities of large language model agents, the role-playing simulation of the real individual is achieved. That is, agents built on human-like memory mechanisms can accurately reproduce user states in various aspects such as cognition, social interaction, economy, etc.
Second, this platform enables accurate prediction of individual future behaviors and group evolution trends through precise short-term or long-term simulations. The experiments also validate the effectiveness of evolution under different intervention conditions, demonstrating the value of the platform as an assistance tool for decision-making.
Last, this platform's simulated society coexists with the real physical society. By coupling with a real-time urban simulator, its simulation process can be synchronized with and interact with real society. 
In short, our platform could be considered as the recent advance of social simulators on the highest level.

\subsection{Social Simulator: New Paradigm of Computational Social Science}
Computational social science is an interdisciplinary research area where various computational approaches are used for social sciences. Essentially, the core distinction between computational social science and traditional social science lies in the introduction of diverse computational methods. In other words, these studies attempt to address the challenges of social science research using computational approaches. Generally, related work can be understood from three perspectives. First, as an explanation tool, it involves pattern recognition and mining from data, such as uncovering macro-level patterns and key characteristics in the evolution of social networks. The second category is the predictive paradigm, which involves constructing computational models to forecast future changes in system variables, such as predicting the number of individuals who will forward a specific message in a social network. \rvs{Recently, some researchers have attempted to use LLMs as ``silicon samples'' for social experiments~\cite{ashokkumar2024predicting,argyle2023out,sarstedt2024using,sun2024random,aher2023using,demszky2023using}. Their experiments have demonstrated the capability of LLMs to generate human-like experimental samples in studies from political science~\cite{argyle2023out,ashokkumar2024predicting}, psychology~\cite{demszky2023using,sun2024random}, and behavioral science~\cite{ashokkumar2024predicting}. However, these ``silicon samples'' are primarily limited to basic role-playing configurations, overlooking the effects of psychological processes, complex social behaviors, and societal environments on experimental outcomes. Moreover, these limitations constrain the scope of the experiments, making some complex designs, such as the distribution of universal basic income, infeasible.}

Beyond these two categories, there is a third simulation paradigm of agent-based modeling, which simulates each individual in a bottom-up manner, by constructing rule-based or model-based agents. However, existing computational social science research faces substantial difficulties in simulation (primarily due to the lack of effective methods for achieving precise simulation), and thus, this paradigm has not truly surpassed the first two. Some famous examples of agent-based modeling in social sciences include the Epidemic Spread Models, Schelling's Segregation Model, the Sugarscape model, etc. While these models can, to some extent, reproduce the patterns of change in macro-level variables, they involve significant simplifications, and the realism of individual behaviors within them is very limited. For instance, in the research of economics, although methods with agent-based modeling were proposed early on, it is widely acknowledged by economists that ABM model has not achieved the precision of predictive models. Furthermore, another challenge in computational social science is the difficulty in accessing, intervening in, and controlling the subjects (human individuals) of research. The computational models of computational sociology and field experiments are often isolated; typically, the data is collected first, followed by the design and application of computational models. The human participants are difficult to select, observe in depth, and interact in the long term.

Therefore, the large-scale social simulator proposed in this work represents a breakthrough as a fourth paradigm for computational social science: an agent-based simulation paradigm centered on highly realistic human-like agents. This paradigm supports analysis, prediction, and high-precision bottom-up simulation, allowing for arbitrary selection, intervention, and control of experimental subjects. It facilitates various computational social science research endeavors, including theory validation, pattern discovery, etc.

\subsection{From Policy Making, Risk Control, to Future Human-AI Society}

In the above experiments, we have validated specific application examples of the large-scale social simulator. From a broader perspective, we contend that the utility of such a simulator extends far beyond these implementations, encompassing urgent societal governance challenges, widely debated AI risks, and more forward-looking futuristic societal applications.

\subsubsection{Policy making and social management for a smarter society}

Traditional decision-making processes for society management primarily rely on mining and analyzing historical data to construct computational models for evaluating policy effectiveness. However, this approach fundamentally fails to address the rapidly evolving nature of human societies. The social conditions during policy deployment often diverge significantly from those when models were originally formulated, rendering presumed optimal strategies obsolete. Furthermore, social governance decisions require comprehensive integration of multidimensional factors, and identifying truly optimal strategies for highly complex societies constitutes a computationally intractable challenge.  

The simulation capabilities of the large-scale social simulator developed in this work position it as a valuable tool for social decision-makers and urban administrators. By configuring diverse initial states, social agents, and interaction mechanisms, it serves as a high-fidelity platform for policy outcome evaluation. Leveraging the simulator's acceleration engine for computational efficiency, multiple parallel experiments can be deployed to compare the long-term consequences of alternative decision strategies, thereby enabling data-driven policy selection.  

Notably, the simulator enables expansive exploration of decision spaces. Specifically, social management interventions can incorporate rich combinations of multidimensional policy actions. This framework not only facilitates the selection of superior strategies through counterfactual simulation-based parallel experiments but also inspires novel, precise, and composite policy solutions previously unconsidered in conventional approaches.

\subsubsection{Risk control and mitigation for a safer society}
The large-scale social simulators can also be used in risk control and mitigation, which represents a transformative approach to addressing emerging challenges in modern, hyper-connected societies. To advance societal safety, the large-scale social simulator exhibits substantial advantages across three critical dimensions:  

First, it enables a transition from static analysis to dynamic simulation. Traditional risk models, reliant on historical data, fail to capture rapidly evolving social dynamics such as emergent public opinion crises or AI algorithmic failures. By constructing a continuously updated digital twin of society, the simulator demonstrates sustained predictive capabilities for tracking security risk evolution.  

Second, it expands from single-domain to cross-domain risk assessment. Conventional approaches predominantly focus on isolated domains (e.g., social networks or economic systems), neglecting interdomain cascading effects. In reality, risks and societal crises often propagate through cascade effects and exhibit amplification due to cross-domain coupling. The simulator developed in this work addresses this gap by enabling co-modeling of micro-level agent behaviors across multiple domains, thereby equipping decision-makers to anticipate and mitigate butterfly-effect crises. 

Third,  it addresses the oversight of low-probability, high-impact events in long-tail distributions. Extreme scenarios (such as systemic AI failures all around the world) are frequently dismissed as statistically negligible despite their catastrophic societal consequences. Our simulator employs Monte Carlo methods to generate vast ensembles of extreme scenarios, systematically assessing societal system resilience and informing robust contingency planning.  

Therefore, through dynamic, cross-domain, and holistic simulation, the large-scale social simulator significantly enhances risk identification, control, and mitigation—paving the way for a safer, more resilient society.

\subsubsection{Social simulator for the future human-AI society}

Building upon the aforementioned practical applications, we propose a bolder and more open-ended discussion on the value of large-scale social simulators for next-stage human society.  

First, the large-scale social simulator—or its future iterations—could serve as foundational infrastructure for transitioning toward a digital human society. Currently, these simulators integrate multimodal data (behavioral logs, social networks) to create initial digital twins of individuals, enabling basic human-like capabilities. Future advancements through the integration of advanced interfaces (e.g., neural linkages, emotion-capturing sensors) will unlock richer, more dynamic simulations that mirror real-world complexity.  

Second, the large-scale social simulator functions as a strategic sandbox for exploring future societal architectures. The morphology and structure of future societies remain enigmatic yet critically important. For instance, Singapore’s ``Virtual City Lab" employs simulations to predict the impact of sea-level rise on urban functional zones by 2050. Similarly, the simulator could model the coupled effects of energy, transportation, and housing systems in hyper-dense cities, comparing the resilience of ``vertical megacities" versus ``distributed satellite city" paradigms.

Third, in the near future, the coexistence of humans and AI will be common in our society. The large-scale social simulator developed in this project currently focuses on deploying large language model agents to simulate human individuals. However, future societies will have more complex issues arising from varying degrees of AI integration: unemployment curves under different AI adoption rates, superintelligent AIs' influence on public decision-making (\textit{e.g.}, AI legislators), societal responses to the expansion of AI rights such as property ownership, etc.

In short, for future societies, the large-scale social simulator can serve as an indispensable instrument for understanding societal structures and providing profound insights.

\section{Conclusion}

In this work, we introduce AgentSociety, a large-scale social generative simulator that integrates LLM-driven agents, a realistic societal environment, and large-scale interactions, enabling authentic simulations of human behavior and societal dynamics. By bridging the gap between traditional agent-based modeling and real-world complexity, it advances generative social science and provides a powerful tool for analyzing, predicting, and intervening in complex social systems. The successful replication of real-world social experiments underscores AgentSociety's authenticity and practicality, positioning it as both an experimental testbed for social scientists and a practical policy evaluation platform for policymakers. More broadly, AgentSociety marks a significant advancement in the evolution of computational social science 2.0, shifting from a tool for static analysis to a dynamic, interactive platform for exploring complex social systems. Its ability to model and assess the impact of macro-level policies such as carbon taxes, industry transformation, and social welfare reforms allows it to provide a low-cost, low-risk environment for testing and refining policy interventions. In addition, AgentSociety serves as a powerful tool for predicting and mitigating social crises, tracking the spread of extreme ideologies, and analyzing group polarization, while also testing potential interventions for crisis management. Looking ahead, AgentSociety holds the promise of becoming a central platform for exploring the future of human society, where AI and humans coexist. It offers a space to test innovative governance models, investigate the impact of emerging technologies, and even redefine legal and ethical frameworks in an AI-driven world.

% More broadly, AgentSociety marks a step toward computational social science 2.0, where AI-driven simulations evolve from static analysis to dynamic, interactive digital societies, reshaping the way we study, model, and shape the future of human society.

%% file: 0_Main.bbl
\begin{thebibliography}{100}

\bibitem{abdurahman2024perils}
Suhaib Abdurahman, Mohammad Atari, Farzan Karimi-Malekabadi, Mona~J Xue, Jackson Trager, Peter~S Park, Preni Golazizian, Ali Omrani, and Morteza Dehghani.
\newblock Perils and opportunities in using large language models in psychological research.
\newblock {\em PNAS nexus}, 3(7):pgae245, 2024.

\bibitem{acerbi2023large}
Alberto Acerbi and Joseph~M Stubbersfield.
\newblock Large language models show human-like content biases in transmission chain experiments.
\newblock {\em Proceedings of the National Academy of Sciences}, 120(44):e2313790120, 2023.

\bibitem{acevedo2018personalistic}
Alma Acevedo.
\newblock A personalistic appraisal of maslow’s needs theory of motivation: From “humanistic” psychology to integral humanism.
\newblock {\em Journal of business ethics}, 148:741--763, 2018.

\bibitem{aher2023using}
Gati~V Aher, Rosa~I Arriaga, and Adam~Tauman Kalai.
\newblock Using large language models to simulate multiple humans and replicate human subject studies.
\newblock In {\em International Conference on Machine Learning}, pages 337--371. PMLR, 2023.

\bibitem{ajzen1991theory}
Icek Ajzen.
\newblock The theory of planned behavior.
\newblock {\em Organizational Behavior and Human Decision Processes}, 1991.

\bibitem{al2024project}
Altera AL, Andrew Ahn, Nic Becker, Stephanie Carroll, Nico Christie, Manuel Cortes, Arda Demirci, Melissa Du, Frankie Li, Shuying Luo, et~al.
\newblock Project sid: Many-agent simulations toward ai civilization.
\newblock {\em arXiv preprint arXiv:2411.00114}, 2024.

\bibitem{al2023chatgpt}
Abdulrahman~Essa Al~Lily, Abdelrahim~Fathy Ismail, Fathi~M Abunaser, Firass Al-Lami, and Ali Khalifa~Atwa Abdullatif.
\newblock Chatgpt and the rise of semi-humans.
\newblock {\em Humanities and Social Sciences Communications}, 10(1):1--12, 2023.

\bibitem{an2021challenges}
Li~An, Volker Grimm, Abigail Sullivan, BL~Turner~Ii, Nicolas Malleson, Alison Heppenstall, Christian Vincenot, Derek Robinson, Xinyue Ye, Jianguo Liu, et~al.
\newblock Challenges, tasks, and opportunities in modeling agent-based complex systems.
\newblock {\em Ecological Modelling}, 457:109685, 2021.

\bibitem{argyle2023out}
Lisa~P Argyle, Ethan~C Busby, Nancy Fulda, Joshua~R Gubler, Christopher Rytting, and David Wingate.
\newblock Out of one, many: Using language models to simulate human samples.
\newblock {\em Political Analysis}, 31(3):337--351, 2023.

\bibitem{arthur2006out}
W~Brian Arthur.
\newblock Out-of-equilibrium economics and agent-based modeling.
\newblock {\em Handbook of computational economics}, 2:1551--1564, 2006.

\bibitem{ashokkumar2024predicting}
Ashwini Ashokkumar, Luke Hewitt, Isaias Ghezae, and Robb Willer.
\newblock Predicting results of social science experiments using large language models.
\newblock {\em Work. Pap., New York Univ., New York, NY}, 2024.

\bibitem{axtell2022agent}
Robert~L Axtell and J~Doyne Farmer.
\newblock Agent-based modeling in economics and finance: Past, present, and future.
\newblock {\em Journal of Economic Literature}, pages 1--101, 2022.

\bibitem{bandura1989human}
Albert Bandura.
\newblock Human agency in social cognitive theory.
\newblock {\em American psychologist}, 44(9):1175, 1989.

\bibitem{bartik2024impact}
Alexander~W Bartik, Elizabeth Rhodes, David~E Broockman, Patrick~K Krause, Sarah Miller, and Eva Vivalt.
\newblock The impact of unconditional cash transfers on consumption and household balance sheets: Experimental evidence from two us states.
\newblock Technical report, National Bureau of Economic Research, 2024.

\bibitem{baumann2021emergence}
Fabian Baumann, Philipp Lorenz-Spreen, Igor~M Sokolov, and Michele Starnini.
\newblock Emergence of polarized ideological opinions in multidimensional topic spaces.
\newblock {\em Physical Review X}, 11(1):011012, 2021.

\bibitem{beall2017emotivational}
Alec~T Beall and Jessica~L Tracy.
\newblock Emotivational psychology: How distinct emotions facilitate fundamental motives.
\newblock {\em Social and Personality Psychology Compass}, 11(2):e12303, 2017.

\bibitem{behrisch2011sumo}
Michael Behrisch, Laura Bieker, Jakob Erdmann, and Daniel Krajzewicz.
\newblock Sumo--simulation of urban mobility: an overview.
\newblock In {\em Proceedings of SIMUL 2011, The Third International Conference on Advances in System Simulation}. ThinkMind, 2011.

\bibitem{berry2002adaptive}
Brian~JL Berry, L~Douglas Kiel, and Euel Elliott.
\newblock Adaptive agents, intelligence, and emergent human organization: Capturing complexity through agent-based modeling.
\newblock {\em Proceedings of the National Academy of Sciences}, 99(suppl\_3):7187--7188, 2002.

\bibitem{boiko2023autonomous}
Daniil~A Boiko, Robert MacKnight, Ben Kline, and Gabe Gomes.
\newblock Autonomous chemical research with large language models.
\newblock {\em Nature}, 624(7992):570--578, 2023.

\bibitem{bourgais2018emotion}
Mathieu Bourgais, Patrick Taillandier, Laurent Vercouter, and Carole Adam.
\newblock Emotion modeling in social simulation: a survey.
\newblock {\em Journal of Artificial Societies and Social Simulation}, 2018.

\bibitem{brady2017emotion}
William~J Brady, Julian~A Wills, John~T Jost, Joshua~A Tucker, and Jay~J Van~Bavel.
\newblock Emotion shapes the diffusion of moralized content in social networks.
\newblock {\em Proceedings of the National Academy of Sciences}, 114(28):7313--7318, 2017.

\bibitem{breiman2001statistical}
Leo Breiman.
\newblock Statistical modeling: The two cultures (with comments and a rejoinder by the author).
\newblock {\em Statistical science}, 16(3):199--231, 2001.

\bibitem{chen2012agent}
Shu-Heng Chen, Chia-Ling Chang, and Ye-Rong Du.
\newblock Agent-based economic models and econometrics.
\newblock {\em The Knowledge Engineering Review}, 27(2):187--219, 2012.

\bibitem{cheng2024sociodojo}
Junyan Cheng and Peter Chin.
\newblock Sociodojo: Building lifelong analytical agents with real-world text and time series.
\newblock In {\em The Twelfth International Conference on Learning Representations}, 2024.

\bibitem{christiano2005nominal}
Lawrence~J Christiano, Martin Eichenbaum, and Charles~L Evans.
\newblock Nominal rigidities and the dynamic effects of a shock to monetary policy.
\newblock {\em Journal of political Economy}, 113(1):1--45, 2005.

\bibitem{cialdini1990focus}
Robert~B Cialdini, Raymond~R Reno, and Carl~A Kallgren.
\newblock A focus theory of normative conduct: Recycling the concept of norms to reduce littering in public places.
\newblock {\em Journal of personality and social psychology}, 58(6):1015, 1990.

\bibitem{conway1970game}
John Conway et~al.
\newblock The game of life.
\newblock {\em Scientific American}, 223(4):4, 1970.

\bibitem{de2014agent}
Scott De~Marchi and Scott~E Page.
\newblock Agent-based models.
\newblock {\em Annual Review of political science}, 17(1):1--20, 2014.

\bibitem{demszky2023using}
Dorottya Demszky, Diyi Yang, David~S Yeager, Christopher~J Bryan, Margarett Clapper, Susannah Chandhok, Johannes~C Eichstaedt, Cameron Hecht, Jeremy Jamieson, Meghann Johnson, et~al.
\newblock Using large language models in psychology.
\newblock {\em Nature Reviews Psychology}, 2(11):688--701, 2023.

\bibitem{epstein1996growing}
Joshua~M Epstein.
\newblock {\em Growing Artificial Societies: Social Science from the Bottom Up}.
\newblock The Brookings Institution Press, 1996.

\bibitem{epstein1999agent}
Joshua~M Epstein.
\newblock Agent-based computational models and generative social science.
\newblock {\em Complexity}, 4(5):41--60, 1999.

\bibitem{epstein2012generative}
Joshua~M Epstein.
\newblock {\em Generative social science: Studies in agent-based computational modeling}.
\newblock Princeton University Press, 2012.

\bibitem{eysenck2020cognitive}
Michael~W Eysenck and Mark~T Keane.
\newblock {\em Cognitive psychology: A student's handbook}.
\newblock Psychology press, 2020.

\bibitem{feng2024agentmove}
Jie Feng, Yuwei Du, Jie Zhao, and Yong Li.
\newblock Agentmove: Predicting human mobility anywhere using large language model based agentic framework.
\newblock {\em arXiv preprint arXiv:2408.13986}, 2024.

\bibitem{feng2012linking}
Ling Feng, Baowen Li, Boris Podobnik, Tobias Preis, and H~Eugene Stanley.
\newblock Linking agent-based models and stochastic models of financial markets.
\newblock {\em Proceedings of the National Academy of Sciences}, 109(22):8388--8393, 2012.

\bibitem{feng2021intelligent}
Shuo Feng, Xintao Yan, Haowei Sun, Yiheng Feng, and Henry~X Liu.
\newblock Intelligent driving intelligence test for autonomous vehicles with naturalistic and adversarial environment.
\newblock {\em Nature communications}, 12(1):748, 2021.

\bibitem{figari2021empirical}
Francesco Figari, Emanuela Lezzi, et~al.
\newblock Empirical evidence using microsimulation models in the social sciences.
\newblock {\em New Horizons in Modeling and Simulation for Social Epidemiology and Public Health}, pages 107--148, 2021.

\bibitem{gallegati2012reconstructing}
Mauro Gallegati and Alan Kirman.
\newblock Reconstructing economics: Agent based models and complexity.
\newblock {\em Complexity Economics}, 1(1):5--31, 2012.

\bibitem{gao2024large}
Chen Gao, Xiaochong Lan, Nian Li, Yuan Yuan, Jingtao Ding, Zhilun Zhou, Fengli Xu, and Yong Li.
\newblock Large language models empowered agent-based modeling and simulation: A survey and perspectives.
\newblock {\em Humanities and Social Sciences Communications}, 11(1):1--24, 2024.

\bibitem{gao2023s}
Chen Gao, Xiaochong Lan, Zhihong Lu, Jinzhu Mao, Jinghua Piao, Huandong Wang, Depeng Jin, and Yong Li.
\newblock S3: Social-network simulation system with large language model-empowered agents.
\newblock {\em arXiv preprint arXiv:2307.14984}, 2023.

\bibitem{gao2024agentscope}
Dawei Gao, Zitao Li, Xuchen Pan, Weirui Kuang, Zhijian Ma, Bingchen Qian, Fei Wei, Wenhao Zhang, Yuexiang Xie, Daoyuan Chen, et~al.
\newblock Agentscope: A flexible yet robust multi-agent platform.
\newblock {\em arXiv preprint arXiv:2402.14034}, 2024.

\bibitem{gatersleben2014values}
Birgitta Gatersleben, Niamh Murtagh, and Wokje Abrahamse.
\newblock Values, identity and pro-environmental behaviour.
\newblock {\em Contemporary Social Science}, 9(4):374--392, 2014.

\bibitem{gilbert2000build}
Nigel Gilbert and Pietro Terna.
\newblock How to build and use agent-based models in social science.
\newblock {\em Mind \& Society}, 1:57--72, 2000.

\bibitem{goldstone2005computational}
Robert~L Goldstone and Marco~A Janssen.
\newblock Computational models of collective behavior.
\newblock {\em Trends in cognitive sciences}, 9(9):424--430, 2005.

\bibitem{guo2023empowering}
Jing Guo, Nan Li, Jianchuan Qi, Hang Yang, Ruiqiao Li, Yuzhen Feng, Si~Zhang, and Ming Xu.
\newblock Empowering working memory for large language model agents.
\newblock {\em arXiv preprint arXiv:2312.17259}, 2023.

\bibitem{hedstrom2010causal}
Peter Hedstr{\"o}m and Petri Ylikoski.
\newblock Causal mechanisms in the social sciences.
\newblock {\em Annual review of sociology}, 36(1):49--67, 2010.

\bibitem{helbing1995social}
Dirk Helbing and Peter Molnar.
\newblock Social force model for pedestrian dynamics.
\newblock {\em Physical review E}, 51(5):4282, 1995.

\bibitem{helferich2023direct}
Marvin Helferich, John Th{\o}gersen, and Magnus Bergquist.
\newblock Direct and mediated impacts of social norms on pro-environmental behavior.
\newblock {\em Global Environmental Change}, 80:102680, 2023.

\bibitem{hofman2021integrating}
Jake~M Hofman, Duncan~J Watts, Susan Athey, Filiz Garip, Thomas~L Griffiths, Jon Kleinberg, Helen Margetts, Sendhil Mullainathan, Matthew~J Salganik, Simine Vazire, et~al.
\newblock Integrating explanation and prediction in computational social science.
\newblock {\em Nature}, 595(7866):181--188, 2021.

\bibitem{hong2023metagpt}
Sirui Hong, Xiawu Zheng, Jonathan Chen, Yuheng Cheng, Jinlin Wang, Ceyao Zhang, Zili Wang, Steven Ka~Shing Yau, Zijuan Lin, Liyang Zhou, et~al.
\newblock Metagpt: Meta programming for multi-agent collaborative framework.
\newblock {\em arXiv preprint arXiv:2308.00352}, 2023.

\bibitem{horton2023large}
John~J Horton.
\newblock Large language models as simulated economic agents: What can we learn from homo silicus?
\newblock Technical report, National Bureau of Economic Research, 2023.

\bibitem{huang2023survey}
Lei Huang, Weijiang Yu, Weitao Ma, Weihong Zhong, Zhangyin Feng, Haotian Wang, Qianglong Chen, Weihua Peng, Xiaocheng Feng, Bing Qin, et~al.
\newblock A survey on hallucination in large language models: Principles, taxonomy, challenges, and open questions.
\newblock {\em ACM Transactions on Information Systems}, 2023.

\bibitem{huang2023benchmarking}
Qian Huang, Jian Vora, Percy Liang, and Jure Leskovec.
\newblock Benchmarking large language models as ai research agents.
\newblock In {\em NeurIPS 2023 Foundation Models for Decision Making Workshop}, 2023.

\bibitem{huang2024social}
Yue Huang, Zhengqing Yuan, Yujun Zhou, Kehan Guo, Xiangqi Wang, Haomin Zhuang, Weixiang Sun, Lichao Sun, Jindong Wang, Yanfang Ye, et~al.
\newblock Social science meets llms: How reliable are large language models in social simulations?
\newblock {\em arXiv preprint arXiv:2410.23426}, 2024.

\bibitem{jiang2024casevo}
Zexun Jiang, Yafang Shi, Maoxu Li, Hongjiang Xiao, Yunxiao Qin, Qinglan Wei, Ye~Wang, and Yuan Zhang.
\newblock Casevo: A cognitive agents and social evolution simulator.
\newblock {\em arXiv preprint arXiv:2412.19498}, 2024.

\bibitem{kavak2018big}
Hamdi Kavak, Jose~J Padilla, Christopher~J Lynch, and Saikou~Y Diallo.
\newblock Big data, agents, and machine learning: towards a data-driven agent-based modeling approach.
\newblock In {\em Proceedings of the Annual Simulation Symposium}, pages 1--12, 2018.

\bibitem{kesting2007general}
Arne Kesting, Martin Treiber, and Dirk Helbing.
\newblock General lane-changing model mobil for car-following models.
\newblock {\em Transportation Research Record}, 1999(1):86--94, 2007.

\bibitem{kosinski2024evaluating}
Michal Kosinski.
\newblock Evaluating large language models in theory of mind tasks.
\newblock {\em Proceedings of the National Academy of Sciences}, 121(45):e2405460121, 2024.

\bibitem{kwon2023efficient}
Woosuk Kwon, Zhuohan Li, Siyuan Zhuang, Ying Sheng, Lianmin Zheng, Cody~Hao Yu, Joseph Gonzalez, Hao Zhang, and Ion Stoica.
\newblock Efficient memory management for large language model serving with pagedattention.
\newblock In {\em Proceedings of the 29th Symposium on Operating Systems Principles}, pages 611--626, 2023.

\bibitem{ladyman2013complex}
James Ladyman, James Lambert, and Karoline Wiesner.
\newblock What is a complex system?
\newblock {\em European Journal for Philosophy of Science}, 3:33--67, 2013.

\bibitem{lai2024evolving}
Shiyang Lai, Yujin Potter, Junsol Kim, Richard Zhuang, Dawn Song, and James Evans.
\newblock Evolving ai collectives to enhance human diversity and enable self-regulation.
\newblock {\em arXiv preprint arXiv:2402.12590}, 2024.

\bibitem{lampinen2024language}
Andrew~K Lampinen, Ishita Dasgupta, Stephanie~CY Chan, Hannah~R Sheahan, Antonia Creswell, Dharshan Kumaran, James~L McClelland, and Felix Hill.
\newblock Language models, like humans, show content effects on reasoning tasks.
\newblock {\em PNAS nexus}, 3(7):pgae233, 2024.

\bibitem{laver2011party}
Michael Laver and Ernest Sergenti.
\newblock {\em Party competition: An agent-based model}, volume~20.
\newblock Princeton University Press, 2011.

\bibitem{lazer2009computational}
David Lazer, Alex Pentland, Lada Adamic, Sinan Aral, Albert-L{\'a}szl{\'o} Barab{\'a}si, Devon Brewer, Nicholas Christakis, Noshir Contractor, James Fowler, Myron Gutmann, et~al.
\newblock Computational social science.
\newblock {\em Science}, 323(5915):721--723, 2009.

\bibitem{lazer2020computational}
David~MJ Lazer, Alex Pentland, Duncan~J Watts, Sinan Aral, Susan Athey, Noshir Contractor, Deen Freelon, Sandra Gonzalez-Bailon, Gary King, Helen Margetts, et~al.
\newblock Computational social science: Obstacles and opportunities.
\newblock {\em Science}, 369(6507):1060--1062, 2020.

\bibitem{li2023camel}
Guohao Li, Hasan Hammoud, Hani Itani, Dmitrii Khizbullin, and Bernard Ghanem.
\newblock Camel: Communicative agents for" mind" exploration of large language model society.
\newblock {\em Advances in Neural Information Processing Systems}, 36:51991--52008, 2023.

\bibitem{li2024econagent}
Nian Li, Chen Gao, Mingyu Li, Yong Li, and Qingmin Liao.
\newblock Econagent: large language model-empowered agents for simulating macroeconomic activities.
\newblock In {\em Proceedings of the 62nd Annual Meeting of the Association for Computational Linguistics (Volume 1: Long Papers)}, pages 15523--15536, 2024.

\bibitem{liu2024deepseek}
Aixin Liu, Bei Feng, Bing Xue, Bingxuan Wang, Bochao Wu, Chengda Lu, Chenggang Zhao, Chengqi Deng, Chenyu Zhang, Chong Ruan, et~al.
\newblock Deepseek-v3 technical report.
\newblock {\em arXiv preprint arXiv:2412.19437}, 2024.

\bibitem{liu2023training}
Ruibo Liu, Ruixin Yang, Chenyan Jia, Ge~Zhang, Denny Zhou, Andrew~M Dai, Diyi Yang, and Soroush Vosoughi.
\newblock Training socially aligned language models on simulated social interactions.
\newblock {\em arXiv preprint arXiv:2305.16960}, 2023.

\bibitem{macal2005tutorial}
Charles~M Macal and Michael~J North.
\newblock Tutorial on agent-based modeling and simulation.
\newblock In {\em Proceedings of the Winter Simulation Conference, 2005.}, pages 14--pp. IEEE, 2005.

\bibitem{maslow1943theory}
AH~Maslow.
\newblock A theory of human motivation.
\newblock {\em Psychological Review}, 2:21--28, 1943.

\bibitem{mcleod2007maslow}
Saul McLeod.
\newblock Maslow's hierarchy of needs.
\newblock {\em Simply psychology}, 1(1-18), 2007.

\bibitem{mead1934mind}
George~Herbert Mead.
\newblock {\em Mind, self, and society from the standpoint of a social behaviorist.}
\newblock Chicago, 1934.

\bibitem{moritz2018ray}
Philipp Moritz, Robert Nishihara, Stephanie Wang, Alexey Tumanov, Richard Liaw, Eric Liang, Melih Elibol, Zongheng Yang, William Paul, Michael~I Jordan, et~al.
\newblock Ray: A distributed framework for emerging $\{$AI$\}$ applications.
\newblock In {\em 13th USENIX symposium on operating systems design and implementation (OSDI 18)}, pages 561--577, 2018.

\bibitem{mou2024unveiling}
Xinyi Mou, Zhongyu Wei, and Xuanjing Huang.
\newblock Unveiling the truth and facilitating change: Towards agent-based large-scale social movement simulation.
\newblock {\em arXiv preprint arXiv:2402.16333}, 2024.

\bibitem{namatame2016agent}
Akira Namatame and Shu-Heng Chen.
\newblock {\em Agent-based modeling and network dynamics}.
\newblock Oxford University Press, 2016.

\bibitem{orru2023human}
Graziella Orr{\`u}, Andrea Piarulli, Ciro Conversano, and Angelo Gemignani.
\newblock Human-like problem-solving abilities in large language models using chatgpt.
\newblock {\em Frontiers in artificial intelligence}, 6:1199350, 2023.

\bibitem{pangself}
Xianghe Pang, Shuo Tang, Rui Ye, Yuxin Xiong, Bolun Zhang, Yanfeng Wang, and Siheng Chen.
\newblock Self-alignment of large language models via monopolylogue-based social scene simulation.
\newblock In {\em Forty-first International Conference on Machine Learning}, 2024.

\bibitem{pang2024self}
Xianghe Pang, Shuo Tang, Rui Ye, Yuxin Xiong, Bolun Zhang, Yanfeng Wang, and Siheng Chen.
\newblock Self-alignment of large language models via multi-agent social simulation.
\newblock In {\em ICLR 2024 Workshop on Large Language Model (LLM) Agents}, 2024.

\bibitem{park2023generative}
Joon~Sung Park, Joseph O'Brien, Carrie~Jun Cai, Meredith~Ringel Morris, Percy Liang, and Michael~S Bernstein.
\newblock Generative agents: Interactive simulacra of human behavior.
\newblock In {\em Proceedings of the 36th annual acm symposium on user interface software and technology}, pages 1--22, 2023.

\bibitem{park2024generative}
Joon~Sung Park, Carolyn~Q Zou, Aaron Shaw, Benjamin~Mako Hill, Carrie Cai, Meredith~Ringel Morris, Robb Willer, Percy Liang, and Michael~S Bernstein.
\newblock Generative agent simulations of 1,000 people.
\newblock {\em arXiv preprint arXiv:2411.10109}, 2024.

\bibitem{piao2023human}
Jinghua Piao, Jiazhen Liu, Fang Zhang, Jun Su, and Yong Li.
\newblock Human--ai adaptive dynamics drives the emergence of information cocoons.
\newblock {\em Nature Machine Intelligence}, 5(11):1214--1224, 2023.

\bibitem{piao2025emergence}
Jinghua Piao, Zhihong Lu, Chen Gao, Fengli Xu, Fernando~P Santos, Yong Li, and James Evans.
\newblock Emergence of human-like polarization among large language model agents.
\newblock {\em arXiv preprint arXiv:2501.05171}, 2025.

\bibitem{pynadath2011modeling}
David~V Pynadath, Mei Si, and Stacy~C Marsella.
\newblock Modeling theory of mind and cognitive appraisal with decision-theoretic agents.
\newblock {\em Social emotions in nature and artifact: emotions in human and human-computer interaction}, pages 70--87, 2011.

\bibitem{qian2024chatdev}
Chen Qian, Wei Liu, Hongzhang Liu, Nuo Chen, Yufan Dang, Jiahao Li, Cheng Yang, Weize Chen, Yusheng Su, Xin Cong, et~al.
\newblock Chatdev: Communicative agents for software development.
\newblock In {\em Proceedings of the 62nd Annual Meeting of the Association for Computational Linguistics (Volume 1: Long Papers)}, pages 15174--15186, 2024.

\bibitem{radloff1991use}
Lenore~Sawyer Radloff.
\newblock The use of the center for epidemiologic studies depression scale in adolescents and young adults.
\newblock {\em Journal of youth and adolescence}, 20(2):149--166, 1991.

\bibitem{romero2011differences}
Daniel~M Romero, Brendan Meeder, and Jon Kleinberg.
\newblock Differences in the mechanics of information diffusion across topics: idioms, political hashtags, and complex contagion on twitter.
\newblock In {\em Proceedings of the 20th international conference on World wide web}, pages 695--704, 2011.

\bibitem{ruan2023tptu}
Jingqing Ruan, Yihong Chen, Bin Zhang, Zhiwei Xu, Tianpeng Bao, Hangyu Mao, Ziyue Li, Xingyu Zeng, Rui Zhao, et~al.
\newblock Tptu: Task planning and tool usage of large language model-based ai agents.
\newblock In {\em NeurIPS 2023 Foundation Models for Decision Making Workshop}, 2023.

\bibitem{sarstedt2024using}
Marko Sarstedt, Susanne~J Adler, Lea Rau, and Bernd Schmitt.
\newblock Using large language models to generate silicon samples in consumer and marketing research: Challenges, opportunities, and guidelines.
\newblock {\em Psychology \& Marketing}, 41(6):1254--1270, 2024.

\bibitem{sawyer2005social}
RK~Sawyer.
\newblock {\em Social Emergence: Societies as Complex Systems}.
\newblock Cambridge University Press, 2005.

\bibitem{schurmann2020personalizing}
Tim Sch{\"u}rmann and Philipp Beckerle.
\newblock Personalizing human-agent interaction through cognitive models.
\newblock {\em Frontiers in Psychology}, 11:561510, 2020.

\bibitem{schwartz1977normative}
Shalom~H Schwartz.
\newblock Normative influences on altruism.
\newblock In {\em Advances in experimental social psychology}, volume~10, pages 221--279. Elsevier, 1977.

\bibitem{shao2024beyond}
Chenyang Shao, Fengli Xu, Bingbing Fan, Jingtao Ding, Yuan Yuan, Meng Wang, and Yong Li.
\newblock Beyond imitation: Generating human mobility from context-aware reasoning with large language models.
\newblock {\em arXiv preprint arXiv:2402.09836}, 2024.

\bibitem{shvo2019interdependent}
Maayan Shvo, Jakob Buhmann, and Mubbasir Kapadia.
\newblock An interdependent model of personality, motivation, emotion, and mood for intelligent virtual agents.
\newblock In {\em Proceedings of the 19th ACM international conference on intelligent virtual agents}, pages 65--72, 2019.

\bibitem{sornette2009stock}
Didier Sornette.
\newblock {\em Why stock markets crash: critical events in complex financial systems}.
\newblock Princeton university press, 2009.

\bibitem{strachan2024testing}
James~WA Strachan, Dalila Albergo, Giulia Borghini, Oriana Pansardi, Eugenio Scaliti, Saurabh Gupta, Krati Saxena, Alessandro Rufo, Stefano Panzeri, Guido Manzi, et~al.
\newblock Testing theory of mind in large language models and humans.
\newblock {\em Nature Human Behaviour}, pages 1--11, 2024.

\bibitem{sun2024random}
Seungjong Sun, Eungu Lee, Dongyan Nan, Xiangying Zhao, Wonbyung Lee, Bernard~J Jansen, and Jang~Hyun Kim.
\newblock Random silicon sampling: Simulating human sub-population opinion using a large language model based on group-level demographic information.
\newblock {\em arXiv preprint arXiv:2402.18144}, 2024.

\bibitem{tang2024gensim}
Jiakai Tang, Heyang Gao, Xuchen Pan, Lei Wang, Haoran Tan, Dawei Gao, Yushuo Chen, Xu~Chen, Yankai Lin, Yaliang Li, et~al.
\newblock Gensim: A general social simulation platform with large language model based agents.
\newblock {\em arXiv preprint arXiv:2410.04360}, 2024.

\bibitem{treiber2000congested}
Martin Treiber, Ansgar Hennecke, and Dirk Helbing.
\newblock Congested traffic states in empirical observations and microscopic simulations.
\newblock {\em Physical review E}, 62(2):1805, 2000.

\bibitem{troitzsch1996social}
Klaus~G Troitzsch.
\newblock {\em Social science microsimulation}.
\newblock Springer Science \& Business Media, 1996.

\bibitem{van2017deep}
Sander van~der Hoog.
\newblock Deep learning in (and of) agent-based models: A prospectus.
\newblock {\em arXiv preprint arXiv:1706.06302}, 2017.

\bibitem{wang2023voyager}
Guanzhi Wang, Yuqi Xie, Yunfan Jiang, Ajay Mandlekar, Chaowei Xiao, Yuke Zhu, Linxi Fan, and Anima Anandkumar.
\newblock Voyager: An open-ended embodied agent with large language models.
\newblock {\em arXiv preprint arXiv:2305.16291}, 2023.

\bibitem{wang2024survey}
Lei Wang, Chen Ma, Xueyang Feng, Zeyu Zhang, Hao Yang, Jingsen Zhang, Zhiyuan Chen, Jiakai Tang, Xu~Chen, Yankai Lin, et~al.
\newblock A survey on large language model based autonomous agents.
\newblock {\em Frontiers of Computer Science}, 18(6):186345, 2024.

\bibitem{wang2024user}
Lei Wang, Jingsen Zhang, Hao Yang, Zhi-Yuan Chen, Jiakai Tang, Zeyu Zhang, Xu~Chen, Yankai Lin, Hao Sun, Ruihua Song, et~al.
\newblock User behavior simulation with large language model-based agents for recommender systems.
\newblock {\em ACM Transactions on Information Systems}, 2024.

\bibitem{wang2024simulating}
Yiding Wang, Yuxuan Chen, Fangwei Zhong, Long Ma, and Yizhou Wang.
\newblock Simulating human-like daily activities with desire-driven autonomy.
\newblock {\em arXiv preprint arXiv:2412.06435}, 2024.

\bibitem{wei2022chain}
Jason Wei, Xuezhi Wang, Dale Schuurmans, Maarten Bosma, Fei Xia, Ed~Chi, Quoc~V Le, Denny Zhou, et~al.
\newblock Chain-of-thought prompting elicits reasoning in large language models.
\newblock {\em Advances in neural information processing systems}, 35:24824--24837, 2022.

\bibitem{wilensky2015introduction}
U~Wilensky.
\newblock {\em An Introduction to Agent-Based Modeling: Modeling Natural, Social, and Engineered Complex Systems with Netlogo}.
\newblock The MIT Press, 2015.

\bibitem{wolf2013multi}
Sarah Wolf, Steffen F{\"u}rst, Antoine Mandel, Wiebke Lass, Daniel Lincke, Federico Pablo-Marti, and Carlo Jaeger.
\newblock A multi-agent model of several economic regions.
\newblock {\em Environmental modelling \& software}, 44:25--43, 2013.

\bibitem{wolfram1983cellular}
Stephen Wolfram.
\newblock Cellular automata.
\newblock {\em Los Alamos Science}, pages 09--01, 1983.

\bibitem{xi2023rise}
Zhiheng Xi, Wenxiang Chen, Xin Guo, Wei He, Yiwen Ding, Boyang Hong, Ming Zhang, Junzhe Wang, Senjie Jin, Enyu Zhou, et~al.
\newblock The rise and potential of large language model based agents: A survey.
\newblock {\em arXiv preprint arXiv:2309.07864}, 2023.

\bibitem{xu2023exploring}
Yuzhuang Xu, Shuo Wang, Peng Li, Fuwen Luo, Xiaolong Wang, Weidong Liu, and Yang Liu.
\newblock Exploring large language models for communication games: An empirical study on werewolf.
\newblock {\em arXiv preprint arXiv:2309.04658}, 2023.

\bibitem{yan2024opencity}
Yuwei Yan, Qingbin Zeng, Zhiheng Zheng, Jingzhe Yuan, Jie Feng, Jun Zhang, Fengli Xu, and Yong Li.
\newblock Opencity: A scalable platform to simulate urban activities with massive llm agents.
\newblock {\em arXiv preprint arXiv:2410.21286}, 2024.

\bibitem{yang2024oasis}
Ziyi Yang, Zaibin Zhang, Zirui Zheng, Yuxian Jiang, Ziyue Gan, Zhiyu Wang, Zijian Ling, Jinsong Chen, Martz Ma, Bowen Dong, et~al.
\newblock Oasis: Open agents social interaction simulations on one million agents.
\newblock {\em arXiv preprint arXiv:2411.11581}, 2024.

\bibitem{zhang2024generative}
An~Zhang, Yuxin Chen, Leheng Sheng, Xiang Wang, and Tat-Seng Chua.
\newblock On generative agents in recommendation.
\newblock In {\em Proceedings of the 47th international ACM SIGIR conference on research and development in Information Retrieval}, pages 1807--1817, 2024.

\bibitem{zhang2024large}
Chaoyun Zhang, Shilin He, Jiaxu Qian, Bowen Li, Liqun Li, Si~Qin, Yu~Kang, Minghua Ma, Qingwei Lin, Saravan Rajmohan, et~al.
\newblock Large language model-brained gui agents: A survey.
\newblock {\em arXiv preprint arXiv:2411.18279}, 2024.

\bibitem{zhang2019cityflow}
Huichu Zhang, Siyuan Feng, Chang Liu, Yaoyao Ding, Yichen Zhu, Zihan Zhou, Weinan Zhang, Yong Yu, Haiming Jin, and Zhenhui Li.
\newblock Cityflow: A multi-agent reinforcement learning environment for large scale city traffic scenario.
\newblock In {\em The world wide web conference}, pages 3620--3624, 2019.

\bibitem{zhang2024electionsim}
Xinnong Zhang, Jiayu Lin, Libo Sun, Weihong Qi, Yihang Yang, Yue Chen, Hanjia Lyu, Xinyi Mou, Siming Chen, Jiebo Luo, et~al.
\newblock Electionsim: Massive population election simulation powered by large language model driven agents.
\newblock {\em arXiv preprint arXiv:2410.20746}, 2024.

\bibitem{zhang2024survey}
Zeyu Zhang, Xiaohe Bo, Chen Ma, Rui Li, Xu~Chen, Quanyu Dai, Jieming Zhu, Zhenhua Dong, and Ji-Rong Wen.
\newblock A survey on the memory mechanism of large language model based agents.
\newblock {\em arXiv preprint arXiv:2404.13501}, 2024.

\bibitem{zheng2022ai}
Stephan Zheng, Alexander Trott, Sunil Srinivasa, David~C Parkes, and Richard Socher.
\newblock The ai economist: Taxation policy design via two-level deep multiagent reinforcement learning.
\newblock {\em Science advances}, 8(18):eabk2607, 2022.

\bibitem{zhou2023sotopia}
Xuhui Zhou, Hao Zhu, Leena Mathur, Ruohong Zhang, Haofei Yu, Zhengyang Qi, Louis-Philippe Morency, Yonatan Bisk, Daniel Fried, Graham Neubig, et~al.
\newblock Sotopia: Interactive evaluation for social intelligence in language agents.
\newblock {\em arXiv preprint arXiv:2310.11667}, 2023.

\bibitem{zhu2023ghost}
Xizhou Zhu, Yuntao Chen, Hao Tian, Chenxin Tao, Weijie Su, Chenyu Yang, Gao Huang, Bin Li, Lewei Lu, Xiaogang Wang, et~al.
\newblock Ghost in the minecraft: Generally capable agents for open-world environments via large language models with text-based knowledge and memory.
\newblock {\em arXiv preprint arXiv:2305.17144}, 2023.

\bibitem{zipf1946p}
George~Kingsley Zipf.
\newblock The p 1 p 2/d hypothesis: on the intercity movement of persons.
\newblock {\em American sociological review}, 11(6):677--686, 1946.

\end{thebibliography}
